\pdfoutput=1
\documentclass[reprint,aps,prl,superscriptaddress,twocolumn,10pt
]{revtex4-2}
\usepackage{graphicx}
\usepackage[svgnames,psnames]{xcolor}
\usepackage[T1]{fontenc}
\usepackage{textcomp}
\usepackage{bm}
\usepackage{amsmath,amssymb,amsthm}
\usepackage{mathtools}
\usepackage{mathrsfs}
\usepackage{comment}
\usepackage[normalem]{ulem}
\usepackage{braket}
\usepackage{siunitx}
\newcommand{\abs}[1]{\lvert #1\rvert}
\newcommand{\Abs}[1]{\left|#1\right|}
\newcommand{\rmd}{{\mathrm{d}}}
\newcommand{\iu}{{\mathrm{i}}}
\usepackage[
colorlinks=true,
citecolor=DarkGreen,
linkcolor=FireBrick,
unicode
]{hyperref}

\allowdisplaybreaks[2]
\begin{document}
\title{
\texorpdfstring{Theory of phonon angular momentum transport \\ across a smooth crystal interface}
{Theory of phonon angular momentum transport across a smooth crystal interface}
}
\author{Yuta Suzuki}
\email{suzuki.y.8cc2@m.isct.ac.jp}
\affiliation{Department of Physics, Institute of Science Tokyo, 2-12-1 Ookayama, Tokyo 152-8551, Japan {(JSPS Research Fellow)}}
\affiliation{Department of Applied Physics, The University of Tokyo, 7-3-1 Hongo, Tokyo 113-0033, Japan}
\author{Shuntaro Sumita}
\affiliation{Department of Basic Science, The University of Tokyo, 3-8-1 Komaba, Tokyo 153-8902, Japan}
\affiliation{Komaba Institute for Science, The University of Tokyo, 3-8-1 Komaba, Tokyo 153-8902, Japan}
\affiliation{Condensed Matter Theory Laboratory, RIKEN CPR, Wako, Saitama 351-0198, Japan}
\author{Yusuke Kato}
\affiliation{Department of Basic Science, The University of Tokyo, 3-8-1 Komaba, Tokyo 153-8902, Japan}
\affiliation{Quantum Research Center for Chirality, Institute for Molecular Science, Okazaki, Aichi 444-8585, Japan}
\affiliation{Department of Physics, Graduate School of Science, The University of Tokyo, 7-3-1 Hongo, Tokyo 113-0033, Japan}

\date{\today}

\begin{abstract}%
We theoretically elucidate the transfer of phonon angular momentum by acoustic modes across a smooth interface between crystals.
We analyze this process, which is difficult to describe with the conventional acoustic mismatch model, 
using a reformulated boundary condition and the Boltzmann theory.
For an interface between a chiral and an achiral crystal, our analysis reveals that thermal gradients in the chiral crystal induce angular momentum, 
which diffuses into the achiral crystal even without heat flow. 
Notably, {within an approximation retaining the chirality-induced population imbalance, }
the density of angular momentum can be enhanced near the interface. 
These findings advance our understanding of phonon transport and its interplay with electron spins.
\end{abstract}

\maketitle

\emph{{Introduction}}.---We encounter angular momentum (AM) in various phenomena, including rotating objects and magnetization. 
Its transport and conversion in solids have become as crucial as energy transfer for spintronic applications~\cite{Julliere1975,Baibich1988,Binasch1989,Miyazaki1995,Butler2001,Yuasa2004,Parkin2004}.
Recently, phonons---the quanta of lattice vibrations---have also been recognized as carriers of AM~\cite{Vonsovskii1962,Levine1962,Portigal1968,Pine1970,Ishii1975,Mclellan1988,Zhang2014,Zhang2015,Kishine2020,Chen2015,
Zhu2018,Zhang2022,AKato2023,Ishito2023a,Ishito2023b,Oishi2024,Wang2024,Tateishi2025,Juraschek2025,Ishizuka2025};
they couple with electron spins and magnons~\cite{Anastassakis1972,Rebane1983,Garanin2015,Nakane2018,Mentink2019,AKato2022,Weissenhofer2023,Shokeen2024,
Hamada2020,Ren2021,Fransson2023,Yao2024,Funato2024,Sano2024,Chaudhary2024,Li2024,Yokoyama2024,Yao2025,
Korenev2016,Holanda2018,Sasaki2021,Jeong2022,Tauchert2022,Kim2023,Ohe2024,Davies2024,Choi2024,Nabei2026}. 
So far, the generation, detection, and conversion of the phonon AM have relied on its transport across crystals with different properties~\cite{Korenev2016,Holanda2018,Sasaki2021,Jeong2022,Kim2023,Ohe2024,Davies2024,Choi2024}. 
For instance, Ohe~\textit{et al.} injected phonons from $\alpha$-quartz into a heavy metal~\cite{Ohe2024}; 
this process converted the phonon AM generated in the quartz into electron spin in the metal.
Because acoustic phonons propagate over long distances with weak attenuation and low reflection, 
their AM transport cannot be explained solely by interfacial exchange with electrons~\cite{Funato2024,Nishimura2025}. 
Thus, there is a pressing need to understand the
direct transmission of phonons carrying AM across the interface.

\begin{figure}[tbp]
\centering
\includegraphics[pagebox=artbox,width=0.90\columnwidth]{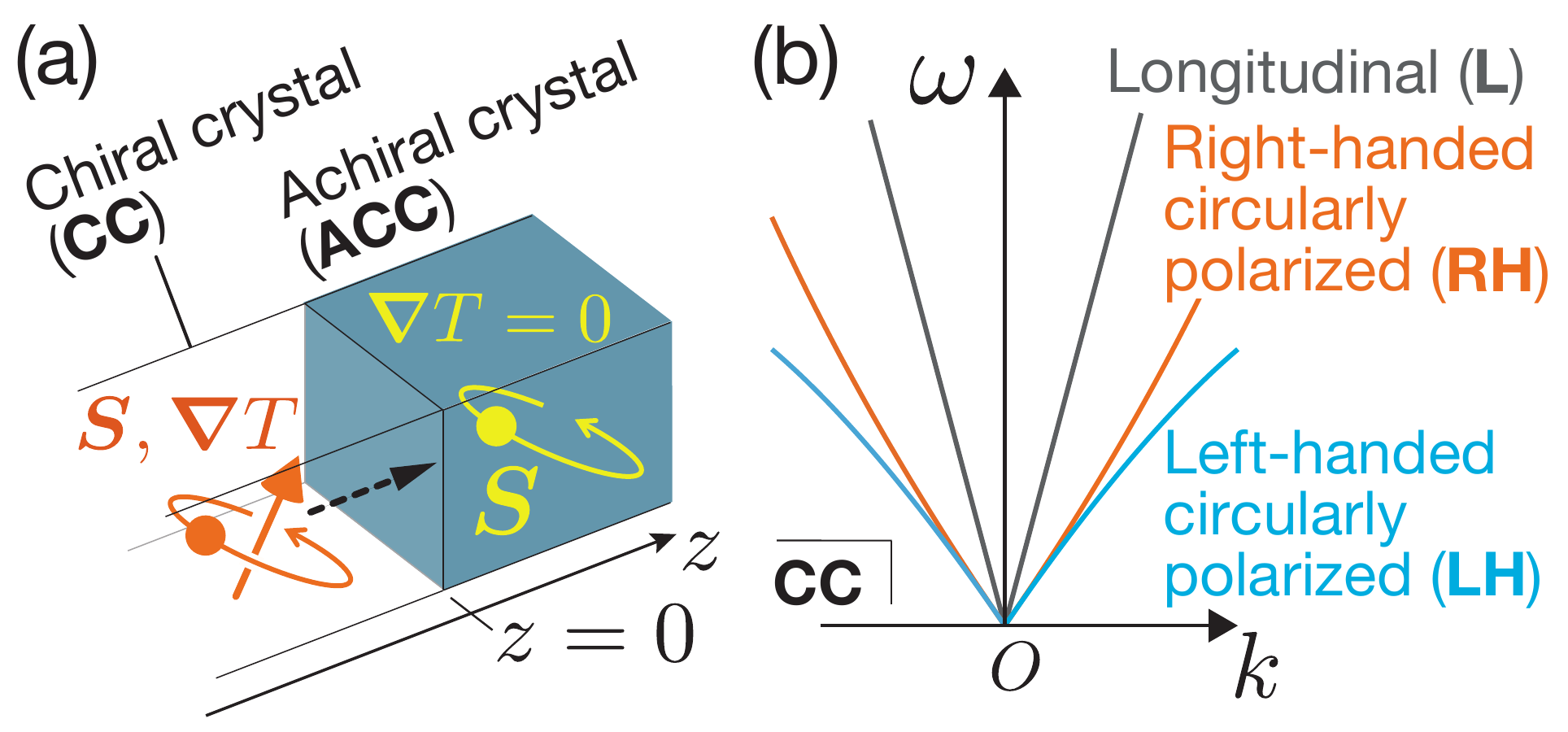}
\caption{
Schematics of a CC/ACC junction and dispersions of the CC as a source of phonon AM $\bm{S}$.
(a)~Diffusion of AM across the interface at $z = 0$. 
(b)~Acoustic modes in the CC.
The AM $\bm{S}$ is generated along the thermal gradient $\bm{\nabla} T$~[(a)] due to the lifting of degeneracy in transverse modes~[(b)].
}
\label{fig: schematic reflec transmis IF}
\end{figure}

In this Letter, we theoretically reveal the interfacial diffusion of phonon AM carried by long-wavelength acoustic modes. 
We focus on a junction between a chiral crystal (CC) and an achiral crystal (ACC). 
As shown in Fig.~\ref{fig: schematic reflec transmis IF}(a), a thermal gradient to the CC generates bulk phonon AM~\cite{Hamada2018,Oiwa2022,Zhang2025}, which is absorbed at the interface with the ACC. 
We assume a smooth interface, where phonon reflection and transmission follow Snell’s law.

{Our main findings are as follows.} 
First, despite the degeneracy of left-handed~(LH) and right-handed~(RH) circularly polarized modes in the ACC, 
finite phonon AM can diffuse into the ACC. 
This AM originates from the imbalance between LH and RH modes, 
which are naturally split in the CC [see Fig.~\ref{fig: schematic reflec transmis IF}(b)].
Second, the diffusion of AM can occur even in the absence of net heat flow across the interface. 
This finding confirms the concept of pure spin current of phonons~\cite{Park2020,Lopez2026},  
analogous to that of electrons~\cite{D'yakonov1971,Dyakonov1971current,Hirsh1999,Murakami2003,Sinova2004}.
{For an illustrative calculation, 
we retain only the population imbalance between the RH and LH modes. 
Under a thermal gradient normal to the interface, the AM density on the ACC side can then be enhanced near the interface within this approximation.}

We have analytically derived these results using the Boltzmann equation and a boundary condition based on elasticity theory.
Notably, we have extended the scope of the acoustic mismatch model~\cite{Little1959,Khalatnikov1965,Swartz1989,Chen2022b}
to cover a wide range of interfacial phonon transport. 
Although the conventional {acoustic mismatch model} is effective for simulating interfacial thermal resistance under small temperature differences, 
it struggles to describe the phonon AM transport when the thermal gradient is applied locally in a junction.
In contrast, our reformulation successfully models the AM transport in various scenarios.
We provide its detailed derivations in our companion paper~\cite{SuzukiSumitaKato2024b}.

\emph{{Chiral/achiral crystal junction}}.---We examine the transmission of phonon AM across the interface at $z=0$, 
separating the CC in $z<0$ from the ACC in $z>0$ [Fig.~\ref{fig: schematic reflec transmis IF}(a)].
We focus on low temperatures $T$ and neglect any excitations other than acoustic phonons.
We also model each crystal as an isotropic medium in the long-wavelength and low-energy limits. 
{Such an approximation, which averages out the crystalline anisotropy, is widely employed in the acoustic mismatch model for interfacial thermal resistance~\cite{Little1959,Swartz1989}. 
For the moment, we ignore the lifting of degeneracy between the transverse modes in the CC shown in Fig.~\ref{fig: schematic reflec transmis IF}(b)
and formulate the interfacial transport problem in the degenerate limit. 
The effect of the splitting will be incorporated later.}

Then, we characterize the phonon states in the CC (or ACC) by wavevectors $\bm{k}$ (or $\bm{q}$) and 
three acoustic modes: a longitudinal mode, denoted by $n = \text{L}$, and two transverse modes, collectively denoted by $n = \text{T}$.
We express the dispersion and sound velocity for each mode as
$\Omega_{\bm{k}n}= v_n k$ and $\bm{v}_{\bm{k}n} = v_n \hat{\bm{k}}$ in the CC, and 
$\omega_{\bm{q}n}= c_n q$ and $\bm{c}_{\bm{q}n} = c_n \hat{\bm{q}}$ in the ACC.
Here $k = \Abs{\bm{k}}$, $q = \Abs{\bm{q}}$, $\hat{\bm{k}} = \bm{k}/k$, and $\hat{\bm{q}} = \bm{q}/q$.

We decompose the transverse modes into RH and LH circularly polarized modes.
{Within the isotropic elastic approximation, the phonon AM has a fixed magnitude $\hbar$, and mixing with the longitudinal mode is neglected
(see Supplemental Material~\cite{SupplementalMaterialLetter} for details).
In the CC, therefore,} the AM of each mode is $\bm{S}_{\bm{k}, \text{RH}} \equiv +\hbar \hat{\bm{k}}$ and 
$\bm{S}_{\bm{k}, \text{LH}} \equiv -\hbar \hat{\bm{k}}$.
In the ACC, it is $\widetilde{\bm{S}}_{\bm{q}, \text{RH}} \equiv +\hbar \hat{\bm{q}}$ and 
$\widetilde{\bm{S}}_{\bm{q}, \text{LH}} \equiv -\hbar \hat{\bm{q}}$.
The longitudinal modes carry no AM.
The AM density in the CC is defined by~\cite{Vonsovskii1962,Levine1962,Ishii1975,Mclellan1988,Zhang2014}:
\begin{equation}
\bm{S}(z<0) \equiv \int \frac{\rmd^3 \bm{k}}{(2\pi)^3} \sum_{n = \text{RH}, \text{LH}}
\bm{S}_{\bm{k}n} \left[F_{\bm{k}n}(z) - F^{(0)}_{\bm{k}n}\right],
\label{eq: local SAM definition in z<0}
\end{equation}
where $F=F_{\bm{k}n}(z)$ is the phonon distribution function in the CC, 
and $F^{(0)}_{\bm{k}n} = 1/\left[\exp({\hbar\Omega_{\bm{k}n}/k_{\text{B}}T}) -1 \right]$ is the equilibrium distribution function.
We assume steady states that are uniform in the direction parallel to the interface, and omitted the $x, y$ and time $t$ dependence of $F$.
We also introduce AM flux in the CC, flowing normal to the interface:
\begin{equation}
\bm{j}^{\text{S}}(z<0)\equiv \int 
\frac{\rmd^3 \bm{k}}{(2\pi)^3} \!
\sum_{n=\text{RH}, \text{LH}} \!
\bm{S}_{\bm{k}n} v^z_{\bm{k}n} \left[F_{\bm{k}n}(z) - F^{(0)}_{\bm{k}n}\right].
\label{eq: definition of SAM flu z<0} 
\end{equation}
We set the group velocities of the RH and LH modes, as introduced above, equal; $\bm{v}_{\mathrm{RH}}=\bm{v}_{\mathrm{LH}}=\bm{v}_{\mathrm{T}}$, 
with a sole exception in deriving Eq.~\eqref{eq: difference between BRH and BLH}.
Likewise, we introduce the distribution function $f=f_{\bm{q}n}(z > 0)$,
the AM density ${\bm{S}}(z>0)$, and the AM flux $\bm{j}^{\text{S}}(z>0)$ in the ACC.

\emph{{Boltzmann theory}}.---{We consider a steady state under a spatially uniform thermal gradient $\bm{\nabla} T$ in the CC. 
The gradient may be either perpendicular or parallel to the interface, and the phonon distributions $F$ and $f$ are assumed to deviate from equilibrium linearly with respect to $\bm{\nabla} T$.}
We can determine the deviations, denoted by $F^{(1)}\simeq F - F^{(0)}$ and $f^{(1)}\simeq f - F^{(0)}$, by the Boltzmann equation:
\begin{subequations}
\begin{align}
 v^z_{\bm{k}n}\frac{\partial F^{(1)}_{\bm{k}n}}{\partial z}
 + \bm{v}_{\bm{k}n}\cdot \bm{\nabla}T \frac{\partial F^{(0)}_{\bm{k}n}}{\partial T}
& = - \frac{F^{(1)}_{\bm{k}n}}{\tau},
&z& < 0, \label{eq: BTE 1}\\
 c^z_{\bm{q}n}\frac{\partial f^{(1)}_{\bm{q}n}}{\partial z} &= - \frac{f^{(1)}_{\bm{q}n}}{\widetilde{\tau}},
&z&>0, \label{eq: BTE 2}
\end{align}
\label{eq: BTE}%
\end{subequations}
where $n = \text{L}, \text{RH}, \text{LH}$.
{We describe all relaxation processes in terms of two relaxation times, $\tau$ and 
$\tilde{\tau}$.}
The solution of Eqs.~\eqref{eq: BTE 1} and \eqref{eq: BTE 2} takes the form~\cite{ZimanTextbookElPh}:
\begin{subequations}
\label{eq: B and C and D terms def}
\begin{align}
 F^{(1)}_{\bm{k}n} &= B_{\bm{k}n} + \Theta_{\text{H}}(-v^z_{\bm{k}n}) \, C_{\bm{k}n}\exp\left[z\Big/({-\tau v^z_{\bm{k}n}})\right]
 \label{eq: B and C terms in F1 },\\ 
 f^{(1)}_{\bm{q}n} &= \Theta_{\text{H}}(c^z_{\bm{q}n}) \, D_{\bm{q}n}\exp\left[-z\Big/{\widetilde{\tau} c^z_{\bm{q}n}}\right]
 \label{eq: D term in f1}  
\end{align}
with the bulk part 
$B_{\bm{k}n}= - (\tau \bm{v}_{\bm{k}n}\cdot \bm{\nabla}T) \, \partial F^{(0)}_{\bm{k}n}/\partial T$
in the CC and the Heaviside step function $\Theta_{\text{H}}$.
The amplitudes near the interface $C_{\bm{k}n}$ and $D_{\bm{q}n}$
are determined by the boundary conditions imposed later [Eqs.~\eqref{eq: bc_for amp 1} and \eqref{eq: bc_for amp 2}].
\end{subequations}

\begin{subequations} 
Multiplying both sides of Eq.~\eqref{eq: BTE 1} [or Eq.~\eqref{eq: BTE 2}] by $\bm{S}_{\bm{k}n}$ [or $\widetilde{\bm{S}}_{\bm{q}n}$]
and summing over all states $(\bm{k}, n)$ [or $(\bm{q}, n)$]
yield a balance between the {decay} of AM flux and the dissipation of AM density:
 \begin{align}
\frac{\partial \bm{j}^{\text{S}} (z)}{\partial z} &= - \frac{\bm{S}(z) - \bm{S}_{0}}{\tau},  &z& < 0,
\label{eq: constitutive SAM 1} \\
\frac{\partial \bm{j}^{\text{S}} (z)}{\partial z} &= - \frac{\bm{S}(z)}{\widetilde{\tau}}, &z& > 0. 
\label{eq: constitutive SAM 2}
 \end{align}
Here we defined the bulk AM density in the CC by
$\bm{S}_0 = \int \rmd^3\bm{k} (2\pi)^{-3} \hbar\hat{\bm{k}} \left(B_{\bm{k},\text{RH}} - B_{\bm{k},\text{LH}}\right)$.
{The time constants $\tau$ and $\widetilde{\tau}$ describe the relaxation of phonon AM due to 
anharmonicity, dephasing of circular polarization~\cite{SuzukiSumitaKato2024b}, and transfer of AM to other excitations.}
\end{subequations}

\emph{{Boundary condition}}.---The phonon distribution incident upon the interface is redistributed into scattered states under specific conditions.
Since we have assumed a smooth interface, both the frequency and the wavevector parallel to the interface are conserved during scattering.
We then express the distribution function of a state $(\bm{k},n)$ as $F_{\bm{k}n} = F_{s,n}(\omega, \bm{k}_{\parallel})$, 
where $s = k_z/\Abs{k_z}$, $\omega = \Omega_{\bm{k}n}$ is the frequency, and $\bm{k}_{\parallel} = (k_x, k_y, 0)$ is the wavevector parallel to the interface. 
Likewise, we write $f_{\bm{q}n} = f_{s,n}(\omega = \omega_{\bm{q}n}, \bm{q}_{\parallel})$ with $s = q_z/\Abs{q_z}$.
The redistribution occurs between $F_{s,n}(\omega, \bm{k}_{\parallel})$ and $f_{s,n}(\omega, \bm{q}_{\parallel})$
with the same $\omega$ and $\bm{k}_{\parallel} = \bm{q}_{\parallel}$.

Now, we provide the conditions for $F_{s, n}(\omega, \bm{k}_{\parallel})$ and $f_{s,n}(\omega, \bm{k}_{\parallel})$ at $z=\pm 0$.
At a fixed value of $\omega$ and $\bm{k}_{\parallel}$, our companion paper~\cite{SuzukiSumitaKato2024b} reveals that
\begin{subequations}
 \begin{align}
 F_{-, n}
 & = \sum_{m = \text{L}, \text{RH}, \text{LH}}\left[
 \mathcal{R}_{nm} F_{+, m}  + \mathcal{T}'_{nm} f_{-, m} \right], \label{eq: boundary condition assump 1}\\
 f_{+, n}
 & = \sum_{m = \text{L}, \text{RH}, \text{LH}}\left[
 \mathcal{T}_{nm} F_{+, m} + \mathcal{R}'_{nm} f_{-, m} \right], \label{eq: boundary condition assump 2}
 \end{align}
\label{eq: boundary condition assump}%
\end{subequations}
where $n = \text{L}, \text{RH}, \text{LH}$.
The coefficients $\mathcal{R}_{nm}(\omega, \bm{k}_{\parallel})$, 
$\mathcal{T}_{nm}(\omega, \bm{k}_{\parallel})$, $\mathcal{R}'_{nm}(\omega, \bm{k}_{\parallel})$, and $\mathcal{T}'_{nm}(\omega, \bm{k}_{\parallel})$ 
denote power reflectance and transmittance of elastic waves:
$\mathcal{R}_{nm}$ (or $\mathcal{R}'_{nm}$) is a reflectance from mode $m$ to $n$ within the CC (or ACC), while
$\mathcal{T}_{nm}$ (or $\mathcal{T}'_{nm}$) is a transmittance from mode $m$ in the CC (or ACC) to mode $n$ in the ACC (or CC).
We determine these coefficients so that both displacement and stress are continuous at the interface~\cite{LandauLifshitzTextbookVol7,SommerfeldTextbookVol2,
Knott1899,Zoeppritz1919,Ewing1957}.
{They depend only on the ratios $v_{\text{L}}/v_{\text{T}}$, $c_{\text{L}}/c_{\text{T}}$,
$c_{\text{T}}/v_{\text{T}}$, and $\zeta_{\text{T}}/Z_{\text{T}}$, where
$Z_{\text{T}}\equiv \rho v_{\text{T}}$ and $\zeta_{\text{T}} \equiv \tilde{\rho} c_{\text{T}}$ are acoustic impedances, 
and $\rho$ ($\widetilde{\rho}$) is the mass density of the CC (ACC).}
In Eqs.~\eqref{eq: boundary condition assump 1} and \eqref{eq: boundary condition assump 2}, 
the phonon distributions of reflected and transmitted states ($F_{-,n}$ and $f_{+,n}$)
are presented as linear combinations of 
those of states incident upon the interface ($F_{+,m}$ and $f_{-,m}$)~\footnote{
We can understand why the coefficients from classical elasticity appear in the phonon boundary condition. 
When the spatial extent of a wave packet is sufficiently broad, it behaves like an elastic plane wave, with energy flux being
reflected or transmitted in a similar manner.
In quantum mechanics, the energy flux is proportional to the number of phonons, i.e., the phonon distribution. 
Thus, Eqs.~\eqref{eq: boundary condition assump 1} and \eqref{eq: boundary condition assump 2} hold 
and involve the elastic reflectance and transmittance in their expressions.
See our companion paper~\cite{SuzukiSumitaKato2024b} for details.}.

{The present boundary condition can be regarded as a nonequilibrium extension of the acoustic mismatch model~\cite{Little1959,Khalatnikov1965,Swartz1989,Chen2022b}. 
Both formulations describe interfacial scattering through elastic-wave reflectance and transmittance, 
but the conventional acoustic mismatch model assumes local thermal equilibrium in each bulk crystal. 
The present formulation instead applies directly to nonequilibrium phonon distribution functions and is therefore suitable for describing phonon AM transport.}

{Using Eqs.~\eqref{eq: boundary condition assump 1} and \eqref{eq: boundary condition assump 2}, }
$C_{\bm{k}n}= C_{-,n}$ and $D_{\bm{q}n}= D_{+,n}$ in Eqs.~\eqref{eq: B and C terms in F1 } and \eqref{eq: D term in f1} are determined by
\begin{subequations}
 \begin{align}
 &B_{-,n}(\omega, \bm{k}_{\parallel}) + C_{-,n}(\omega, \bm{k}_{\parallel})\nonumber\\
&\qquad = \sum_m \mathcal{R}_{nm}(\omega, \bm{k}_{\parallel}) B_{+, m}(\omega, \bm{k}_{\parallel}), \label{eq: bc_for amp 1}\\
 &D_{+, n}(\omega, \bm{k}_{\parallel})= \sum_m \mathcal{T}_{nm}(\omega, \bm{k}_{\parallel}) B_{+, m}(\omega, \bm{k}_{\parallel}).
\label{eq: bc_for amp 2}
 \end{align}
\label{eq: bc_for amp}%
\end{subequations}
Here we consider only the deviation of the distribution functions from equilibrium, since Eqs.~\eqref{eq: boundary condition assump 1} and \eqref{eq: boundary condition assump 2} are automatically satisfied in equilibrium~\cite{SuzukiSumitaKato2024b}.

\emph{{Lifting of degeneracy}}.---Let us recall the splitting between the dispersions of RH and LH modes in the CC
[Fig.~\ref{fig: schematic reflec transmis IF}(b)].
Following studies on the splitting in non-centrosymmetric crystals~\cite{Kluge1965,Portigal1968,Joffrin1970,Pine1970,Pine1971,Kishine2020,Tsunetsugu2022}, 
we introduce a quadratic splitting term $\chi k^2$ and dispersions 
$\Omega_{\bm{k},\text{RH}} \simeq v_{\text{T}}k + {\chi k^2}/{2}$
and $\Omega_{\bm{k},\text{LH}} \simeq v_{\text{T}}k - {\chi k^2}/{2}$.
These dispersions depend only on the modulus $k$, reflecting the isotropy of the CC. 
The sign of the scalar $\chi$ depends on the handedness of the CC. 

This splitting leads to the difference in nonequilibrium distribution (see Supplemental Material~\cite{SupplementalMaterialLetter}):
 \begin{equation}
 B_{\bm{k}, \text{RH}} - B_{\bm{k}, \text{LH}}
\simeq \frac{\tau\chi \bm{k}\cdot\bm{\nabla}T}{T}
\left.\frac{\coth w - \frac{3}{2w}}{\sinh^2 w/w^2}\right|_{w=\frac{\hbar v_{\text{T}} k}{2k_{\text{B}}T}}.
\label{eq: difference between BRH and BLH}
\end{equation}
Here we assume $\abs{\chi} \, k_{\text{B}}T/\hbar v_{\text{T}}\ll v_{\text{T}}$~\footnote{
This condition is valid for $T\ll \SI{100}{K}$ in $\alpha$-quartz, 
where we used the magnitude of the splitting $\abs{\chi}/v_{\text{T}}\sim \SI{3e-10}{m}$ observed
in the direction close to the $k_z$ axis~\cite{Joffrin1970,Pine1970,Pine1971}.
}. 
{For the quantitative analysis below, we retain only the chirality-induced population imbalance between the RH and LH modes, 
while neglecting other $O(\chi)$ effects of the mode splitting.}
Effects on the group velocities and on the interfacial reflection and transmission processes are neglected, 
and the two modes are therefore treated as having a common velocity $v_{\mathrm T}$. 
{
Within this approximation, the resulting AM profiles should therefore be viewed as representative rather than quantitatively exact.}
A more general treatment retaining all first-order effects of the splitting is presented in {Eq.~(S41b) of} Supplemental Material~\cite{SupplementalMaterialLetter}.

The population imbalance yields the AM density in the bulk of the CC, i.e., phonon Edelstein effect~\cite{Hamada2018}:
\begin{equation}
\bm{S}_0 = \frac{4\pi^2}{45}\hbar \, \tau\chi \left(\frac{k_{\text{B}}T}{\hbar v_{\text{T}}}\right)^4 \frac{\bm{\nabla}T}{T}
\equiv  \alpha(T) \bm{\nabla}T.
\label{eq: bulk SAM and alpha def}
\end{equation}
The temperature dependence, $\alpha(T) \propto T^3$, stems from the $O(k^2)$ splitting between transverse modes~\footnote{
For $O(k^4)$ splitting, as suggested for tellurium~\cite{Tsunetsugu2022}, we expect $\alpha(T) \propto T^5$. 
The specific form of $\alpha(T)$, however, does not affect subsequent analysis.}.

\emph{{Diffusion of phonon AM}}.---{We now describe the diffusion of the AM generated in the CC under the thermal gradient} 
[Fig.~\ref{fig: schematic reflec transmis IF}(a)].
First, we clarify AM flux $\bm{j}^{\text{S}}(z)$. 
We remark that the reflectance and transmittance under the same $\omega$ and $\bm{k}_{\parallel}$ satisfy symmetries:
$\mathcal{R}_{\text{RH}, \text{RH}}= \mathcal{R}_{\text{LH}, \text{LH}}$, $\mathcal{T}_{\text{RH}, \text{RH}}= \mathcal{T}_{\text{LH}, \text{LH}}$,
$\mathcal{R}_{\text{RH}, \text{LH}}= \mathcal{R}_{\text{LH}, \text{RH}}$, 
$\mathcal{T}_{\text{RH}, \text{LH}}= \mathcal{T}_{\text{LH}, \text{RH}}$,
$\mathcal{R}_{\text{RH}, \text{L}}= \mathcal{R}_{\text{LH}, \text{L}}$, and
$\mathcal{T}_{\text{RH}, \text{L}}= \mathcal{T}_{\text{LH}, \text{L}}$.
Substituting these relations, the boundary conditions~\eqref{eq: bc_for amp 1} and \eqref{eq: bc_for amp 2}, 
and the imbalance~\eqref{eq: difference between BRH and BLH}
into the definition~\eqref{eq: definition of SAM flu z<0}, 
we obtain $j^{\text{S}}_i = \alpha (T) v_{\text{T}} \beta_i \, \partial T/\partial r_i$ with $i = x, y, z$.
Here we used $\alpha (T)$ in Eq.~\eqref{eq: bulk SAM and alpha def} and $\beta_i (z)$ defined by
\begin{subequations}
 \begin{align}
\beta_z &= 
\begin{cases}
\frac{3}{2} I_{3,1}\left[1 + \Delta \mathcal{R}(\theta); 1, \frac{z}{\tau v_{\text{T}}}\right] & z < 0\\
\frac{3}{2} I_{3,1}\left[\Delta \mathcal{T}(\theta); \frac{c_{\text{T}}}{v_{\text{T}}}, \frac{z}{\widetilde{\tau} c_{\text{T}}}\right] & z > 0
\end{cases}, \label{eq: set of J SAM expressions 1}\\
\beta_x &= \beta_y =
\begin{cases}
 \frac{3}{4} I_{1,0}\left[1 - \Delta \mathcal{R}(\theta); 1, \frac{z}{\tau v_{\text{T}}}\right] & z < 0\\
 \frac{3}{4} I_{1,0}\left[\Delta \mathcal{T}(\theta); \frac{c_{\text{T}}}{v_{\text{T}}}, \frac{z}{\widetilde{\tau} c_{\text{T}}}\right] & z > 0
\label{eq: why J SAM in z >0 is finite}
\end{cases}.
 \end{align}
\label{eq: set of J SAM expressions}%
\end{subequations}
See Supplemental Material~\cite{SupplementalMaterialLetter} for derivations.
We introduced
$\Delta\mathcal{R}(\theta)
\equiv \mathcal{R}_{\text{RH},\text{RH}}(\omega, \bm{k}_{\parallel}) - \mathcal{R}_{\text{LH},\text{RH}}(\omega, \bm{k}_{\parallel})$, 
$\Delta\mathcal{T}(\theta)
\equiv \mathcal{T}_{\text{RH},\text{RH}}(\omega, \bm{k}_{\parallel}) - \mathcal{T}_{\text{LH},\text{RH}}(\omega, \bm{k}_{\parallel})$
as functions of the angle $\theta$ of incidence measured from the interface normal. 
{The angle $\theta$ of incidence is given as $\sin \theta = \abs{\bm{k}_{\parallel}}/k = v_{\text{T}} \abs{\bm{k}_{\parallel}}/\omega$.}
We also defined an integral
\begin{align}
&I_{m, n}[g(\theta); \gamma, w]
\equiv \gamma^{1-n}\int^{\pi/2}_0 \rmd\theta~\sin^{4-m}\theta\cos^{m-n}\theta \nonumber\\
&\times \left[1- \gamma^2\sin^2\theta\right]^{n/2} g(\theta)
\exp\left[- \frac{\abs{w}}{\sqrt{1- \gamma^2\sin^2\theta}}\right].\label{eq: a convenient integral for AM refl and trans}
\end{align}
The AM density $\bm{S}(z)$ also follows from the continuity equation~\eqref{eq: constitutive SAM 1} and \eqref{eq: constitutive SAM 2}
(see Supplemental Material~\cite{SupplementalMaterialLetter}).

\begin{figure*}
 \centering
\includegraphics[pagebox=artbox,width=0.82\textwidth]{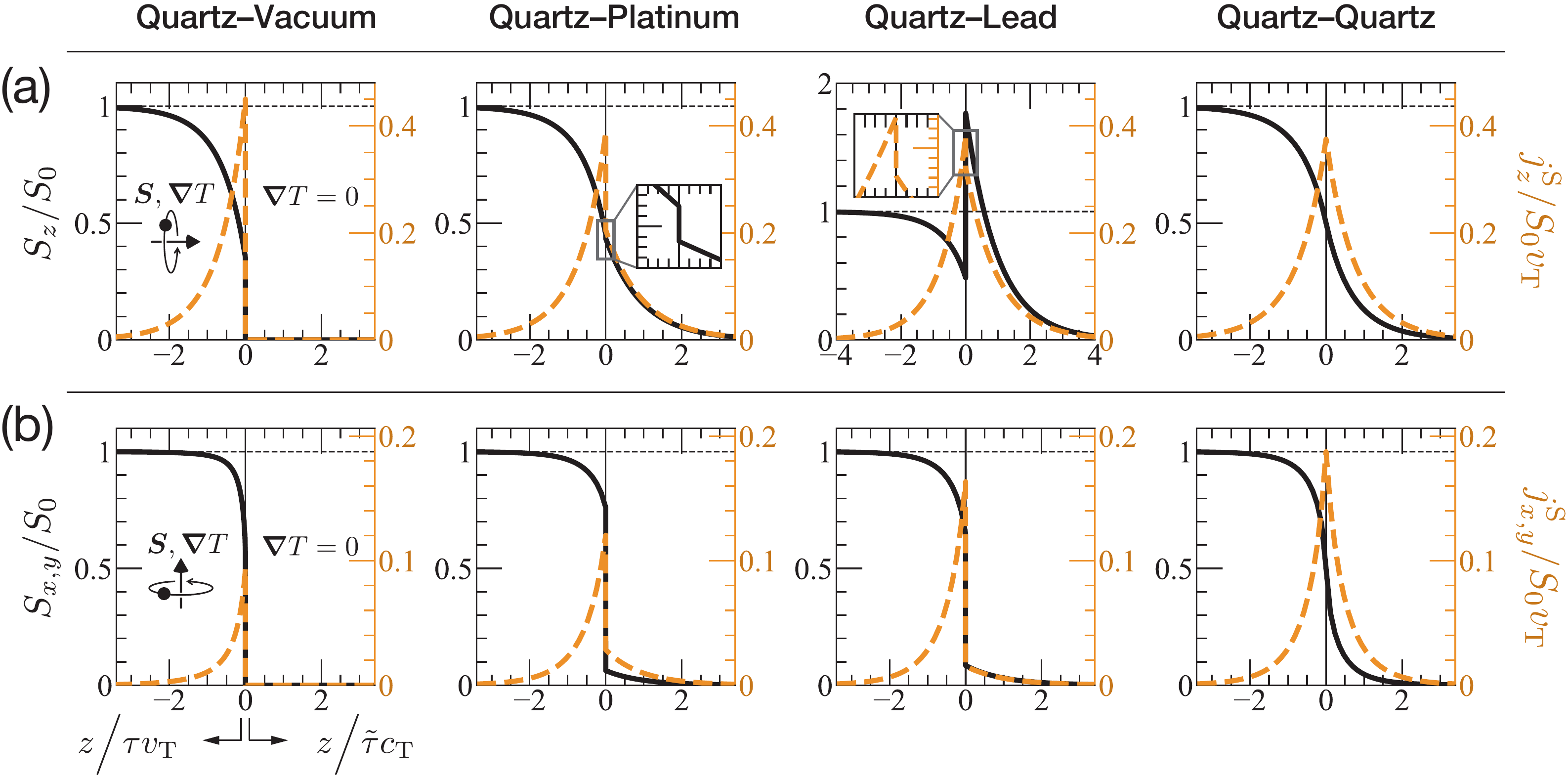}
\caption{
Diffusion of phonon-AM density $\bm{S}$ (solid curves) and its flux density $\bm{j}^{\text{S}}$ (dashed curves), 
illustrated for four different interfaces.
Each row denotes the case when both the thermal gradient $\bm{\nabla} T$ and polarization of AM are in the direction 
(a)~normal and (b)~parallel to the interface.
{The inset in the Quartz--Platinum panel of (a) enlarges the region $[-0.02, 0.02]\times [0.42, 0.46]$. 
The inset in the Quartz--Lead panel enlarges the region $[-0.2, 0.2]\times [0.3, 0.385]$.}
We set parameters as $v_{\text{L}}/v_{\text{T}} = 1.59$ and 
[Quartz--Vacuum]~$\zeta_{\text{T}}/Z_{\text{T}} = 0$,
[Quartz--Platinum]~$c_{\text{L}}/c_{\text{T}} = 2.25$,
$c_{\text{T}}/v_{\text{T}} = 0.493$, and $\zeta_{\text{T}}/Z_{\text{T}} = 3.99$,
[Quartz--Lead]~$c_{\text{L}}/c_{\text{T}} = 2.84$,
$c_{\text{T}}/v_{\text{T}} = 0.183$, and $\zeta_{\text{T}}/Z_{\text{T}} = 0.947$,
[Quartz--Quartz]~$c_{\text{L}}/v_{\text{T}} = 1.59$ and 
$c_{\text{T}}/v_{\text{T}} = \zeta_{\text{T}}/Z_{\text{T}} = 1$~\cite{Rikanempyo2024}.
{All curves are calculated within the population-imbalance approximation described in the text.}
}
\label{fig: AM spatial distribution} %
\end{figure*}

The finite coefficients $\beta_{x, y, z} (z >0 )$ in Eqs.~\eqref{eq: set of J SAM expressions 1} and \eqref{eq: why J SAM in z >0 is finite}
predict that the AM flows into the ACC, irrespective of the polarized direction of AM.
Notably, even when no heat flux flows normal to the interface, i.e., $\bm{\nabla} T \perp \hat{\bm{z}}$,
finite AM can diffuse across the interface: $\beta_{x, y} (z >0 )\neq 0$.
This pure spin current of phonons arises from the difference between the spin-conserving $\mathcal{T}_{\text{RH},\text{RH}}$
and spin-flip $\mathcal{T}_{\text{LH},\text{RH}}$ transmittances.
Their difference $\Delta \mathcal{T}$ enables the transfer of imbalance in phonon distribution between the RH and LH modes 
from the bulk of the CC into the ACC [see Eqs.~\eqref{eq: set of J SAM expressions 1} and \eqref{eq: why J SAM in z >0 is finite}].

{Note that when $\bm{\nabla} T \perp \hat{\bm z}$, a heat current along the interface may be induced in the ACC. 
This heat current, however, does not drive the AM transfer, because symmetry forbids the linear coupling between the heat current (a polar vector) and phonon AM (an axial vector) in an ACC~\cite{Hamada2018}.
The AM transfer here is also distinct from the phonon AM Hall effect~\cite{Park2020,Lopez2026}, which describes a bulk transverse transport of phonon AM. 
Instead, the present AM transfer originates from the leakage of excess phonon AM accumulated in the CC.}

{Within the population-imbalance approximation,}
we can analytically evaluate the upper limit of the AM density $\bm{S}$ and flux $\bm{j}^{\text{S}}$ diffusing into the ACC, 
which represents the maximum efficiency of AM transfer:
\begin{subequations}
 \begin{align}
\frac{\displaystyle S_i(z>0)}{\displaystyle S_{0, i}}& \leq
 \begin{cases}
 v_{\text{T}}/(2 c_{\text{T}}) & i = z\\
 1/2 & i= x, y
 \end{cases}, \label{eq: S_i upper bound}\\
 \frac{\displaystyle j^{\text{S}}_i(z>0)}{ \displaystyle S_{0, i} v_{\text{T}}} & = \beta_i \leq 
 \begin{cases}
 1/2 & i =z\\
 3/16 & i =x, y
 \end{cases}.
 \end{align}
\end{subequations}
{The upper limits follow solely from $\Delta \mathcal{T} \le 1$ (see Supplemental Material~\cite{SupplementalMaterialLetter}).}
In particular, $S_{x, y} (z= + 0)$ and $j^{\text{S}}_{x, y}(z = +0)$ are maximized
under perfect transmittance of transverse modes $\Delta \mathcal{T}(\theta) \equiv 1$. 
Equation~\eqref{eq: S_i upper bound} raises the possibility of the amplification of $S_z (z >0)$
by selecting an ACC with low sound velocity $c_{\text{T}}$ ($\ll v_{\text{T}}$).

\emph{{Examples of the interface}}.---In Figs.~\ref{fig: AM spatial distribution}(a) and \ref{fig: AM spatial distribution}(b),
we demonstrate the diffusion of AM through the junction.
We select quartz as the CC in $z<0$ and an ACC---vacuum (free end), platinum, or lead---in $z > 0$.
{For reference, we also show the case labeled Quartz--Quartz, in which a thermal gradient is applied only to the chiral quartz in $z<0$, while the achiral medium in $z>0$ is assigned the same isotropic elastic parameters as quartz.}
Notably, both $\bm{S}$ and $\bm{j}^{\text{S}}$, irrespective of its direction, diffuse across the interface, unless the ACC is vacuum.
In particular, Fig.~\ref{fig: AM spatial distribution}(b) confirms the diffusion of AM  when no heat flux flows normal to the interface.

In the quartz/lead junction with relatively large $v_{\text{T}}/c_{\text{T}}$,
the AM density $S_z$ near the interface in the ACC exceeds the bulk AM density $S_0$ in the CC.
This excess manifests the amplification of $S_z (z >0 )$, as suggested in Eq.~\eqref{eq: S_i upper bound}.

\emph{{Discontinuity in the AM flux}}.---The interface we consider is rotationally symmetric about the $z$ axis.
Thus, there is no net source or sink of AM polarized normal to the interface. 
{Besides the discontinuity in AM density, which is not constrained by the conservation law,} 
the discontinuity in AM flux at $z=0$, shown in Fig.~\ref{fig: AM spatial distribution}(a), appears to contradict the AM conservation.

We find that this contradiction is resolved by introducing the \emph{orbital} AM of phonons~\cite{Nakane2018}.
Here we define the orbital AM of a wave packet as the vector product of its center coordinate $\bm{r}$ and momentum $\hbar \bm{k}$ or $\hbar \bm{q}$,
weighted by the distribution function $F_{\bm{k}n}(\bm{r})$ or $f_{\bm{q}n}(\bm{r})$~\footnote{
We focus on extrinsic orbital AM and exclude intrinsic orbital AM arising from acoustic vortex beams~\cite{Hefner1999,Thomas2003,Ayub2011,Wang2021}.
}.
The orbital AM is independent of the \emph{spin} AM $\bm{S}$ of phonons, which arises from circular polarizations.
We stress that the sum of spin and orbital AM is conserved~\cite{Nakane2018,Bliokh2006}.
The discontinuity in spin AM flux at the interface [Fig.~\ref{fig: AM spatial distribution}(a)]
is thus compensated by that in orbital AM flux.
See our companion paper~\cite{SuzukiSumitaKato2024b} for detailed descriptions.

\emph{{Discussions and conclusions}}.---We have revealed the diffusion of accumulated phonon AM across crystals.
Notably, this diffusion occurs even without net heat flow across the interface.
We have discovered that its mechanism lies in the difference, $\Delta \mathcal{T}$, between spin-conserving and spin-flip transmittance 
of acoustic power at the interface. 
This difference enables the transfer of imbalance in the acoustic-phonon distribution of RH and LH circularly polarized modes.  
Surprisingly, even centrosymmetric crystals, which 
cannot generate the AM on their own under thermal gradients, 
can acquire finite AM through this diffusion. 
{For the quantitative analysis, we retain the chirality-induced population imbalance and neglect other $O(\chi)$ effects of the mode splitting.
Within this approximation, the AM density can be enhanced near the interface.}
{If such an enhancement is confirmed by calculations beyond the present approximation, its implications for the phonon thermal Hall effect~\cite{Oh2025} and chiral-phonon-based dark-matter detection~\cite{Romao2023,Matas2025} would merit investigation.}
{While the AM density profiles and upper-bound estimates are representative within this approximation, the general interfacial-transfer mechanism is independent of this simplification.}

In materials with strong spin--orbit coupling, phonon AM can be partially transferred to electron spins~\cite{Hamada2020,Juraschek2022,Fransson2023,Yao2024,Funato2024,Chaudhary2024,Li2024,Nishimura2025}. 
We therefore expect that its interfacial diffusion generates a spin current in heavy metals.
This expectation is consistent with the inverse spin Hall voltage observed by Ohe~\textit{et al.} in a platinum electrode adjacent to $\alpha$-quartz under a thermal gradient $\bm{\nabla} T$~\cite{Ohe2024}. 
Their experimental configuration, with $\bm{\nabla}T$ parallel to the interface, corresponds to the situation shown in Fig.~\ref{fig: AM spatial distribution}(b).  
The observed voltages therefore indicate interfacial transmission of phonon AM.  
{Related experimental situations have also been discussed in Refs.~\cite{Kim2023,Nabei2026}.}
Together with studies on interfacial spin conversion~\cite{Funato2024,Nishimura2025}, 
our findings offer a reliable explanation of the AM transport across crystal interfaces.

\begin{acknowledgments}
\emph{{Acknowledgments}}.---We wish to thank M.~Kato, H.~Matsuura, S.~Murakami, E.~Saitoh, H.~Shishido, J.~Kishine, Y.~Togawa,
and H.~Kusunose for helpful discussions.
In particular, we thank J.~Kishine for information on earlier phonon angular momentum studies, 
H.~Matsuura for communication on boundary condition,
and S.~Murakami for his comments on total angular momentum conservation at the interface.
This work was supported by JSPS KAKENHI Grants No.~JP20K03855, No.~JP21H01032, No.~JP22KJ0856, No.~JP23K03333,
No.~JP24KJ1036, and No.~JP25H02113.
This research was also supported by Joint Research by the Institute for Molecular Science (IMS program No.~23IMS1101). 
This research was also supported by the grant of OML Project by the National Institutes of Natural Sciences (NINS program No.~OML012301).
\end{acknowledgments}
\bibliography{reference}

\begin{thebibliography}{96}%
\makeatletter
\providecommand \@ifxundefined [1]{%
 \@ifx{#1\undefined}
}%
\providecommand \@ifnum [1]{%
 \ifnum #1\expandafter \@firstoftwo
 \else \expandafter \@secondoftwo
 \fi
}%
\providecommand \@ifx [1]{%
 \ifx #1\expandafter \@firstoftwo
 \else \expandafter \@secondoftwo
 \fi
}%
\providecommand \natexlab [1]{#1}%
\providecommand \enquote  [1]{``#1''}%
\providecommand \bibnamefont  [1]{#1}%
\providecommand \bibfnamefont [1]{#1}%
\providecommand \citenamefont [1]{#1}%
\providecommand \href@noop [0]{\@secondoftwo}%
\providecommand \href [0]{\begingroup \@sanitize@url \@href}%
\providecommand \@href[1]{\@@startlink{#1}\@@href}%
\providecommand \@@href[1]{\endgroup#1\@@endlink}%
\providecommand \@sanitize@url [0]{\catcode `\\12\catcode `\$12\catcode
  `\&12\catcode `\#12\catcode `\^12\catcode `\_12\catcode `\%12\relax}%
\providecommand \@@startlink[1]{}%
\providecommand \@@endlink[0]{}%
\providecommand \url  [0]{\begingroup\@sanitize@url \@url }%
\providecommand \@url [1]{\endgroup\@href {#1}{\urlprefix }}%
\providecommand \urlprefix  [0]{URL }%
\providecommand \Eprint [0]{\href }%
\providecommand \doibase [0]{https://doi.org/}%
\providecommand \selectlanguage [0]{\@gobble}%
\providecommand \bibinfo  [0]{\@secondoftwo}%
\providecommand \bibfield  [0]{\@secondoftwo}%
\providecommand \translation [1]{[#1]}%
\providecommand \BibitemOpen [0]{}%
\providecommand \bibitemStop [0]{}%
\providecommand \bibitemNoStop [0]{.\EOS\space}%
\providecommand \EOS [0]{\spacefactor3000\relax}%
\providecommand \BibitemShut  [1]{\csname bibitem#1\endcsname}%
\let\auto@bib@innerbib\@empty
\bibitem [{\citenamefont {Julliere}(1975)}]{Julliere1975}%
  \BibitemOpen
  \bibfield  {author} {\bibinfo {author} {\bibfnamefont {M.}~\bibnamefont
  {Julliere}},\ }\href {https://doi.org/10.1016/0375-9601(75)90174-7}
  {\bibfield  {journal} {\bibinfo  {journal} {Phys. Lett. A}\ }\textbf
  {\bibinfo {volume} {54}},\ \bibinfo {pages} {225} (\bibinfo {year}
  {1975})}\BibitemShut {NoStop}%
\bibitem [{\citenamefont {Baibich}\ \emph {et~al.}(1988)\citenamefont
  {Baibich}, \citenamefont {Broto}, \citenamefont {Fert}, \citenamefont
  {Van~Dau}, \citenamefont {Petroff}, \citenamefont {Etienne}, \citenamefont
  {Creuzet}, \citenamefont {Friederich},\ and\ \citenamefont
  {Chazelas}}]{Baibich1988}%
  \BibitemOpen
  \bibfield  {author} {\bibinfo {author} {\bibfnamefont {M.~N.}\ \bibnamefont
  {Baibich}}, \bibinfo {author} {\bibfnamefont {J.~M.}\ \bibnamefont {Broto}},
  \bibinfo {author} {\bibfnamefont {A.}~\bibnamefont {Fert}}, \bibinfo {author}
  {\bibfnamefont {F.~N.}\ \bibnamefont {Van~Dau}}, \bibinfo {author}
  {\bibfnamefont {F.}~\bibnamefont {Petroff}}, \bibinfo {author} {\bibfnamefont
  {P.}~\bibnamefont {Etienne}}, \bibinfo {author} {\bibfnamefont
  {G.}~\bibnamefont {Creuzet}}, \bibinfo {author} {\bibfnamefont
  {A.}~\bibnamefont {Friederich}},\ and\ \bibinfo {author} {\bibfnamefont
  {J.}~\bibnamefont {Chazelas}},\ }\href
  {https://doi.org/10.1103/PhysRevLett.61.2472} {\bibfield  {journal} {\bibinfo
   {journal} {Phys. Rev. Lett.}\ }\textbf {\bibinfo {volume} {61}},\ \bibinfo
  {pages} {2472} (\bibinfo {year} {1988})}\BibitemShut {NoStop}%
\bibitem [{\citenamefont {Binasch}\ \emph {et~al.}(1989)\citenamefont
  {Binasch}, \citenamefont {Gr{\"{u}}nberg}, \citenamefont {Saurenbach},\ and\
  \citenamefont {Zinn}}]{Binasch1989}%
  \BibitemOpen
  \bibfield  {author} {\bibinfo {author} {\bibfnamefont {G.}~\bibnamefont
  {Binasch}}, \bibinfo {author} {\bibfnamefont {P.}~\bibnamefont
  {Gr{\"{u}}nberg}}, \bibinfo {author} {\bibfnamefont {F.}~\bibnamefont
  {Saurenbach}},\ and\ \bibinfo {author} {\bibfnamefont {W.}~\bibnamefont
  {Zinn}},\ }\href {https://doi.org/10.1103/PhysRevB.39.4828} {\bibfield
  {journal} {\bibinfo  {journal} {Phys. Rev. B}\ }\textbf {\bibinfo {volume}
  {39}},\ \bibinfo {pages} {4828} (\bibinfo {year} {1989})}\BibitemShut
  {NoStop}%
\bibitem [{\citenamefont {Miyazaki}\ and\ \citenamefont
  {Tezuka}(1995)}]{Miyazaki1995}%
  \BibitemOpen
  \bibfield  {author} {\bibinfo {author} {\bibfnamefont {T.}~\bibnamefont
  {Miyazaki}}\ and\ \bibinfo {author} {\bibfnamefont {N.}~\bibnamefont
  {Tezuka}},\ }\href {https://doi.org/10.1016/0304-8853(95)90001-2} {\bibfield
  {journal} {\bibinfo  {journal} {J. Magn. Magn. Mater.}\ }\textbf {\bibinfo
  {volume} {139}},\ \bibinfo {pages} {L231} (\bibinfo {year}
  {1995})}\BibitemShut {NoStop}%
\bibitem [{\citenamefont {Butler}\ \emph {et~al.}(2001)\citenamefont {Butler},
  \citenamefont {Zhang}, \citenamefont {Schulthess},\ and\ \citenamefont
  {MacLaren}}]{Butler2001}%
  \BibitemOpen
  \bibfield  {author} {\bibinfo {author} {\bibfnamefont {W.~H.}\ \bibnamefont
  {Butler}}, \bibinfo {author} {\bibfnamefont {X.-G.}\ \bibnamefont {Zhang}},
  \bibinfo {author} {\bibfnamefont {T.~C.}\ \bibnamefont {Schulthess}},\ and\
  \bibinfo {author} {\bibfnamefont {J.~M.}\ \bibnamefont {MacLaren}},\ }\href
  {https://doi.org/10.1103/PhysRevB.63.054416} {\bibfield  {journal} {\bibinfo
  {journal} {Phys. Rev. B}\ }\textbf {\bibinfo {volume} {63}},\ \bibinfo
  {pages} {054416} (\bibinfo {year} {2001})}\BibitemShut {NoStop}%
\bibitem [{\citenamefont {Yuasa}\ \emph {et~al.}(2004)\citenamefont {Yuasa},
  \citenamefont {Nagahama}, \citenamefont {Fukushima}, \citenamefont {Suzuki},\
  and\ \citenamefont {Ando}}]{Yuasa2004}%
  \BibitemOpen
  \bibfield  {author} {\bibinfo {author} {\bibfnamefont {S.}~\bibnamefont
  {Yuasa}}, \bibinfo {author} {\bibfnamefont {T.}~\bibnamefont {Nagahama}},
  \bibinfo {author} {\bibfnamefont {A.}~\bibnamefont {Fukushima}}, \bibinfo
  {author} {\bibfnamefont {Y.}~\bibnamefont {Suzuki}},\ and\ \bibinfo {author}
  {\bibfnamefont {K.}~\bibnamefont {Ando}},\ }\href
  {https://doi.org/10.1038/nmat1257} {\bibfield  {journal} {\bibinfo  {journal}
  {Nat. Mater.}\ }\textbf {\bibinfo {volume} {3}},\ \bibinfo {pages} {868}
  (\bibinfo {year} {2004})}\BibitemShut {NoStop}%
\bibitem [{\citenamefont {Parkin}\ \emph {et~al.}(2004)\citenamefont {Parkin},
  \citenamefont {Kaiser}, \citenamefont {Panchula}, \citenamefont {Rice},
  \citenamefont {Hughes}, \citenamefont {Samant},\ and\ \citenamefont
  {Yang}}]{Parkin2004}%
  \BibitemOpen
  \bibfield  {author} {\bibinfo {author} {\bibfnamefont {S.~S.~P.}\
  \bibnamefont {Parkin}}, \bibinfo {author} {\bibfnamefont {C.}~\bibnamefont
  {Kaiser}}, \bibinfo {author} {\bibfnamefont {A.}~\bibnamefont {Panchula}},
  \bibinfo {author} {\bibfnamefont {P.~M.}\ \bibnamefont {Rice}}, \bibinfo
  {author} {\bibfnamefont {B.}~\bibnamefont {Hughes}}, \bibinfo {author}
  {\bibfnamefont {M.}~\bibnamefont {Samant}},\ and\ \bibinfo {author}
  {\bibfnamefont {S.-H.}\ \bibnamefont {Yang}},\ }\href
  {https://doi.org/10.1038/nmat1256} {\bibfield  {journal} {\bibinfo  {journal}
  {Nat. Mater.}\ }\textbf {\bibinfo {volume} {3}},\ \bibinfo {pages} {862}
  (\bibinfo {year} {2004})}\BibitemShut {NoStop}%
\bibitem [{\citenamefont {Vonsovskii}\ and\ \citenamefont
  {Svirskii}(1961)}]{Vonsovskii1962}%
  \BibitemOpen
  \bibfield  {author} {\bibinfo {author} {\bibfnamefont {S.~V.}\ \bibnamefont
  {Vonsovskii}}\ and\ \bibinfo {author} {\bibfnamefont {M.~S.}\ \bibnamefont
  {Svirskii}},\ }\href@noop {} {\bibfield  {journal} {\bibinfo  {journal} {Fiz.
  Tverd. Tela}\ }\textbf {\bibinfo {volume} {3}},\ \bibinfo {pages} {2160}
  (\bibinfo {year} {1961})},\ \bibinfo {note} {[Sov. Phys. Solid State
  \textbf{3}, 7 ({1962})]}\BibitemShut {NoStop}%
\bibitem [{\citenamefont {Levine}(1962)}]{Levine1962}%
  \BibitemOpen
  \bibfield  {author} {\bibinfo {author} {\bibfnamefont {A.~D.}\ \bibnamefont
  {Levine}},\ }\href {https://doi.org/10.1007/BF02754355} {\bibfield  {journal}
  {\bibinfo  {journal} {Nuovo Cimento}\ }\textbf {\bibinfo {volume} {26}},\
  \bibinfo {pages} {190} (\bibinfo {year} {1962})}\BibitemShut {NoStop}%
\bibitem [{\citenamefont {Portigal}\ and\ \citenamefont
  {Burstein}(1968)}]{Portigal1968}%
  \BibitemOpen
  \bibfield  {author} {\bibinfo {author} {\bibfnamefont {D.~L.}\ \bibnamefont
  {Portigal}}\ and\ \bibinfo {author} {\bibfnamefont {E.}~\bibnamefont
  {Burstein}},\ }\href {https://doi.org/10.1103/PhysRev.170.673} {\bibfield
  {journal} {\bibinfo  {journal} {Phys. Rev.}\ }\textbf {\bibinfo {volume}
  {170}},\ \bibinfo {pages} {673} (\bibinfo {year} {1968})}\BibitemShut
  {NoStop}%
\bibitem [{\citenamefont {Pine}(1970)}]{Pine1970}%
  \BibitemOpen
  \bibfield  {author} {\bibinfo {author} {\bibfnamefont {A.~S.}\ \bibnamefont
  {Pine}},\ }\href {https://doi.org/10.1103/PhysRevB.2.2049} {\bibfield
  {journal} {\bibinfo  {journal} {Phys. Rev. B}\ }\textbf {\bibinfo {volume}
  {2}},\ \bibinfo {pages} {2049} (\bibinfo {year} {1970})}\BibitemShut
  {NoStop}%
\bibitem [{\citenamefont {Ishii}(1975)}]{Ishii1975}%
  \BibitemOpen
  \bibfield  {author} {\bibinfo {author} {\bibfnamefont {T.}~\bibnamefont
  {Ishii}},\ }\href {http://hdl.handle.net/2433/88912} {\bibfield  {journal}
  {\bibinfo  {journal} {Bussei-Kenkyu}\ }\textbf {\bibinfo {volume} {23}},\
  \bibinfo {pages} {217} (\bibinfo {year} {1975})}\BibitemShut {NoStop}%
\bibitem [{\citenamefont {McLellan}(1988)}]{Mclellan1988}%
  \BibitemOpen
  \bibfield  {author} {\bibinfo {author} {\bibfnamefont {A.}~\bibnamefont
  {McLellan}},\ }\href {https://doi.org/10.1088/0022-3719/21/7/009} {\bibfield
  {journal} {\bibinfo  {journal} {J. Phys. C}\ }\textbf {\bibinfo {volume}
  {21}},\ \bibinfo {pages} {1177} (\bibinfo {year} {1988})}\BibitemShut
  {NoStop}%
\bibitem [{\citenamefont {Zhang}\ and\ \citenamefont {Niu}(2014)}]{Zhang2014}%
  \BibitemOpen
  \bibfield  {author} {\bibinfo {author} {\bibfnamefont {L.}~\bibnamefont
  {Zhang}}\ and\ \bibinfo {author} {\bibfnamefont {Q.}~\bibnamefont {Niu}},\
  }\href {https://doi.org/10.1103/PhysRevLett.112.085503} {\bibfield  {journal}
  {\bibinfo  {journal} {Phys. Rev. Lett.}\ }\textbf {\bibinfo {volume} {112}},\
  \bibinfo {pages} {085503} (\bibinfo {year} {2014})}\BibitemShut {NoStop}%
\bibitem [{\citenamefont {Zhang}\ and\ \citenamefont {Niu}(2015)}]{Zhang2015}%
  \BibitemOpen
  \bibfield  {author} {\bibinfo {author} {\bibfnamefont {L.}~\bibnamefont
  {Zhang}}\ and\ \bibinfo {author} {\bibfnamefont {Q.}~\bibnamefont {Niu}},\
  }\href {https://doi.org/10.1103/PhysRevLett.115.115502} {\bibfield  {journal}
  {\bibinfo  {journal} {Phys. Rev. Lett.}\ }\textbf {\bibinfo {volume} {115}},\
  \bibinfo {pages} {115502} (\bibinfo {year} {2015})}\BibitemShut {NoStop}%
\bibitem [{\citenamefont {Kishine}\ \emph {et~al.}(2020)\citenamefont
  {Kishine}, \citenamefont {Ovchinnikov},\ and\ \citenamefont
  {Tereshchenko}}]{Kishine2020}%
  \BibitemOpen
  \bibfield  {author} {\bibinfo {author} {\bibfnamefont {J.}~\bibnamefont
  {Kishine}}, \bibinfo {author} {\bibfnamefont {A.~S.}\ \bibnamefont
  {Ovchinnikov}},\ and\ \bibinfo {author} {\bibfnamefont {A.~A.}\ \bibnamefont
  {Tereshchenko}},\ }\href {https://doi.org/10.1103/PhysRevLett.125.245302}
  {\bibfield  {journal} {\bibinfo  {journal} {Phys. Rev. Lett.}\ }\textbf
  {\bibinfo {volume} {125}},\ \bibinfo {pages} {245302} (\bibinfo {year}
  {2020})}\BibitemShut {NoStop}%
\bibitem [{\citenamefont {Chen}\ \emph {et~al.}(2015)\citenamefont {Chen},
  \citenamefont {Zheng}, \citenamefont {Fuhrer},\ and\ \citenamefont
  {Yan}}]{Chen2015}%
  \BibitemOpen
  \bibfield  {author} {\bibinfo {author} {\bibfnamefont {S.-Y.}\ \bibnamefont
  {Chen}}, \bibinfo {author} {\bibfnamefont {C.}~\bibnamefont {Zheng}},
  \bibinfo {author} {\bibfnamefont {M.~S.}\ \bibnamefont {Fuhrer}},\ and\
  \bibinfo {author} {\bibfnamefont {J.}~\bibnamefont {Yan}},\ }\href
  {https://doi.org/10.1021/acs.nanolett.5b00092} {\bibfield  {journal}
  {\bibinfo  {journal} {Nano Lett.}\ }\textbf {\bibinfo {volume} {15}},\
  \bibinfo {pages} {2526} (\bibinfo {year} {2015})}\BibitemShut {NoStop}%
\bibitem [{\citenamefont {Zhu}\ \emph {et~al.}(2018)\citenamefont {Zhu},
  \citenamefont {Yi}, \citenamefont {Li}, \citenamefont {Xiao}, \citenamefont
  {Zhang}, \citenamefont {Yang}, \citenamefont {Kaindl}, \citenamefont {Li},
  \citenamefont {Wang},\ and\ \citenamefont {Zhang}}]{Zhu2018}%
  \BibitemOpen
  \bibfield  {author} {\bibinfo {author} {\bibfnamefont {H.}~\bibnamefont
  {Zhu}}, \bibinfo {author} {\bibfnamefont {J.}~\bibnamefont {Yi}}, \bibinfo
  {author} {\bibfnamefont {M.-Y.}\ \bibnamefont {Li}}, \bibinfo {author}
  {\bibfnamefont {J.}~\bibnamefont {Xiao}}, \bibinfo {author} {\bibfnamefont
  {L.}~\bibnamefont {Zhang}}, \bibinfo {author} {\bibfnamefont {C.-W.}\
  \bibnamefont {Yang}}, \bibinfo {author} {\bibfnamefont {R.~A.}\ \bibnamefont
  {Kaindl}}, \bibinfo {author} {\bibfnamefont {L.-J.}\ \bibnamefont {Li}},
  \bibinfo {author} {\bibfnamefont {Y.}~\bibnamefont {Wang}},\ and\ \bibinfo
  {author} {\bibfnamefont {X.}~\bibnamefont {Zhang}},\ }\href
  {https://doi.org/10.1126/science.aar2711} {\bibfield  {journal} {\bibinfo
  {journal} {Science}\ }\textbf {\bibinfo {volume} {359}},\ \bibinfo {pages}
  {579} (\bibinfo {year} {2018})}\BibitemShut {NoStop}%
\bibitem [{\citenamefont {Zhang}\ and\ \citenamefont
  {Murakami}(2022)}]{Zhang2022}%
  \BibitemOpen
  \bibfield  {author} {\bibinfo {author} {\bibfnamefont {T.}~\bibnamefont
  {Zhang}}\ and\ \bibinfo {author} {\bibfnamefont {S.}~\bibnamefont
  {Murakami}},\ }\href {https://doi.org/10.1103/PhysRevResearch.4.L012024}
  {\bibfield  {journal} {\bibinfo  {journal} {Phys. Rev. Res.}\ }\textbf
  {\bibinfo {volume} {4}},\ \bibinfo {pages} {L012024} (\bibinfo {year}
  {2022})}\BibitemShut {NoStop}%
\bibitem [{\citenamefont {Kato}\ and\ \citenamefont
  {Kishine}(2023)}]{AKato2023}%
  \BibitemOpen
  \bibfield  {author} {\bibinfo {author} {\bibfnamefont {A.}~\bibnamefont
  {Kato}}\ and\ \bibinfo {author} {\bibfnamefont {J.-i.}\ \bibnamefont
  {Kishine}},\ }\href {https://doi.org/10.7566/JPSJ.92.075002} {\bibfield
  {journal} {\bibinfo  {journal} {J. Phys. Soc. Jpn.}\ }\textbf {\bibinfo
  {volume} {92}},\ \bibinfo {pages} {075002} (\bibinfo {year}
  {2023})}\BibitemShut {NoStop}%
\bibitem [{\citenamefont {Ishito}\ \emph
  {et~al.}(2023{\natexlab{a}})\citenamefont {Ishito}, \citenamefont {Mao},
  \citenamefont {Kousaka}, \citenamefont {Togawa}, \citenamefont {Iwasaki},
  \citenamefont {Zhang}, \citenamefont {Murakami}, \citenamefont {Kishine},\
  and\ \citenamefont {Satoh}}]{Ishito2023a}%
  \BibitemOpen
  \bibfield  {author} {\bibinfo {author} {\bibfnamefont {K.}~\bibnamefont
  {Ishito}}, \bibinfo {author} {\bibfnamefont {H.}~\bibnamefont {Mao}},
  \bibinfo {author} {\bibfnamefont {Y.}~\bibnamefont {Kousaka}}, \bibinfo
  {author} {\bibfnamefont {Y.}~\bibnamefont {Togawa}}, \bibinfo {author}
  {\bibfnamefont {S.}~\bibnamefont {Iwasaki}}, \bibinfo {author} {\bibfnamefont
  {T.}~\bibnamefont {Zhang}}, \bibinfo {author} {\bibfnamefont
  {S.}~\bibnamefont {Murakami}}, \bibinfo {author} {\bibfnamefont {J.-i.}\
  \bibnamefont {Kishine}},\ and\ \bibinfo {author} {\bibfnamefont
  {T.}~\bibnamefont {Satoh}},\ }\href
  {https://doi.org/10.1038/s41567-022-01790-x} {\bibfield  {journal} {\bibinfo
  {journal} {Nat. Phys.}\ }\textbf {\bibinfo {volume} {19}},\ \bibinfo {pages}
  {35} (\bibinfo {year} {2023}{\natexlab{a}})}\BibitemShut {NoStop}%
\bibitem [{\citenamefont {Ishito}\ \emph
  {et~al.}(2023{\natexlab{b}})\citenamefont {Ishito}, \citenamefont {Mao},
  \citenamefont {Kobayashi}, \citenamefont {Kousaka}, \citenamefont {Togawa},
  \citenamefont {Kusunose}, \citenamefont {Kishine},\ and\ \citenamefont
  {Satoh}}]{Ishito2023b}%
  \BibitemOpen
  \bibfield  {author} {\bibinfo {author} {\bibfnamefont {K.}~\bibnamefont
  {Ishito}}, \bibinfo {author} {\bibfnamefont {H.}~\bibnamefont {Mao}},
  \bibinfo {author} {\bibfnamefont {K.}~\bibnamefont {Kobayashi}}, \bibinfo
  {author} {\bibfnamefont {Y.}~\bibnamefont {Kousaka}}, \bibinfo {author}
  {\bibfnamefont {Y.}~\bibnamefont {Togawa}}, \bibinfo {author} {\bibfnamefont
  {H.}~\bibnamefont {Kusunose}}, \bibinfo {author} {\bibfnamefont
  {J.}~\bibnamefont {Kishine}},\ and\ \bibinfo {author} {\bibfnamefont
  {T.}~\bibnamefont {Satoh}},\ }\href {https://doi.org/10.1002/chir.23544}
  {\bibfield  {journal} {\bibinfo  {journal} {Chirality}\ }\textbf {\bibinfo
  {volume} {35}},\ \bibinfo {pages} {338} (\bibinfo {year}
  {2023}{\natexlab{b}})}\BibitemShut {NoStop}%
\bibitem [{\citenamefont {Oishi}\ \emph {et~al.}(2024)\citenamefont {Oishi},
  \citenamefont {Fujii},\ and\ \citenamefont {Koreeda}}]{Oishi2024}%
  \BibitemOpen
  \bibfield  {author} {\bibinfo {author} {\bibfnamefont {E.}~\bibnamefont
  {Oishi}}, \bibinfo {author} {\bibfnamefont {Y.}~\bibnamefont {Fujii}},\ and\
  \bibinfo {author} {\bibfnamefont {A.}~\bibnamefont {Koreeda}},\ }\href
  {https://doi.org/10.1103/PhysRevB.109.104306} {\bibfield  {journal} {\bibinfo
   {journal} {Phys. Rev. B}\ }\textbf {\bibinfo {volume} {109}},\ \bibinfo
  {pages} {104306} (\bibinfo {year} {2024})}\BibitemShut {NoStop}%
\bibitem [{\citenamefont {Wang}\ \emph {et~al.}(2024)\citenamefont {Wang},
  \citenamefont {Sun}, \citenamefont {Li},\ and\ \citenamefont
  {Zhang}}]{Wang2024}%
  \BibitemOpen
  \bibfield  {author} {\bibinfo {author} {\bibfnamefont {T.}~\bibnamefont
  {Wang}}, \bibinfo {author} {\bibfnamefont {H.}~\bibnamefont {Sun}}, \bibinfo
  {author} {\bibfnamefont {X.}~\bibnamefont {Li}},\ and\ \bibinfo {author}
  {\bibfnamefont {L.}~\bibnamefont {Zhang}},\ }\href
  {https://doi.org/10.1021/acs.nanolett.4c00606} {\bibfield  {journal}
  {\bibinfo  {journal} {Nano Lett.}\ }\textbf {\bibinfo {volume} {24}},\
  \bibinfo {pages} {4311} (\bibinfo {year} {2024})}\BibitemShut {NoStop}%
\bibitem [{\citenamefont {Tateishi}\ \emph {et~al.}(2025)\citenamefont
  {Tateishi}, \citenamefont {Kato},\ and\ \citenamefont
  {Kishine}}]{Tateishi2025}%
  \BibitemOpen
  \bibfield  {author} {\bibinfo {author} {\bibfnamefont {T.}~\bibnamefont
  {Tateishi}}, \bibinfo {author} {\bibfnamefont {A.}~\bibnamefont {Kato}},\
  and\ \bibinfo {author} {\bibfnamefont {J.-i.}\ \bibnamefont {Kishine}},\
  }\href {https://doi.org/10.7566/JPSJ.94.053601} {\bibfield  {journal}
  {\bibinfo  {journal} {J. Phys. Soc. Jpn.}\ }\textbf {\bibinfo {volume}
  {94}},\ \bibinfo {pages} {053601} (\bibinfo {year} {2025})}\BibitemShut
  {NoStop}%
\bibitem [{\citenamefont {Juraschek}\ \emph {et~al.}(2025)\citenamefont
  {Juraschek}, \citenamefont {Geilhufe}, \citenamefont {Zhu}, \citenamefont
  {Basini}, \citenamefont {Baum}, \citenamefont {Baydin}, \citenamefont
  {Chaudhary}, \citenamefont {Fechner}, \citenamefont {Flebus}, \citenamefont
  {Grissonnanche}, \citenamefont {Kirilyuk}, \citenamefont {Lemeshko},
  \citenamefont {Maehrlein}, \citenamefont {Mignolet}, \citenamefont
  {Murakami}, \citenamefont {Niu}, \citenamefont {Nowak}, \citenamefont
  {Romao}, \citenamefont {Rostami}, \citenamefont {Satoh}, \citenamefont
  {Spaldin}, \citenamefont {Ueda},\ and\ \citenamefont
  {Zhang}}]{Juraschek2025}%
  \BibitemOpen
  \bibfield  {author} {\bibinfo {author} {\bibfnamefont {D.~M.}\ \bibnamefont
  {Juraschek}}, \bibinfo {author} {\bibfnamefont {R.~M.}\ \bibnamefont
  {Geilhufe}}, \bibinfo {author} {\bibfnamefont {H.}~\bibnamefont {Zhu}},
  \bibinfo {author} {\bibfnamefont {M.}~\bibnamefont {Basini}}, \bibinfo
  {author} {\bibfnamefont {P.}~\bibnamefont {Baum}}, \bibinfo {author}
  {\bibfnamefont {A.}~\bibnamefont {Baydin}}, \bibinfo {author} {\bibfnamefont
  {S.}~\bibnamefont {Chaudhary}}, \bibinfo {author} {\bibfnamefont
  {M.}~\bibnamefont {Fechner}}, \bibinfo {author} {\bibfnamefont
  {B.}~\bibnamefont {Flebus}}, \bibinfo {author} {\bibfnamefont
  {G.}~\bibnamefont {Grissonnanche}}, \bibinfo {author} {\bibfnamefont {A.~I.}\
  \bibnamefont {Kirilyuk}}, \bibinfo {author} {\bibfnamefont {M.}~\bibnamefont
  {Lemeshko}}, \bibinfo {author} {\bibfnamefont {S.~F.}\ \bibnamefont
  {Maehrlein}}, \bibinfo {author} {\bibfnamefont {M.}~\bibnamefont {Mignolet}},
  \bibinfo {author} {\bibfnamefont {S.}~\bibnamefont {Murakami}}, \bibinfo
  {author} {\bibfnamefont {Q.}~\bibnamefont {Niu}}, \bibinfo {author}
  {\bibfnamefont {U.}~\bibnamefont {Nowak}}, \bibinfo {author} {\bibfnamefont
  {C.~P.}\ \bibnamefont {Romao}}, \bibinfo {author} {\bibfnamefont
  {H.}~\bibnamefont {Rostami}}, \bibinfo {author} {\bibfnamefont
  {T.}~\bibnamefont {Satoh}}, \bibinfo {author} {\bibfnamefont {N.~A.}\
  \bibnamefont {Spaldin}}, \bibinfo {author} {\bibfnamefont {H.}~\bibnamefont
  {Ueda}},\ and\ \bibinfo {author} {\bibfnamefont {L.}~\bibnamefont {Zhang}},\
  }\href {https://doi.org/10.1038/s41567-025-03001-9} {\bibfield  {journal}
  {\bibinfo  {journal} {Nat. Phys.}\ }\textbf {\bibinfo {volume} {21}},\
  \bibinfo {pages} {1532} (\bibinfo {year} {2025})}\BibitemShut {NoStop}%
\bibitem [{\citenamefont {Ishizuka}\ and\ \citenamefont
  {Sato}(2025)}]{Ishizuka2025}%
  \BibitemOpen
  \bibfield  {author} {\bibinfo {author} {\bibfnamefont {H.}~\bibnamefont
  {Ishizuka}}\ and\ \bibinfo {author} {\bibfnamefont {M.}~\bibnamefont
  {Sato}},\ }\href {https://doi.org/10.48550/arXiv.2505.05313} {\bibfield
  {journal} {\bibinfo  {journal} {arXiv preprint}\ } (\bibinfo {year}
  {2025})},\ \Eprint {https://arxiv.org/abs/2505.05313} {arXiv:2505.05313
  [cond-mat.mes-hall]} \BibitemShut {NoStop}%
\bibitem [{\citenamefont {Anastassakis}\ \emph {et~al.}(1972)\citenamefont
  {Anastassakis}, \citenamefont {Burstein}, \citenamefont {Maradudin},\ and\
  \citenamefont {Minnick}}]{Anastassakis1972}%
  \BibitemOpen
  \bibfield  {author} {\bibinfo {author} {\bibfnamefont {E.}~\bibnamefont
  {Anastassakis}}, \bibinfo {author} {\bibfnamefont {E.}~\bibnamefont
  {Burstein}}, \bibinfo {author} {\bibfnamefont {A.}~\bibnamefont
  {Maradudin}},\ and\ \bibinfo {author} {\bibfnamefont {R.}~\bibnamefont
  {Minnick}},\ }\href {https://doi.org/10.1016/0022-3697(72)90034-0} {\bibfield
   {journal} {\bibinfo  {journal} {J. Phys. Chem. Solids}\ }\textbf {\bibinfo
  {volume} {33}},\ \bibinfo {pages} {519} (\bibinfo {year} {1972})}\BibitemShut
  {NoStop}%
\bibitem [{\citenamefont {Rebane}(1983)}]{Rebane1983}%
  \BibitemOpen
  \bibfield  {author} {\bibinfo {author} {\bibfnamefont {Y.~T.}\ \bibnamefont
  {Rebane}},\ }\href@noop {} {\bibfield  {journal} {\bibinfo  {journal} {Zh.
  Eksp. Teor. Fiz.}\ }\textbf {\bibinfo {volume} {84}},\ \bibinfo {pages}
  {2323} (\bibinfo {year} {1983})},\ \bibinfo {note} {[{Sov. Phys. JETP}
  \textbf{57}, {1356} (1983)]}\BibitemShut {NoStop}%
\bibitem [{\citenamefont {Garanin}\ and\ \citenamefont
  {Chudnovsky}(2015)}]{Garanin2015}%
  \BibitemOpen
  \bibfield  {author} {\bibinfo {author} {\bibfnamefont {D.~A.}\ \bibnamefont
  {Garanin}}\ and\ \bibinfo {author} {\bibfnamefont {E.~M.}\ \bibnamefont
  {Chudnovsky}},\ }\href {https://doi.org/10.1103/PhysRevB.92.024421}
  {\bibfield  {journal} {\bibinfo  {journal} {Phys. Rev. B}\ }\textbf {\bibinfo
  {volume} {92}},\ \bibinfo {pages} {024421} (\bibinfo {year}
  {2015})}\BibitemShut {NoStop}%
\bibitem [{\citenamefont {Nakane}\ and\ \citenamefont
  {Kohno}(2018)}]{Nakane2018}%
  \BibitemOpen
  \bibfield  {author} {\bibinfo {author} {\bibfnamefont {J.~J.}\ \bibnamefont
  {Nakane}}\ and\ \bibinfo {author} {\bibfnamefont {H.}~\bibnamefont {Kohno}},\
  }\href {https://doi.org/10.1103/PhysRevB.97.174403} {\bibfield  {journal}
  {\bibinfo  {journal} {Phys. Rev. B}\ }\textbf {\bibinfo {volume} {97}},\
  \bibinfo {pages} {174403} (\bibinfo {year} {2018})}\BibitemShut {NoStop}%
\bibitem [{\citenamefont {Mentink}\ \emph {et~al.}(2019)\citenamefont
  {Mentink}, \citenamefont {Katsnelson},\ and\ \citenamefont
  {Lemeshko}}]{Mentink2019}%
  \BibitemOpen
  \bibfield  {author} {\bibinfo {author} {\bibfnamefont {J.~H.}\ \bibnamefont
  {Mentink}}, \bibinfo {author} {\bibfnamefont {M.~I.}\ \bibnamefont
  {Katsnelson}},\ and\ \bibinfo {author} {\bibfnamefont {M.}~\bibnamefont
  {Lemeshko}},\ }\href {https://doi.org/10.1103/PhysRevB.99.064428} {\bibfield
  {journal} {\bibinfo  {journal} {Phys. Rev. B}\ }\textbf {\bibinfo {volume}
  {99}},\ \bibinfo {pages} {064428} (\bibinfo {year} {2019})}\BibitemShut
  {NoStop}%
\bibitem [{\citenamefont {Kato}\ \emph {et~al.}(2022)\citenamefont {Kato},
  \citenamefont {Yamamoto},\ and\ \citenamefont {Kishine}}]{AKato2022}%
  \BibitemOpen
  \bibfield  {author} {\bibinfo {author} {\bibfnamefont {A.}~\bibnamefont
  {Kato}}, \bibinfo {author} {\bibfnamefont {H.~M.}\ \bibnamefont {Yamamoto}},\
  and\ \bibinfo {author} {\bibfnamefont {J.-i.}\ \bibnamefont {Kishine}},\
  }\href {https://doi.org/10.1103/PhysRevB.105.195117} {\bibfield  {journal}
  {\bibinfo  {journal} {Phys. Rev. B}\ }\textbf {\bibinfo {volume} {105}},\
  \bibinfo {pages} {195117} (\bibinfo {year} {2022})}\BibitemShut {NoStop}%
\bibitem [{\citenamefont {Wei\ss{}enhofer}\ \emph {et~al.}(2023)\citenamefont
  {Wei\ss{}enhofer}, \citenamefont {Lange}, \citenamefont {Kamra},
  \citenamefont {Mankovsky}, \citenamefont {Polesya}, \citenamefont {Ebert},\
  and\ \citenamefont {Nowak}}]{Weissenhofer2023}%
  \BibitemOpen
  \bibfield  {author} {\bibinfo {author} {\bibfnamefont {M.}~\bibnamefont
  {Wei\ss{}enhofer}}, \bibinfo {author} {\bibfnamefont {H.}~\bibnamefont
  {Lange}}, \bibinfo {author} {\bibfnamefont {A.}~\bibnamefont {Kamra}},
  \bibinfo {author} {\bibfnamefont {S.}~\bibnamefont {Mankovsky}}, \bibinfo
  {author} {\bibfnamefont {S.}~\bibnamefont {Polesya}}, \bibinfo {author}
  {\bibfnamefont {H.}~\bibnamefont {Ebert}},\ and\ \bibinfo {author}
  {\bibfnamefont {U.}~\bibnamefont {Nowak}},\ }\href
  {https://doi.org/10.1103/PhysRevB.108.L060404} {\bibfield  {journal}
  {\bibinfo  {journal} {Phys. Rev. B}\ }\textbf {\bibinfo {volume} {108}},\
  \bibinfo {pages} {L060404} (\bibinfo {year} {2023})}\BibitemShut {NoStop}%
\bibitem [{\citenamefont {Shokeen}\ \emph {et~al.}(2024)\citenamefont
  {Shokeen}, \citenamefont {Heber}, \citenamefont {Kutnyakhov}, \citenamefont
  {Wang}, \citenamefont {Yaroslavtsev}, \citenamefont {Maldonado},
  \citenamefont {Berritta}, \citenamefont {Wind}, \citenamefont {Wenthaus},
  \citenamefont {Pressacco}, \citenamefont {Min}, \citenamefont {Nissen},
  \citenamefont {Mahatha}, \citenamefont {Dziarzhytski}, \citenamefont
  {Oppeneer}, \citenamefont {Rossnagel}, \citenamefont {Elmers}, \citenamefont
  {Sch{\"o}nhense},\ and\ \citenamefont {D{\"u}rr}}]{Shokeen2024}%
  \BibitemOpen
  \bibfield  {author} {\bibinfo {author} {\bibfnamefont {V.}~\bibnamefont
  {Shokeen}}, \bibinfo {author} {\bibfnamefont {M.}~\bibnamefont {Heber}},
  \bibinfo {author} {\bibfnamefont {D.}~\bibnamefont {Kutnyakhov}}, \bibinfo
  {author} {\bibfnamefont {X.}~\bibnamefont {Wang}}, \bibinfo {author}
  {\bibfnamefont {A.}~\bibnamefont {Yaroslavtsev}}, \bibinfo {author}
  {\bibfnamefont {P.}~\bibnamefont {Maldonado}}, \bibinfo {author}
  {\bibfnamefont {M.}~\bibnamefont {Berritta}}, \bibinfo {author}
  {\bibfnamefont {N.}~\bibnamefont {Wind}}, \bibinfo {author} {\bibfnamefont
  {L.}~\bibnamefont {Wenthaus}}, \bibinfo {author} {\bibfnamefont
  {F.}~\bibnamefont {Pressacco}}, \bibinfo {author} {\bibfnamefont {C.-H.}\
  \bibnamefont {Min}}, \bibinfo {author} {\bibfnamefont {M.}~\bibnamefont
  {Nissen}}, \bibinfo {author} {\bibfnamefont {S.~K.}\ \bibnamefont {Mahatha}},
  \bibinfo {author} {\bibfnamefont {S.}~\bibnamefont {Dziarzhytski}}, \bibinfo
  {author} {\bibfnamefont {P.~M.}\ \bibnamefont {Oppeneer}}, \bibinfo {author}
  {\bibfnamefont {K.}~\bibnamefont {Rossnagel}}, \bibinfo {author}
  {\bibfnamefont {H.-J.}\ \bibnamefont {Elmers}}, \bibinfo {author}
  {\bibfnamefont {G.}~\bibnamefont {Sch{\"o}nhense}},\ and\ \bibinfo {author}
  {\bibfnamefont {H.~A.}\ \bibnamefont {D{\"u}rr}},\ }\href
  {https://doi.org/10.1126/sciadv.adj2407} {\bibfield  {journal} {\bibinfo
  {journal} {Sci. Adv.}\ }\textbf {\bibinfo {volume} {10}},\ \bibinfo {pages}
  {eadj2407} (\bibinfo {year} {2024})}\BibitemShut {NoStop}%
\bibitem [{\citenamefont {Hamada}\ and\ \citenamefont
  {Murakami}(2020)}]{Hamada2020}%
  \BibitemOpen
  \bibfield  {author} {\bibinfo {author} {\bibfnamefont {M.}~\bibnamefont
  {Hamada}}\ and\ \bibinfo {author} {\bibfnamefont {S.}~\bibnamefont
  {Murakami}},\ }\href {https://doi.org/10.1103/PhysRevResearch.2.023275}
  {\bibfield  {journal} {\bibinfo  {journal} {Phys. Rev. Res.}\ }\textbf
  {\bibinfo {volume} {2}},\ \bibinfo {pages} {023275} (\bibinfo {year}
  {2020})}\BibitemShut {NoStop}%
\bibitem [{\citenamefont {Ren}\ \emph {et~al.}(2021)\citenamefont {Ren},
  \citenamefont {Xiao}, \citenamefont {Saparov},\ and\ \citenamefont
  {Niu}}]{Ren2021}%
  \BibitemOpen
  \bibfield  {author} {\bibinfo {author} {\bibfnamefont {Y.}~\bibnamefont
  {Ren}}, \bibinfo {author} {\bibfnamefont {C.}~\bibnamefont {Xiao}}, \bibinfo
  {author} {\bibfnamefont {D.}~\bibnamefont {Saparov}},\ and\ \bibinfo {author}
  {\bibfnamefont {Q.}~\bibnamefont {Niu}},\ }\href
  {https://doi.org/10.1103/PhysRevLett.127.186403} {\bibfield  {journal}
  {\bibinfo  {journal} {Phys. Rev. Lett.}\ }\textbf {\bibinfo {volume} {127}},\
  \bibinfo {pages} {186403} (\bibinfo {year} {2021})}\BibitemShut {NoStop}%
\bibitem [{\citenamefont {Fransson}(2023)}]{Fransson2023}%
  \BibitemOpen
  \bibfield  {author} {\bibinfo {author} {\bibfnamefont {J.}~\bibnamefont
  {Fransson}},\ }\href {https://doi.org/10.1103/PhysRevResearch.5.L022039}
  {\bibfield  {journal} {\bibinfo  {journal} {Phys. Rev. Res.}\ }\textbf
  {\bibinfo {volume} {5}},\ \bibinfo {pages} {L022039} (\bibinfo {year}
  {2023})}\BibitemShut {NoStop}%
\bibitem [{\citenamefont {Yao}\ and\ \citenamefont {Murakami}(2024)}]{Yao2024}%
  \BibitemOpen
  \bibfield  {author} {\bibinfo {author} {\bibfnamefont {D.}~\bibnamefont
  {Yao}}\ and\ \bibinfo {author} {\bibfnamefont {S.}~\bibnamefont {Murakami}},\
  }\href {https://doi.org/10.7566/JPSJ.93.034708} {\bibfield  {journal}
  {\bibinfo  {journal} {J. Phys. Soc. Jpn.}\ }\textbf {\bibinfo {volume}
  {93}},\ \bibinfo {pages} {034708} (\bibinfo {year} {2024})}\BibitemShut
  {NoStop}%
\bibitem [{\citenamefont {Funato}\ \emph {et~al.}(2024)\citenamefont {Funato},
  \citenamefont {Matsuo},\ and\ \citenamefont {Kato}}]{Funato2024}%
  \BibitemOpen
  \bibfield  {author} {\bibinfo {author} {\bibfnamefont {T.}~\bibnamefont
  {Funato}}, \bibinfo {author} {\bibfnamefont {M.}~\bibnamefont {Matsuo}},\
  and\ \bibinfo {author} {\bibfnamefont {T.}~\bibnamefont {Kato}},\ }\href
  {https://doi.org/10.1103/PhysRevLett.132.236201} {\bibfield  {journal}
  {\bibinfo  {journal} {Phys. Rev. Lett.}\ }\textbf {\bibinfo {volume} {132}},\
  \bibinfo {pages} {236201} (\bibinfo {year} {2024})}\BibitemShut {NoStop}%
\bibitem [{\citenamefont {Sano}\ and\ \citenamefont {Kato}(2024)}]{Sano2024}%
  \BibitemOpen
  \bibfield  {author} {\bibinfo {author} {\bibfnamefont {R.}~\bibnamefont
  {Sano}}\ and\ \bibinfo {author} {\bibfnamefont {T.}~\bibnamefont {Kato}},\
  }\href {https://doi.org/10.48550/arXiv.2404.19000} {\bibfield  {journal}
  {\bibinfo  {journal} {arXiv preprint}\ } (\bibinfo {year} {2024})},\ \Eprint
  {https://arxiv.org/abs/2404.19000} {arXiv:2404.19000 [cond-mat.mes-hall]}
  \BibitemShut {NoStop}%
\bibitem [{\citenamefont {Chaudhary}\ \emph {et~al.}(2024)\citenamefont
  {Chaudhary}, \citenamefont {Juraschek}, \citenamefont {Rodriguez-Vega},\ and\
  \citenamefont {Fiete}}]{Chaudhary2024}%
  \BibitemOpen
  \bibfield  {author} {\bibinfo {author} {\bibfnamefont {S.}~\bibnamefont
  {Chaudhary}}, \bibinfo {author} {\bibfnamefont {D.~M.}\ \bibnamefont
  {Juraschek}}, \bibinfo {author} {\bibfnamefont {M.}~\bibnamefont
  {Rodriguez-Vega}},\ and\ \bibinfo {author} {\bibfnamefont {G.~A.}\
  \bibnamefont {Fiete}},\ }\href {https://doi.org/10.1103/PhysRevB.110.094401}
  {\bibfield  {journal} {\bibinfo  {journal} {Phys. Rev. B}\ }\textbf {\bibinfo
  {volume} {110}},\ \bibinfo {pages} {094401} (\bibinfo {year}
  {2024})}\BibitemShut {NoStop}%
\bibitem [{\citenamefont {Li}\ \emph {et~al.}(2024)\citenamefont {Li},
  \citenamefont {Zhong}, \citenamefont {Cheng}, \citenamefont {Chen},
  \citenamefont {Wang}, \citenamefont {Liu}, \citenamefont {Sun}, \citenamefont
  {Zhang},\ and\ \citenamefont {Zhou}}]{Li2024}%
  \BibitemOpen
  \bibfield  {author} {\bibinfo {author} {\bibfnamefont {X.}~\bibnamefont
  {Li}}, \bibinfo {author} {\bibfnamefont {J.}~\bibnamefont {Zhong}}, \bibinfo
  {author} {\bibfnamefont {J.}~\bibnamefont {Cheng}}, \bibinfo {author}
  {\bibfnamefont {H.}~\bibnamefont {Chen}}, \bibinfo {author} {\bibfnamefont
  {H.}~\bibnamefont {Wang}}, \bibinfo {author} {\bibfnamefont {J.}~\bibnamefont
  {Liu}}, \bibinfo {author} {\bibfnamefont {D.}~\bibnamefont {Sun}}, \bibinfo
  {author} {\bibfnamefont {L.}~\bibnamefont {Zhang}},\ and\ \bibinfo {author}
  {\bibfnamefont {J.}~\bibnamefont {Zhou}},\ }\href
  {https://doi.org/10.1007/s11433-023-2281-x} {\bibfield  {journal} {\bibinfo
  {journal} {Sci. China Phys., Mech. Astron.}\ }\textbf {\bibinfo {volume}
  {67}},\ \bibinfo {pages} {237511} (\bibinfo {year} {2024})}\BibitemShut
  {NoStop}%
\bibitem [{\citenamefont {Yokoyama}(2024)}]{Yokoyama2024}%
  \BibitemOpen
  \bibfield  {author} {\bibinfo {author} {\bibfnamefont {T.}~\bibnamefont
  {Yokoyama}},\ }\href {https://doi.org/10.7566/JPSJ.93.123705} {\bibfield
  {journal} {\bibinfo  {journal} {J. Phys. Soc. Jpn.}\ }\textbf {\bibinfo
  {volume} {93}},\ \bibinfo {pages} {123705} (\bibinfo {year}
  {2024})}\BibitemShut {NoStop}%
\bibitem [{\citenamefont {Yao}\ and\ \citenamefont {Murakami}(2025)}]{Yao2025}%
  \BibitemOpen
  \bibfield  {author} {\bibinfo {author} {\bibfnamefont {D.}~\bibnamefont
  {Yao}}\ and\ \bibinfo {author} {\bibfnamefont {S.}~\bibnamefont {Murakami}},\
  }\href {https://doi.org/10.1103/PhysRevB.111.134414} {\bibfield  {journal}
  {\bibinfo  {journal} {Phys. Rev. B}\ }\textbf {\bibinfo {volume} {111}},\
  \bibinfo {pages} {134414} (\bibinfo {year} {2025})}\BibitemShut {NoStop}%
\bibitem [{\citenamefont {Korenev}\ \emph {et~al.}(2016)\citenamefont
  {Korenev}, \citenamefont {Salewski}, \citenamefont {Akimov}, \citenamefont
  {Sapega}, \citenamefont {Langer}, \citenamefont {Kalitukha}, \citenamefont
  {Debus}, \citenamefont {Dzhioev}, \citenamefont {Yakovlev}, \citenamefont
  {M{\"{u}}ller}, \citenamefont {Schr{\"{o}}der}, \citenamefont {H{\"{o}}vel},
  \citenamefont {Karczewski}, \citenamefont {Wiater}, \citenamefont
  {Wojtowicz}, \citenamefont {Kusrayev},\ and\ \citenamefont
  {Bayer}}]{Korenev2016}%
  \BibitemOpen
  \bibfield  {author} {\bibinfo {author} {\bibfnamefont {V.~L.}\ \bibnamefont
  {Korenev}}, \bibinfo {author} {\bibfnamefont {M.}~\bibnamefont {Salewski}},
  \bibinfo {author} {\bibfnamefont {I.~A.}\ \bibnamefont {Akimov}}, \bibinfo
  {author} {\bibfnamefont {V.~F.}\ \bibnamefont {Sapega}}, \bibinfo {author}
  {\bibfnamefont {L.}~\bibnamefont {Langer}}, \bibinfo {author} {\bibfnamefont
  {I.~V.}\ \bibnamefont {Kalitukha}}, \bibinfo {author} {\bibfnamefont
  {J.}~\bibnamefont {Debus}}, \bibinfo {author} {\bibfnamefont {R.~I.}\
  \bibnamefont {Dzhioev}}, \bibinfo {author} {\bibfnamefont {D.~R.}\
  \bibnamefont {Yakovlev}}, \bibinfo {author} {\bibfnamefont {D.}~\bibnamefont
  {M{\"{u}}ller}}, \bibinfo {author} {\bibfnamefont {C.}~\bibnamefont
  {Schr{\"{o}}der}}, \bibinfo {author} {\bibfnamefont {H.}~\bibnamefont
  {H{\"{o}}vel}}, \bibinfo {author} {\bibfnamefont {G.}~\bibnamefont
  {Karczewski}}, \bibinfo {author} {\bibfnamefont {M.}~\bibnamefont {Wiater}},
  \bibinfo {author} {\bibfnamefont {T.}~\bibnamefont {Wojtowicz}}, \bibinfo
  {author} {\bibfnamefont {Y.~G.}\ \bibnamefont {Kusrayev}},\ and\ \bibinfo
  {author} {\bibfnamefont {M.}~\bibnamefont {Bayer}},\ }\href
  {https://doi.org/10.1038/nphys3497} {\bibfield  {journal} {\bibinfo
  {journal} {Nat. Phys.}\ }\textbf {\bibinfo {volume} {12}},\ \bibinfo {pages}
  {85} (\bibinfo {year} {2016})}\BibitemShut {NoStop}%
\bibitem [{\citenamefont {Holanda}\ \emph {et~al.}(2018)\citenamefont
  {Holanda}, \citenamefont {Maior}, \citenamefont {Azevedo},\ and\
  \citenamefont {Rezende}}]{Holanda2018}%
  \BibitemOpen
  \bibfield  {author} {\bibinfo {author} {\bibfnamefont {J.}~\bibnamefont
  {Holanda}}, \bibinfo {author} {\bibfnamefont {D.}~\bibnamefont {Maior}},
  \bibinfo {author} {\bibfnamefont {A.}~\bibnamefont {Azevedo}},\ and\ \bibinfo
  {author} {\bibfnamefont {S.}~\bibnamefont {Rezende}},\ }\href
  {https://doi.org/10.1038/s41567-018-0079-y} {\bibfield  {journal} {\bibinfo
  {journal} {Nat. Phys.}\ }\textbf {\bibinfo {volume} {14}},\ \bibinfo {pages}
  {500} (\bibinfo {year} {2018})}\BibitemShut {NoStop}%
\bibitem [{\citenamefont {Sasaki}\ \emph {et~al.}(2021)\citenamefont {Sasaki},
  \citenamefont {Nii},\ and\ \citenamefont {Onose}}]{Sasaki2021}%
  \BibitemOpen
  \bibfield  {author} {\bibinfo {author} {\bibfnamefont {R.}~\bibnamefont
  {Sasaki}}, \bibinfo {author} {\bibfnamefont {Y.}~\bibnamefont {Nii}},\ and\
  \bibinfo {author} {\bibfnamefont {Y.}~\bibnamefont {Onose}},\ }\href
  {https://doi.org/10.1038/s41467-021-22728-6} {\bibfield  {journal} {\bibinfo
  {journal} {Nat. Commun.}\ }\textbf {\bibinfo {volume} {12}},\ \bibinfo
  {pages} {2599} (\bibinfo {year} {2021})}\BibitemShut {NoStop}%
\bibitem [{\citenamefont {Jeong}\ \emph {et~al.}(2022)\citenamefont {Jeong},
  \citenamefont {Kim}, \citenamefont {Seo}, \citenamefont {Park}, \citenamefont
  {Jeong}, \citenamefont {Kim}, \citenamefont {Lauter}, \citenamefont {Egami},
  \citenamefont {Han},\ and\ \citenamefont {Choi}}]{Jeong2022}%
  \BibitemOpen
  \bibfield  {author} {\bibinfo {author} {\bibfnamefont {S.~G.}\ \bibnamefont
  {Jeong}}, \bibinfo {author} {\bibfnamefont {J.}~\bibnamefont {Kim}}, \bibinfo
  {author} {\bibfnamefont {A.}~\bibnamefont {Seo}}, \bibinfo {author}
  {\bibfnamefont {S.}~\bibnamefont {Park}}, \bibinfo {author} {\bibfnamefont
  {H.~Y.}\ \bibnamefont {Jeong}}, \bibinfo {author} {\bibfnamefont {Y.-M.}\
  \bibnamefont {Kim}}, \bibinfo {author} {\bibfnamefont {V.}~\bibnamefont
  {Lauter}}, \bibinfo {author} {\bibfnamefont {T.}~\bibnamefont {Egami}},
  \bibinfo {author} {\bibfnamefont {J.~H.}\ \bibnamefont {Han}},\ and\ \bibinfo
  {author} {\bibfnamefont {W.~S.}\ \bibnamefont {Choi}},\ }\href
  {https://doi.org/10.1126/sciadv.abm4005} {\bibfield  {journal} {\bibinfo
  {journal} {Sci. Adv.}\ }\textbf {\bibinfo {volume} {8}},\ \bibinfo {pages}
  {4} (\bibinfo {year} {2022})}\BibitemShut {NoStop}%
\bibitem [{\citenamefont {Tauchert}\ \emph {et~al.}(2022)\citenamefont
  {Tauchert}, \citenamefont {Volkov}, \citenamefont {Ehberger}, \citenamefont
  {Kazenwadel}, \citenamefont {Evers}, \citenamefont {Lange}, \citenamefont
  {Donges}, \citenamefont {Book}, \citenamefont {Kreuzpaintner}, \citenamefont
  {Nowak},\ and\ \citenamefont {Baum}}]{Tauchert2022}%
  \BibitemOpen
  \bibfield  {author} {\bibinfo {author} {\bibfnamefont {S.~R.}\ \bibnamefont
  {Tauchert}}, \bibinfo {author} {\bibfnamefont {M.}~\bibnamefont {Volkov}},
  \bibinfo {author} {\bibfnamefont {D.}~\bibnamefont {Ehberger}}, \bibinfo
  {author} {\bibfnamefont {D.}~\bibnamefont {Kazenwadel}}, \bibinfo {author}
  {\bibfnamefont {M.}~\bibnamefont {Evers}}, \bibinfo {author} {\bibfnamefont
  {H.}~\bibnamefont {Lange}}, \bibinfo {author} {\bibfnamefont
  {A.}~\bibnamefont {Donges}}, \bibinfo {author} {\bibfnamefont
  {A.}~\bibnamefont {Book}}, \bibinfo {author} {\bibfnamefont {W.}~\bibnamefont
  {Kreuzpaintner}}, \bibinfo {author} {\bibfnamefont {U.}~\bibnamefont
  {Nowak}},\ and\ \bibinfo {author} {\bibfnamefont {P.}~\bibnamefont {Baum}},\
  }\href {https://doi.org/10.1038/s41586-021-04306-4} {\bibfield  {journal}
  {\bibinfo  {journal} {Nature}\ }\textbf {\bibinfo {volume} {602}},\ \bibinfo
  {pages} {73} (\bibinfo {year} {2022})}\BibitemShut {NoStop}%
\bibitem [{\citenamefont {Kim}\ \emph {et~al.}(2023)\citenamefont {Kim},
  \citenamefont {Vetter}, \citenamefont {Yan}, \citenamefont {Yang},
  \citenamefont {Wang}, \citenamefont {Sun}, \citenamefont {Yang},
  \citenamefont {Comstock}, \citenamefont {Li}, \citenamefont {Zhou},
  \citenamefont {Zhang}, \citenamefont {You}, \citenamefont {Sun},\ and\
  \citenamefont {Liu}}]{Kim2023}%
  \BibitemOpen
  \bibfield  {author} {\bibinfo {author} {\bibfnamefont {K.}~\bibnamefont
  {Kim}}, \bibinfo {author} {\bibfnamefont {E.}~\bibnamefont {Vetter}},
  \bibinfo {author} {\bibfnamefont {L.}~\bibnamefont {Yan}}, \bibinfo {author}
  {\bibfnamefont {C.}~\bibnamefont {Yang}}, \bibinfo {author} {\bibfnamefont
  {Z.}~\bibnamefont {Wang}}, \bibinfo {author} {\bibfnamefont {R.}~\bibnamefont
  {Sun}}, \bibinfo {author} {\bibfnamefont {Y.}~\bibnamefont {Yang}}, \bibinfo
  {author} {\bibfnamefont {A.~H.}\ \bibnamefont {Comstock}}, \bibinfo {author}
  {\bibfnamefont {X.}~\bibnamefont {Li}}, \bibinfo {author} {\bibfnamefont
  {J.}~\bibnamefont {Zhou}}, \bibinfo {author} {\bibfnamefont {L.}~\bibnamefont
  {Zhang}}, \bibinfo {author} {\bibfnamefont {W.}~\bibnamefont {You}}, \bibinfo
  {author} {\bibfnamefont {D.}~\bibnamefont {Sun}},\ and\ \bibinfo {author}
  {\bibfnamefont {J.}~\bibnamefont {Liu}},\ }\href
  {https://doi.org/10.1038/s41563-023-01473-9} {\bibfield  {journal} {\bibinfo
  {journal} {Nat. Mater.}\ }\textbf {\bibinfo {volume} {22}},\ \bibinfo {pages}
  {322} (\bibinfo {year} {2023})}\BibitemShut {NoStop}%
\bibitem [{\citenamefont {Ohe}\ \emph {et~al.}(2024)\citenamefont {Ohe},
  \citenamefont {Shishido}, \citenamefont {Kato}, \citenamefont {Utsumi},
  \citenamefont {Matsuura},\ and\ \citenamefont {Togawa}}]{Ohe2024}%
  \BibitemOpen
  \bibfield  {author} {\bibinfo {author} {\bibfnamefont {K.}~\bibnamefont
  {Ohe}}, \bibinfo {author} {\bibfnamefont {H.}~\bibnamefont {Shishido}},
  \bibinfo {author} {\bibfnamefont {M.}~\bibnamefont {Kato}}, \bibinfo {author}
  {\bibfnamefont {S.}~\bibnamefont {Utsumi}}, \bibinfo {author} {\bibfnamefont
  {H.}~\bibnamefont {Matsuura}},\ and\ \bibinfo {author} {\bibfnamefont
  {Y.}~\bibnamefont {Togawa}},\ }\href
  {https://doi.org/10.1103/PhysRevLett.132.056302} {\bibfield  {journal}
  {\bibinfo  {journal} {Phys. Rev. Lett.}\ }\textbf {\bibinfo {volume} {132}},\
  \bibinfo {pages} {056302} (\bibinfo {year} {2024})}\BibitemShut {NoStop}%
\bibitem [{\citenamefont {Davies}\ \emph {et~al.}(2024)\citenamefont {Davies},
  \citenamefont {Fennema}, \citenamefont {Tsukamoto}, \citenamefont
  {Razdolski}, \citenamefont {Kimel},\ and\ \citenamefont
  {Kirilyuk}}]{Davies2024}%
  \BibitemOpen
  \bibfield  {author} {\bibinfo {author} {\bibfnamefont {C.}~\bibnamefont
  {Davies}}, \bibinfo {author} {\bibfnamefont {F.}~\bibnamefont {Fennema}},
  \bibinfo {author} {\bibfnamefont {A.}~\bibnamefont {Tsukamoto}}, \bibinfo
  {author} {\bibfnamefont {I.}~\bibnamefont {Razdolski}}, \bibinfo {author}
  {\bibfnamefont {A.}~\bibnamefont {Kimel}},\ and\ \bibinfo {author}
  {\bibfnamefont {A.}~\bibnamefont {Kirilyuk}},\ }\href
  {https://doi.org/10.1038/s41586-024-07200-x} {\bibfield  {journal} {\bibinfo
  {journal} {Nature}\ }\textbf {\bibinfo {volume} {628}},\ \bibinfo {pages}
  {540} (\bibinfo {year} {2024})}\BibitemShut {NoStop}%
\bibitem [{\citenamefont {Choi}\ \emph {et~al.}(2024)\citenamefont {Choi},
  \citenamefont {Jeong}, \citenamefont {Song}, \citenamefont {Park},
  \citenamefont {Shin}, \citenamefont {Choi},\ and\ \citenamefont
  {Lee}}]{Choi2024}%
  \BibitemOpen
  \bibfield  {author} {\bibinfo {author} {\bibfnamefont {I.~H.}\ \bibnamefont
  {Choi}}, \bibinfo {author} {\bibfnamefont {S.~G.}\ \bibnamefont {Jeong}},
  \bibinfo {author} {\bibfnamefont {S.}~\bibnamefont {Song}}, \bibinfo {author}
  {\bibfnamefont {S.}~\bibnamefont {Park}}, \bibinfo {author} {\bibfnamefont
  {D.~B.}\ \bibnamefont {Shin}}, \bibinfo {author} {\bibfnamefont {W.~S.}\
  \bibnamefont {Choi}},\ and\ \bibinfo {author} {\bibfnamefont {J.~S.}\
  \bibnamefont {Lee}},\ }\href {https://doi.org/10.1038/s41565-024-01719-w}
  {\bibfield  {journal} {\bibinfo  {journal} {Nat. Nanotechnol.}\ }\textbf
  {\bibinfo {volume} {19}},\ \bibinfo {pages} {1277} (\bibinfo {year}
  {2024})}\BibitemShut {NoStop}%
\bibitem [{\citenamefont {Nabei}\ \emph {et~al.}(2026)\citenamefont {Nabei},
  \citenamefont {Yang}, \citenamefont {Sun}, \citenamefont {Jones},
  \citenamefont {Mai}, \citenamefont {Wang}, \citenamefont {Bodin},
  \citenamefont {Pandey}, \citenamefont {Wang}, \citenamefont {Xiong},
  \citenamefont {Comstock}, \citenamefont {Ewing}, \citenamefont {Bingen},
  \citenamefont {Sun}, \citenamefont {Smirnov}, \citenamefont {Zhang},
  \citenamefont {Hoffmann}, \citenamefont {Rao}, \citenamefont {Hu},
  \citenamefont {Vardeny}, \citenamefont {Yan}, \citenamefont {Li},
  \citenamefont {Zhou}, \citenamefont {Liu},\ and\ \citenamefont
  {Sun}}]{Nabei2026}%
  \BibitemOpen
  \bibfield  {author} {\bibinfo {author} {\bibfnamefont {Y.}~\bibnamefont
  {Nabei}}, \bibinfo {author} {\bibfnamefont {C.}~\bibnamefont {Yang}},
  \bibinfo {author} {\bibfnamefont {H.}~\bibnamefont {Sun}}, \bibinfo {author}
  {\bibfnamefont {H.}~\bibnamefont {Jones}}, \bibinfo {author} {\bibfnamefont
  {T.}~\bibnamefont {Mai}}, \bibinfo {author} {\bibfnamefont {T.}~\bibnamefont
  {Wang}}, \bibinfo {author} {\bibfnamefont {R.}~\bibnamefont {Bodin}},
  \bibinfo {author} {\bibfnamefont {B.}~\bibnamefont {Pandey}}, \bibinfo
  {author} {\bibfnamefont {Z.}~\bibnamefont {Wang}}, \bibinfo {author}
  {\bibfnamefont {Y.}~\bibnamefont {Xiong}}, \bibinfo {author} {\bibfnamefont
  {A.~H.}\ \bibnamefont {Comstock}}, \bibinfo {author} {\bibfnamefont
  {B.}~\bibnamefont {Ewing}}, \bibinfo {author} {\bibfnamefont
  {J.}~\bibnamefont {Bingen}}, \bibinfo {author} {\bibfnamefont
  {R.}~\bibnamefont {Sun}}, \bibinfo {author} {\bibfnamefont {D.}~\bibnamefont
  {Smirnov}}, \bibinfo {author} {\bibfnamefont {W.}~\bibnamefont {Zhang}},
  \bibinfo {author} {\bibfnamefont {A.}~\bibnamefont {Hoffmann}}, \bibinfo
  {author} {\bibfnamefont {R.}~\bibnamefont {Rao}}, \bibinfo {author}
  {\bibfnamefont {M.}~\bibnamefont {Hu}}, \bibinfo {author} {\bibfnamefont
  {Z.~V.}\ \bibnamefont {Vardeny}}, \bibinfo {author} {\bibfnamefont
  {B.}~\bibnamefont {Yan}}, \bibinfo {author} {\bibfnamefont {X.}~\bibnamefont
  {Li}}, \bibinfo {author} {\bibfnamefont {J.}~\bibnamefont {Zhou}}, \bibinfo
  {author} {\bibfnamefont {J.}~\bibnamefont {Liu}},\ and\ \bibinfo {author}
  {\bibfnamefont {D.}~\bibnamefont {Sun}},\ }\href
  {https://doi.org/10.1038/s41567-025-03134-x} {\bibfield  {journal} {\bibinfo
  {journal} {Nat. Phys.}\ }\textbf {\bibinfo {volume} {22}},\ \bibinfo {pages}
  {245} (\bibinfo {year} {2026})}\BibitemShut {NoStop}%
\bibitem [{\citenamefont {Nishimura}\ \emph {et~al.}(2025)\citenamefont
  {Nishimura}, \citenamefont {Funato}, \citenamefont {Matsuo},\ and\
  \citenamefont {Kato}}]{Nishimura2025}%
  \BibitemOpen
  \bibfield  {author} {\bibinfo {author} {\bibfnamefont {N.}~\bibnamefont
  {Nishimura}}, \bibinfo {author} {\bibfnamefont {T.}~\bibnamefont {Funato}},
  \bibinfo {author} {\bibfnamefont {M.}~\bibnamefont {Matsuo}},\ and\ \bibinfo
  {author} {\bibfnamefont {T.}~\bibnamefont {Kato}},\ }\href
  {https://doi.org/10.1016/j.jmmm.2025.173386} {\bibfield  {journal} {\bibinfo
  {journal} {J. Magn. Magn. Mater.}\ }\textbf {\bibinfo {volume} {630}},\
  \bibinfo {pages} {173386} (\bibinfo {year} {2025})}\BibitemShut {NoStop}%
\bibitem [{\citenamefont {Hamada}\ \emph {et~al.}(2018)\citenamefont {Hamada},
  \citenamefont {Minamitani}, \citenamefont {Hirayama},\ and\ \citenamefont
  {Murakami}}]{Hamada2018}%
  \BibitemOpen
  \bibfield  {author} {\bibinfo {author} {\bibfnamefont {M.}~\bibnamefont
  {Hamada}}, \bibinfo {author} {\bibfnamefont {E.}~\bibnamefont {Minamitani}},
  \bibinfo {author} {\bibfnamefont {M.}~\bibnamefont {Hirayama}},\ and\
  \bibinfo {author} {\bibfnamefont {S.}~\bibnamefont {Murakami}},\ }\href
  {https://doi.org/10.1103/PhysRevLett.121.175301} {\bibfield  {journal}
  {\bibinfo  {journal} {Phys. Rev. Lett.}\ }\textbf {\bibinfo {volume} {121}},\
  \bibinfo {pages} {175301} (\bibinfo {year} {2018})}\BibitemShut {NoStop}%
\bibitem [{\citenamefont {Oiwa}\ and\ \citenamefont
  {Kusunose}(2022)}]{Oiwa2022}%
  \BibitemOpen
  \bibfield  {author} {\bibinfo {author} {\bibfnamefont {R.}~\bibnamefont
  {Oiwa}}\ and\ \bibinfo {author} {\bibfnamefont {H.}~\bibnamefont
  {Kusunose}},\ }\href {https://doi.org/10.1103/PhysRevLett.129.116401}
  {\bibfield  {journal} {\bibinfo  {journal} {Phys. Rev. Lett.}\ }\textbf
  {\bibinfo {volume} {129}},\ \bibinfo {pages} {116401} (\bibinfo {year}
  {2022})}\BibitemShut {NoStop}%
\bibitem [{\citenamefont {Zhang}\ \emph {et~al.}(2025)\citenamefont {Zhang},
  \citenamefont {Peshcherenko}, \citenamefont {Yang}, \citenamefont {Ward},
  \citenamefont {Raghuvanshi}, \citenamefont {Lindsay}, \citenamefont {Felser},
  \citenamefont {Zhang}, \citenamefont {Yan},\ and\ \citenamefont
  {Miao}}]{Zhang2025}%
  \BibitemOpen
  \bibfield  {author} {\bibinfo {author} {\bibfnamefont {H.}~\bibnamefont
  {Zhang}}, \bibinfo {author} {\bibfnamefont {N.}~\bibnamefont {Peshcherenko}},
  \bibinfo {author} {\bibfnamefont {F.}~\bibnamefont {Yang}}, \bibinfo {author}
  {\bibfnamefont {T.}~\bibnamefont {Ward}}, \bibinfo {author} {\bibfnamefont
  {P.}~\bibnamefont {Raghuvanshi}}, \bibinfo {author} {\bibfnamefont
  {L.}~\bibnamefont {Lindsay}}, \bibinfo {author} {\bibfnamefont
  {C.}~\bibnamefont {Felser}}, \bibinfo {author} {\bibfnamefont
  {Y.}~\bibnamefont {Zhang}}, \bibinfo {author} {\bibfnamefont {J.-Q.}\
  \bibnamefont {Yan}},\ and\ \bibinfo {author} {\bibfnamefont {H.}~\bibnamefont
  {Miao}},\ }\href {https://doi.org/10.1038/s41567-025-02952-3} {\bibfield
  {journal} {\bibinfo  {journal} {Nat. Phys.}\ }\textbf {\bibinfo {volume}
  {21}},\ \bibinfo {pages} {1387} (\bibinfo {year} {2025})}\BibitemShut
  {NoStop}%
\bibitem [{\citenamefont {Park}\ and\ \citenamefont {Yang}(2020)}]{Park2020}%
  \BibitemOpen
  \bibfield  {author} {\bibinfo {author} {\bibfnamefont {S.}~\bibnamefont
  {Park}}\ and\ \bibinfo {author} {\bibfnamefont {B.-J.}\ \bibnamefont
  {Yang}},\ }\href {https://doi.org/10.1021/acs.nanolett.0c03220} {\bibfield
  {journal} {\bibinfo  {journal} {Nano Lett.}\ }\textbf {\bibinfo {volume}
  {20}},\ \bibinfo {pages} {7694} (\bibinfo {year} {2020})}\BibitemShut
  {NoStop}%
\bibitem [{\citenamefont {Lopez}\ \emph {et~al.}(2026)\citenamefont {Lopez},
  \citenamefont {Brehm},\ and\ \citenamefont {Juraschek}}]{Lopez2026}%
  \BibitemOpen
  \bibfield  {author} {\bibinfo {author} {\bibfnamefont {D.~A.~B.}\
  \bibnamefont {Lopez}}, \bibinfo {author} {\bibfnamefont {V.}~\bibnamefont
  {Brehm}},\ and\ \bibinfo {author} {\bibfnamefont {D.~M.}\ \bibnamefont
  {Juraschek}},\ }\href {https://doi.org/10.48550/arXiv.2604.01899} {\bibfield
  {journal} {\bibinfo  {journal} {arXiv preprint}\ } (\bibinfo {year}
  {2026})},\ \Eprint {https://arxiv.org/abs/2604.01899} {arXiv:2604.01899
  [cond-mat.mtrl-sci]} \BibitemShut {NoStop}%
\bibitem [{\citenamefont {D'yakonov}\ and\ \citenamefont
  {Perel'}(1971)}]{D'yakonov1971}%
  \BibitemOpen
  \bibfield  {author} {\bibinfo {author} {\bibfnamefont {M.~I.}\ \bibnamefont
  {D'yakonov}}\ and\ \bibinfo {author} {\bibfnamefont {V.~I.}\ \bibnamefont
  {Perel'}},\ }\href@noop {} {\bibfield  {journal} {\bibinfo  {journal} {ZhETF
  Pis. Red.}\ }\textbf {\bibinfo {volume} {13}},\ \bibinfo {pages} {657}
  (\bibinfo {year} {1971})},\ \bibinfo {note} {[{JETP Lett.} \textbf{13}, {467}
  (1971)]}\BibitemShut {NoStop}%
\bibitem [{\citenamefont {Dyakonov}\ and\ \citenamefont
  {Perel}(1971)}]{Dyakonov1971current}%
  \BibitemOpen
  \bibfield  {author} {\bibinfo {author} {\bibfnamefont {M.~I.}\ \bibnamefont
  {Dyakonov}}\ and\ \bibinfo {author} {\bibfnamefont {V.}~\bibnamefont
  {Perel}},\ }\href {https://doi.org/10.1016/0375-9601(71)90196-4} {\bibfield
  {journal} {\bibinfo  {journal} {Phys. Lett. A}\ }\textbf {\bibinfo {volume}
  {35}},\ \bibinfo {pages} {459} (\bibinfo {year} {1971})}\BibitemShut
  {NoStop}%
\bibitem [{\citenamefont {Hirsch}(1999)}]{Hirsh1999}%
  \BibitemOpen
  \bibfield  {author} {\bibinfo {author} {\bibfnamefont {J.~E.}\ \bibnamefont
  {Hirsch}},\ }\href {https://doi.org/10.1103/PhysRevLett.83.1834} {\bibfield
  {journal} {\bibinfo  {journal} {Phys. Rev. Lett.}\ }\textbf {\bibinfo
  {volume} {83}},\ \bibinfo {pages} {1834} (\bibinfo {year}
  {1999})}\BibitemShut {NoStop}%
\bibitem [{\citenamefont {Murakami}\ \emph {et~al.}(2003)\citenamefont
  {Murakami}, \citenamefont {Nagaosa},\ and\ \citenamefont
  {Zhang}}]{Murakami2003}%
  \BibitemOpen
  \bibfield  {author} {\bibinfo {author} {\bibfnamefont {S.}~\bibnamefont
  {Murakami}}, \bibinfo {author} {\bibfnamefont {N.}~\bibnamefont {Nagaosa}},\
  and\ \bibinfo {author} {\bibfnamefont {S.-C.}\ \bibnamefont {Zhang}},\ }\href
  {https://doi.org/10.1126/science.1087128} {\bibfield  {journal} {\bibinfo
  {journal} {Science}\ }\textbf {\bibinfo {volume} {301}},\ \bibinfo {pages}
  {1348} (\bibinfo {year} {2003})}\BibitemShut {NoStop}%
\bibitem [{\citenamefont {Sinova}\ \emph {et~al.}(2004)\citenamefont {Sinova},
  \citenamefont {Culcer}, \citenamefont {Niu}, \citenamefont {Sinitsyn},
  \citenamefont {Jungwirth},\ and\ \citenamefont {MacDonald}}]{Sinova2004}%
  \BibitemOpen
  \bibfield  {author} {\bibinfo {author} {\bibfnamefont {J.}~\bibnamefont
  {Sinova}}, \bibinfo {author} {\bibfnamefont {D.}~\bibnamefont {Culcer}},
  \bibinfo {author} {\bibfnamefont {Q.}~\bibnamefont {Niu}}, \bibinfo {author}
  {\bibfnamefont {N.~A.}\ \bibnamefont {Sinitsyn}}, \bibinfo {author}
  {\bibfnamefont {T.}~\bibnamefont {Jungwirth}},\ and\ \bibinfo {author}
  {\bibfnamefont {A.~H.}\ \bibnamefont {MacDonald}},\ }\href
  {https://doi.org/10.1103/PhysRevLett.92.126603} {\bibfield  {journal}
  {\bibinfo  {journal} {Phys. Rev. Lett.}\ }\textbf {\bibinfo {volume} {92}},\
  \bibinfo {pages} {126603} (\bibinfo {year} {2004})}\BibitemShut {NoStop}%
\bibitem [{\citenamefont {Little}(1959)}]{Little1959}%
  \BibitemOpen
  \bibfield  {author} {\bibinfo {author} {\bibfnamefont {W.~A.}\ \bibnamefont
  {Little}},\ }\href {https://doi.org/10.1139/p59-037} {\bibfield  {journal}
  {\bibinfo  {journal} {Can. J. Phys.}\ }\textbf {\bibinfo {volume} {37}},\
  \bibinfo {pages} {334} (\bibinfo {year} {1959})}\BibitemShut {NoStop}%
\bibitem [{\citenamefont {Khalatnikov}(1965)}]{Khalatnikov1965}%
  \BibitemOpen
  \bibfield  {author} {\bibinfo {author} {\bibfnamefont {I.~M.}\ \bibnamefont
  {Khalatnikov}},\ }\href@noop {} {\emph {\bibinfo {title} {An Introduction to
  the Theory of Superfluidity}}},\ Frontiers in Physics\ (\bibinfo  {publisher}
  {W.A. Benjamin},\ \bibinfo {address} {New York},\ \bibinfo {year} {1965})\
  \bibinfo {note} {translated by P.C. Hohenberg}\BibitemShut {NoStop}%
\bibitem [{\citenamefont {Swartz}\ and\ \citenamefont
  {Pohl}(1989)}]{Swartz1989}%
  \BibitemOpen
  \bibfield  {author} {\bibinfo {author} {\bibfnamefont {E.~T.}\ \bibnamefont
  {Swartz}}\ and\ \bibinfo {author} {\bibfnamefont {R.~O.}\ \bibnamefont
  {Pohl}},\ }\href {https://doi.org/10.1103/RevModPhys.61.605} {\bibfield
  {journal} {\bibinfo  {journal} {Rev. Mod. Phys.}\ }\textbf {\bibinfo {volume}
  {61}},\ \bibinfo {pages} {605} (\bibinfo {year} {1989})}\BibitemShut
  {NoStop}%
\bibitem [{\citenamefont {Chen}\ \emph {et~al.}(2022)\citenamefont {Chen},
  \citenamefont {Xu}, \citenamefont {Zhou},\ and\ \citenamefont
  {Li}}]{Chen2022b}%
  \BibitemOpen
  \bibfield  {author} {\bibinfo {author} {\bibfnamefont {J.}~\bibnamefont
  {Chen}}, \bibinfo {author} {\bibfnamefont {X.}~\bibnamefont {Xu}}, \bibinfo
  {author} {\bibfnamefont {J.}~\bibnamefont {Zhou}},\ and\ \bibinfo {author}
  {\bibfnamefont {B.}~\bibnamefont {Li}},\ }\href
  {https://doi.org/10.1103/RevModPhys.94.025002} {\bibfield  {journal}
  {\bibinfo  {journal} {Rev. Mod. Phys.}\ }\textbf {\bibinfo {volume} {94}},\
  \bibinfo {pages} {025002} (\bibinfo {year} {2022})}\BibitemShut {NoStop}%
\bibitem [{Suz()}]{SuzukiSumitaKato2024b}%
  \BibitemOpen
  \href@noop {} {}\bibinfo {note} {Y.~Suzuki, S.~Sumita, and Y.~Kato (full
  paper), jointly submitted as a companion paper.}\BibitemShut {Stop}%
\bibitem [{Sup()}]{SupplementalMaterialLetter}%
  \BibitemOpen
  \href@noop {} {}\bibinfo {note} {See Supplemental Material at [url] for
  further details, including discussions of the splitting of long-wavelength
  acoustic phonon modes and the associated angular momentum in chiral crystals,
  a derivation of interfacial phonon angular-momentum transfer retaining all
  first-order effects of the chirality-induced splitting, upper-bound estimates
  for the transmitted angular momentum, and transport across interfaces between
  chiral crystals with the same or opposite handedness.}\BibitemShut {Stop}%
\bibitem [{\citenamefont {Ziman}(1960)}]{ZimanTextbookElPh}%
  \BibitemOpen
  \bibfield  {author} {\bibinfo {author} {\bibfnamefont {J.~M.}\ \bibnamefont
  {Ziman}},\ }\href {https://doi.org/10.1093/acprof:oso/9780198507796.001.0001}
  {\emph {\bibinfo {title} {Electrons and Phonons: the Theory of Transport
  Phenomena in Solids}}},\ International Series of Monographs on Physics\
  (\bibinfo  {publisher} {Clarendon Press},\ \bibinfo {address} {Oxford, UK},\
  \bibinfo {year} {1960})\ \bibinfo {note} {{Chapter 11}}\BibitemShut {NoStop}%
\bibitem [{\citenamefont {Landau}\ \emph {et~al.}(1995)\citenamefont {Landau},
  \citenamefont {Lifshitz}, \citenamefont {Sykes}, \citenamefont {Reid},
  \citenamefont {Kosevich},\ and\ \citenamefont
  {Pitaevskii}}]{LandauLifshitzTextbookVol7}%
  \BibitemOpen
  \bibfield  {author} {\bibinfo {author} {\bibfnamefont {L.~D.}\ \bibnamefont
  {Landau}}, \bibinfo {author} {\bibfnamefont {E.~M.}\ \bibnamefont
  {Lifshitz}}, \bibinfo {author} {\bibfnamefont {J.~B.}\ \bibnamefont {Sykes}},
  \bibinfo {author} {\bibfnamefont {W.~H.}\ \bibnamefont {Reid}}, \bibinfo
  {author} {\bibfnamefont {A.~M.}\ \bibnamefont {Kosevich}},\ and\ \bibinfo
  {author} {\bibfnamefont {L.~P.}\ \bibnamefont {Pitaevskii}},\ }\href@noop {}
  {\emph {\bibinfo {title} {Theory of Elasticity}}},\ \bibinfo {edition} {3rd}\
  ed.,\ \bibinfo {series} {Course of Theoretical Physics}, Vol.~\bibinfo
  {volume} {7}\ (\bibinfo  {publisher} {Butterworth-Heinemann},\ \bibinfo
  {address} {Oxford, UK},\ \bibinfo {year} {1995})\BibitemShut {NoStop}%
\bibitem [{\citenamefont {Sommerfeld}(1964)}]{SommerfeldTextbookVol2}%
  \BibitemOpen
  \bibfield  {author} {\bibinfo {author} {\bibfnamefont {A.}~\bibnamefont
  {Sommerfeld}},\ }\href@noop {} {\emph {\bibinfo {title} {Mechanics of
  Deformable Bodies}}},\ \bibinfo {series} {Lectures on Theoretical Physics},
  Vol.~\bibinfo {volume} {2}\ (\bibinfo  {publisher} {Academic Press},\
  \bibinfo {address} {New York},\ \bibinfo {year} {1964})\BibitemShut {NoStop}%
\bibitem [{\citenamefont {Knott}(1899)}]{Knott1899}%
  \BibitemOpen
  \bibfield  {author} {\bibinfo {author} {\bibfnamefont {C.~G.}\ \bibnamefont
  {Knott}},\ }\href {https://doi.org/10.1080/14786449908621305} {\bibfield
  {journal} {\bibinfo  {journal} {Lond. Edinb. Dubl. Phil. Mag.}\ }\textbf
  {\bibinfo {volume} {48}},\ \bibinfo {pages} {64} (\bibinfo {year}
  {1899})}\BibitemShut {NoStop}%
\bibitem [{\citenamefont {Zoeppritz}(1919)}]{Zoeppritz1919}%
  \BibitemOpen
  \bibfield  {author} {\bibinfo {author} {\bibfnamefont {K.}~\bibnamefont
  {Zoeppritz}},\ }\href
  {https://www.digizeitschriften.de/dms/img/?PID=GDZPPN002505290} {\bibfield
  {journal} {\bibinfo  {journal} {Nachrichten von der Gesellschaft der
  Wissenschaften zu G{\"o}ttingen, Mathematisch-Physikalische Klasse}\ }\textbf
  {\bibinfo {volume} {1}},\ \bibinfo {pages} {66} (\bibinfo {year}
  {1919})}\BibitemShut {NoStop}%
\bibitem [{\citenamefont {Ewing}\ \emph {et~al.}(1957)\citenamefont {Ewing},
  \citenamefont {Jardetzky},\ and\ \citenamefont {Press}}]{Ewing1957}%
  \BibitemOpen
  \bibfield  {author} {\bibinfo {author} {\bibfnamefont {W.}~\bibnamefont
  {Ewing}}, \bibinfo {author} {\bibfnamefont {W.~S.}\ \bibnamefont
  {Jardetzky}},\ and\ \bibinfo {author} {\bibfnamefont {F.}~\bibnamefont
  {Press}},\ }\href@noop {} {\emph {\bibinfo {title} {Elastic Waves in Layered
  Media}}},\ \bibinfo {series} {Lamont Geological Observatory Contribution}\
  No.\ \bibinfo {number} {189}\ (\bibinfo  {publisher} {McGraw-Hill},\ \bibinfo
  {address} {New York},\ \bibinfo {year} {1957})\BibitemShut {NoStop}%
\bibitem [{Note1()}]{Note1}%
  \BibitemOpen
  \bibinfo {note} {We can understand why the coefficients from classical
  elasticity appear in the phonon boundary condition. When the spatial extent
  of a wave packet is sufficiently broad, it behaves like an elastic plane
  wave, with energy flux being reflected or transmitted in a similar manner. In
  quantum mechanics, the energy flux is proportional to the number of phonons,
  i.e., the phonon distribution. Thus, Eqs.~\protect \textup {\hbox
  {\mathsurround \z@ \protect \normalfont (\ignorespaces \ref {eq: boundary
  condition assump 1}\unskip \@@italiccorr )}} and \protect \textup {\hbox
  {\mathsurround \z@ \protect \normalfont (\ignorespaces \ref {eq: boundary
  condition assump 2}\unskip \@@italiccorr )}} hold and involve the elastic
  reflectance and transmittance in their expressions. See our companion
  paper~\cite {SuzukiSumitaKato2024b} for details.}\BibitemShut {Stop}%
\bibitem [{\citenamefont {Kluge}\ and\ \citenamefont
  {Scholz}(1965)}]{Kluge1965}%
  \BibitemOpen
  \bibfield  {author} {\bibinfo {author} {\bibfnamefont {G.}~\bibnamefont
  {Kluge}}\ and\ \bibinfo {author} {\bibfnamefont {G.}~\bibnamefont {Scholz}},\
  }\href
  {http://www.ingentaconnect.com/contentone/dav/aaua/1965/00000016/00000001/art00006}
  {\bibfield  {journal} {\bibinfo  {journal} {Acta Acust. United Ac.}\ }\textbf
  {\bibinfo {volume} {16}},\ \bibinfo {pages} {60} (\bibinfo {year}
  {1965})}\BibitemShut {NoStop}%
\bibitem [{\citenamefont {Joffrin}\ and\ \citenamefont
  {Levelut}(1970)}]{Joffrin1970}%
  \BibitemOpen
  \bibfield  {author} {\bibinfo {author} {\bibfnamefont {J.}~\bibnamefont
  {Joffrin}}\ and\ \bibinfo {author} {\bibfnamefont {A.}~\bibnamefont
  {Levelut}},\ }\href
  {https://doi.org/https://doi.org/10.1016/0038-1098(70)90611-3} {\bibfield
  {journal} {\bibinfo  {journal} {Solid State Commun.}\ }\textbf {\bibinfo
  {volume} {8}},\ \bibinfo {pages} {1573} (\bibinfo {year} {1970})}\BibitemShut
  {NoStop}%
\bibitem [{\citenamefont {Pine}(1971)}]{Pine1971}%
  \BibitemOpen
  \bibfield  {author} {\bibinfo {author} {\bibfnamefont {A.~S.}\ \bibnamefont
  {Pine}},\ }\href {https://doi.org/10.1121/1.1912444} {\bibfield  {journal}
  {\bibinfo  {journal} {J. Acoust. Soc. Am.}\ }\textbf {\bibinfo {volume}
  {49}},\ \bibinfo {pages} {1026} (\bibinfo {year} {1971})}\BibitemShut
  {NoStop}%
\bibitem [{\citenamefont {Tsunetsugu}\ and\ \citenamefont
  {Kusunose}(2023)}]{Tsunetsugu2022}%
  \BibitemOpen
  \bibfield  {author} {\bibinfo {author} {\bibfnamefont {H.}~\bibnamefont
  {Tsunetsugu}}\ and\ \bibinfo {author} {\bibfnamefont {H.}~\bibnamefont
  {Kusunose}},\ }\href {https://doi.org/10.7566/JPSJ.92.023601} {\bibfield
  {journal} {\bibinfo  {journal} {J. Phys. Soc. Jpn.}\ }\textbf {\bibinfo
  {volume} {92}},\ \bibinfo {pages} {023601} (\bibinfo {year}
  {2023})}\BibitemShut {NoStop}%
\bibitem [{Note2()}]{Note2}%
  \BibitemOpen
  \bibinfo {note} {This condition is valid for $T\ll \SI {100}{K}$ in $\alpha
  $-quartz, where we used the magnitude of the splitting $\delimiter 69640972
  \chi \delimiter 86418188 /v_{\protect \text {T}}\sim \SI {3e-10}{m}$ observed
  in the direction close to the $k_z$ axis~\cite
  {Joffrin1970,Pine1970,Pine1971}.}\BibitemShut {Stop}%
\bibitem [{Note3()}]{Note3}%
  \BibitemOpen
  \bibinfo {note} {For $O(k^4)$ splitting, as suggested for tellurium~\cite
  {Tsunetsugu2022}, we expect $\alpha (T) \propto T^5$. The specific form of
  $\alpha (T)$, however, does not affect subsequent analysis.}\BibitemShut
  {Stop}%
\bibitem [{\citenamefont {{National Astronomical Observatory of
  Japan}}(2024)}]{Rikanempyo2024}%
  \BibitemOpen
  \bibfield  {author} {\bibinfo {author} {\bibnamefont {{National Astronomical
  Observatory of Japan}}},\ }\href {https://official.rikanenpyo.jp} {\emph
  {\bibinfo {title} {Chronological Scientific Tables 2025}}}\ (\bibinfo
  {publisher} {Maruzen},\ \bibinfo {address} {Tokyo},\ \bibinfo {year}
  {2024})\BibitemShut {NoStop}%
\bibitem [{Note4()}]{Note4}%
  \BibitemOpen
  \bibinfo {note} {We focus on extrinsic orbital AM and exclude intrinsic
  orbital AM arising from acoustic vortex beams~\cite
  {Hefner1999,Thomas2003,Ayub2011,Wang2021}.}\BibitemShut {Stop}%
\bibitem [{\citenamefont {Bliokh}\ and\ \citenamefont
  {Freilikher}(2006)}]{Bliokh2006}%
  \BibitemOpen
  \bibfield  {author} {\bibinfo {author} {\bibfnamefont {K.~Y.}\ \bibnamefont
  {Bliokh}}\ and\ \bibinfo {author} {\bibfnamefont {V.~D.}\ \bibnamefont
  {Freilikher}},\ }\href {https://doi.org/10.1103/PhysRevB.74.174302}
  {\bibfield  {journal} {\bibinfo  {journal} {Phys. Rev. B}\ }\textbf {\bibinfo
  {volume} {74}},\ \bibinfo {pages} {174302} (\bibinfo {year}
  {2006})}\BibitemShut {NoStop}%
\bibitem [{\citenamefont {Oh}\ and\ \citenamefont {Nagaosa}(2025)}]{Oh2025}%
  \BibitemOpen
  \bibfield  {author} {\bibinfo {author} {\bibfnamefont {T.}~\bibnamefont
  {Oh}}\ and\ \bibinfo {author} {\bibfnamefont {N.}~\bibnamefont {Nagaosa}},\
  }\href {https://doi.org/10.1103/PhysRevX.15.011036} {\bibfield  {journal}
  {\bibinfo  {journal} {Phys. Rev. X}\ }\textbf {\bibinfo {volume} {15}},\
  \bibinfo {pages} {011036} (\bibinfo {year} {2025})}\BibitemShut {NoStop}%
\bibitem [{\citenamefont {Romao}\ \emph {et~al.}(2023)\citenamefont {Romao},
  \citenamefont {Catena}, \citenamefont {Spaldin},\ and\ \citenamefont
  {Matas}}]{Romao2023}%
  \BibitemOpen
  \bibfield  {author} {\bibinfo {author} {\bibfnamefont {C.~P.}\ \bibnamefont
  {Romao}}, \bibinfo {author} {\bibfnamefont {R.}~\bibnamefont {Catena}},
  \bibinfo {author} {\bibfnamefont {N.~A.}\ \bibnamefont {Spaldin}},\ and\
  \bibinfo {author} {\bibfnamefont {M.}~\bibnamefont {Matas}},\ }\href
  {https://doi.org/10.1103/PhysRevResearch.5.043262} {\bibfield  {journal}
  {\bibinfo  {journal} {Phys. Rev. Res.}\ }\textbf {\bibinfo {volume} {5}},\
  \bibinfo {pages} {043262} (\bibinfo {year} {2023})}\BibitemShut {NoStop}%
\bibitem [{\citenamefont {Matas}\ \emph {et~al.}(2025)\citenamefont {Matas},
  \citenamefont {Krizek},\ and\ \citenamefont {Romao}}]{Matas2025}%
  \BibitemOpen
  \bibfield  {author} {\bibinfo {author} {\bibfnamefont {M.}~\bibnamefont
  {Matas}}, \bibinfo {author} {\bibfnamefont {F.}~\bibnamefont {Krizek}},\ and\
  \bibinfo {author} {\bibfnamefont {C.~P.}\ \bibnamefont {Romao}},\ }\href
  {https://arxiv.org/abs/2511.20461} {\bibfield  {journal} {\bibinfo  {journal}
  {arXiv preprint}\ } (\bibinfo {year} {2025})},\ \Eprint
  {https://arxiv.org/abs/2511.20461} {arXiv:2511.20461 [hep-ph]} \BibitemShut
  {NoStop}%
\bibitem [{\citenamefont {Juraschek}\ \emph {et~al.}(2022)\citenamefont
  {Juraschek}, \citenamefont {Neuman},\ and\ \citenamefont
  {Narang}}]{Juraschek2022}%
  \BibitemOpen
  \bibfield  {author} {\bibinfo {author} {\bibfnamefont {D.~M.}\ \bibnamefont
  {Juraschek}}, \bibinfo {author} {\bibfnamefont {T.~c.~v.}\ \bibnamefont
  {Neuman}},\ and\ \bibinfo {author} {\bibfnamefont {P.}~\bibnamefont
  {Narang}},\ }\href {https://doi.org/10.1103/PhysRevResearch.4.013129}
  {\bibfield  {journal} {\bibinfo  {journal} {Phys. Rev. Res.}\ }\textbf
  {\bibinfo {volume} {4}},\ \bibinfo {pages} {013129} (\bibinfo {year}
  {2022})}\BibitemShut {NoStop}%
\bibitem [{\citenamefont {Hefner}\ and\ \citenamefont
  {Marston}(1999)}]{Hefner1999}%
  \BibitemOpen
  \bibfield  {author} {\bibinfo {author} {\bibfnamefont {B.~T.}\ \bibnamefont
  {Hefner}}\ and\ \bibinfo {author} {\bibfnamefont {P.~L.}\ \bibnamefont
  {Marston}},\ }\href {https://doi.org/10.1121/1.428184} {\bibfield  {journal}
  {\bibinfo  {journal} {J. Acoust. Soc. Am.}\ }\textbf {\bibinfo {volume}
  {106}},\ \bibinfo {pages} {3313} (\bibinfo {year} {1999})}\BibitemShut
  {NoStop}%
\bibitem [{\citenamefont {Thomas}\ and\ \citenamefont
  {Marchiano}(2003)}]{Thomas2003}%
  \BibitemOpen
  \bibfield  {author} {\bibinfo {author} {\bibfnamefont {J.-L.}\ \bibnamefont
  {Thomas}}\ and\ \bibinfo {author} {\bibfnamefont {R.}~\bibnamefont
  {Marchiano}},\ }\href {https://doi.org/10.1103/PhysRevLett.91.244302}
  {\bibfield  {journal} {\bibinfo  {journal} {Phys. Rev. Lett.}\ }\textbf
  {\bibinfo {volume} {91}},\ \bibinfo {pages} {244302} (\bibinfo {year}
  {2003})}\BibitemShut {NoStop}%
\bibitem [{\citenamefont {Ayub}\ \emph {et~al.}(2011)\citenamefont {Ayub},
  \citenamefont {Ali},\ and\ \citenamefont {Mendonca}}]{Ayub2011}%
  \BibitemOpen
  \bibfield  {author} {\bibinfo {author} {\bibfnamefont {M.~K.}\ \bibnamefont
  {Ayub}}, \bibinfo {author} {\bibfnamefont {S.}~\bibnamefont {Ali}},\ and\
  \bibinfo {author} {\bibfnamefont {J.~T.}\ \bibnamefont {Mendonca}},\ }\href
  {https://doi.org/10.1063/1.3655429} {\bibfield  {journal} {\bibinfo
  {journal} {Phys. Plasmas}\ }\textbf {\bibinfo {volume} {18}},\ \bibinfo
  {pages} {102117} (\bibinfo {year} {2011})}\BibitemShut {NoStop}%
\bibitem [{\citenamefont {Wang}\ \emph {et~al.}(2021)\citenamefont {Wang},
  \citenamefont {Tan}, \citenamefont {Liang}, \citenamefont {Ma}, \citenamefont
  {Wang},\ and\ \citenamefont {Cheng}}]{Wang2021}%
  \BibitemOpen
  \bibfield  {author} {\bibinfo {author} {\bibfnamefont {W.}~\bibnamefont
  {Wang}}, \bibinfo {author} {\bibfnamefont {Y.}~\bibnamefont {Tan}}, \bibinfo
  {author} {\bibfnamefont {B.}~\bibnamefont {Liang}}, \bibinfo {author}
  {\bibfnamefont {G.}~\bibnamefont {Ma}}, \bibinfo {author} {\bibfnamefont
  {S.}~\bibnamefont {Wang}},\ and\ \bibinfo {author} {\bibfnamefont
  {J.}~\bibnamefont {Cheng}},\ }\href
  {https://doi.org/10.1103/PhysRevB.104.174301} {\bibfield  {journal} {\bibinfo
   {journal} {Phys. Rev. B}\ }\textbf {\bibinfo {volume} {104}},\ \bibinfo
  {pages} {174301} (\bibinfo {year} {2021})}\BibitemShut {NoStop}%
\end{thebibliography}%

\clearpage
\renewcommand{\thesection}{S\arabic{section}}
\renewcommand{\theequation}{S\arabic{equation}}
\setcounter{equation}{0}
\renewcommand{\thefigure}{S\arabic{figure}}
\setcounter{figure}{0}
\renewcommand{\thetable}{S\arabic{table}}
\setcounter{table}{0}
\makeatletter
\c@secnumdepth = 3
\makeatother
\onecolumngrid
\begin{center}
 {\large \textmd{Supplemental Materials for:} \\[0.3em] {\bfseries 
Theory of phonon angular momentum transport \\ across a smooth crystal interface
}}
\end{center}
\setcounter{page}{1}
\setcounter{section}{0}

\bigskip

In the main text, several technical aspects of the interfacial transport of phonon angular momentum (AM) were omitted for brevity.
We provide additional discussions and derivations in this Supplemental Material.
We first discuss the splitting of phonon modes and the AM carried by them in chiral crystals.
This discussion justifies the assumptions on mode splitting and phonon AM employed in the main text within the isotropic elastic approximation.

We next present a derivation that retains all first-order contributions arising from the chirality-induced splitting of phonon modes. 
The resulting expression identifies the various contributions to interfacial AM transport and clarifies the approximations adopted in the main text. 
The retained contributions and neglected first-order corrections are explicitly separated. 
{We also assess the effect of the approximation in Eq.~\eqref{105616_5Oct26} on the resulting AM density. }

\section{Long-wavelength acoustic phonons in chiral crystals}

In this section, we summarize general properties of long-wavelength acoustic phonons in nonmagnetic chiral crystals.
Chirality produces two distinct effects on such phonons:
(i)~an $O(k^2)$ splitting of degenerate transverse modes accompanied by phonon AM $\pm\hbar$, and
(ii)~hybridization among nondegenerate acoustic modes, which generates wavevector-dependent phonon AM.

In the main text, we adopted the isotropic-body approximation for the crystal. In this case, only effect (i) needs to be considered, while effect (ii) can be neglected.

\subsection{General formulation}

In the long-wavelength limit $k = |\bm{k}| \to 0$, the dynamical matrix of lattice vibrations can be expanded as
\begin{equation}
\underline{D}(\bm{k})
= \underline{D}^{(0)}
 + \underline{D}^{(1)}_i k_i
 + \underline{D}^{(2)}_{ij} k_i k_j
 + \underline{D}^{(3)}_{ijl} k_i k_j k_l + O(k^4), \label{185807_5Jun26}
\end{equation}
where the underlined quantities denote matrices acting on both displacement and sublattice degrees of freedom. 
Repeated Cartesian indices are summed over $i,j,l=x,y,z$.

The dynamical matrix satisfies the following properties:
\begin{itemize}
\item Hermiticity:
\begin{equation}
\underline{D}^{\dagger}(\bm{k})=\underline{D}(\bm{k}),
\end{equation}
\item Reciprocity:
\begin{equation}
\underline{D}^{\top}(\bm{k})=\underline{D}(-\bm{k}),
\end{equation}
\item For centrosymmetric crystals, inversion symmetry further requires
\begin{equation}
\underline{D}(\bm{k})=\underline{D}(-\bm{k}).
\end{equation}
\end{itemize}
Therefore, the cubic-order coefficient $\underline{D}^{(3)}_{ijl}$
is antisymmetric and exists only in non-centrosymmetric crystals, such as chiral crystals.

Since acoustic and optical phonons are well separated in energy near $\bm{k}=0$, 
we neglect their hybridization and focus on the effective $3\times 3$ dynamical matrix for acoustic modes alone. 
The phonon dispersion and polarization vectors are determined from
\begin{equation}
 \underline{D}(\bm{k}) \bm{e}_{\bm{k}n} = \omega_{\bm{k}n}^2 \bm{e}_{\bm{k}n}. \label{190028_5Jun26}
\end{equation}
Since $\underline{D}(\bm{k})^*=\underline{D}(-\bm{k})$, the phase convention may be chosen such that
\begin{equation}
\bm{e}_{\bm{k}n}^*=\bm{e}_{-\bm{k}, n}.
\end{equation}

Following Ref.~\cite{Zhang2014}, the phonon (spin) AM is defined as
\begin{equation}
 \bm{S}_{\bm{k}n} = -\iu\hbar\bm{e}^*_{\bm{k}n}\times \bm{e}_{\bm{k}n}
= -\iu\hbar\bm{e}_{-\bm{k}, n}\times \bm{e}_{\bm{k}n}
= - \bm{S}_{-\bm{k}, n}, 
\end{equation}
which is odd in $\bm{k}$.

\subsection{Leading-order acoustic phonons}

Because acoustic phonons are Nambu--Goldstone modes, their eigenvalues satisfy $\omega_{\bm{k}n}^2 = O(k^2)$.
Therefore, the leading contribution to the acoustic-sector dynamical matrix is
$\underline{D}(\bm{k}) \simeq \underline{D}^{(2)}_{ij} k_i k_j$. 
Since $\underline{D}^{(2)}_{ij}$ is real and symmetric, 
the corresponding lowest-order polarization vectors $\bm{e}^{(0)}_{\bm{k}n}$ can be chosen real.
Consequently, the phonon AM vanishes at this order:
\begin{equation}
 \bm{S}^{(0)}_{\bm{k}n} = -\iu\hbar\bm{e}^{(0) *}_{\bm{k}n}\times \bm{e}^{(0)}_{\bm{k}n}
= -\iu\hbar\bm{e}^{(0)}_{\bm{k}n}\times \bm{e}^{(0)}_{\bm{k}n} = 0.
\end{equation}
Thus, finite phonon AM in chiral crystals originates from the cubic-order term
$\underline{D}^{(3)}_{ijl} k_i k_j k_l$.

\subsection{Effects of the cubic-order term}
\label{010350_23Jun26}

We now consider the effects of the perturbation $\underline{D}^{(3)}_{ijl}k_i k_j k_l$. 
Its physical consequences differ depending on whether the acoustic modes are degenerate or nondegenerate.

\subsubsection{Degenerate transverse modes}

We first consider a doubly degenerate transverse acoustic mode, as realized in the isotropic-body approximation adopted in the main text. 
Let the two degenerate transverse modes be labeled by $\mathrm{T1}$ and $\mathrm{T2}$, with common frequency
$\omega^{(0)}_{\bm{k}, \text{T}}\equiv \omega^{(0)}_{\bm{k}, \text{T1}} = \omega^{(0)}_{\bm{k}, \text{T2}}$. 
The unperturbed polarization vectors $\bm{e}^{(0)}_{\bm{k}\mathrm{T1}}$ and $\bm{e}^{(0)}_{\bm{k}\mathrm{T2}}$
are taken to be real and orthonormal.

Within the degenerate subspace, the perturbation reduces to the $2\times2$ matrix
\begin{equation}
M \equiv 
\begin{bmatrix}
\bm{e}_{\bm{k}, \text{T1}}^{(0)\, \dagger} \\[6pt] \bm{e}_{\bm{k}, \text{T2}}^{(0)\, \dagger} 
\end{bmatrix}
 \underline{D}^{(3)}_{ijl} k_i k_j k_l
\begin{bmatrix}
\bm{e}^{(0)}_{\bm{k}, \text{T1}} & \bm{e}^{(0)}_{\bm{k}, \text{T2}} 
\end{bmatrix}. 
\end{equation}
Because $\underline{D}^{(3)}_{ijl}$ is antisymmetric and Hermitian,
\begin{equation}
 M = k_i k_j k_l
\begin{bmatrix}
 0 & -i\Delta_{ijl} \\ i\Delta_{ijl} & 0
\end{bmatrix}, 
\end{equation}
where $\Delta_{ijl}$ is real.
The perturbed eigenmodes are therefore circularly polarized combinations,
\begin{equation}
 \bm{e}_{\bm{k}, \text{T}\pm}
\equiv \frac{\bm{e}^{(0)}_{\bm{k}, \text{T1}}\pm i \mathrm{sgn}\, (\Delta_{ijl} k_i k_j k_l) \bm{e}^{(0)}_{\bm{k}, \text{T2}}}{\sqrt{2}}
\end{equation}
with dispersions
\begin{equation}
 \omega_{\bm{k}, \text{T}\pm} = 
\sqrt{\left(\omega^{(0)}_{\bm{k}\text{T}}\right)^2 \pm \Abs{\Delta_{ijl} k_i k_j k_l}}
\simeq \omega^{(0)}_{\bm{k}\text{T}} \pm 
\frac{\Abs{\Delta_{ijl} k_i k_j k_l}}{2\omega^{(0)}_{\bm{k}\text{T}}}. \label{000216_6Jun26}
\end{equation}
Since $\omega^{(0)}_{\bm{k}\mathrm{T}}=O(k)$, the splitting is of order $O(k^2)$.

The two circularly polarized modes carry opposite phonon AM,
\begin{align}
\bm{S}_{\bm{k}, \text{T}\pm} 
&= -\iu\hbar \bm{e}^{*}_{\bm{k}, \text{T}\pm}\times \bm{e}_{\bm{k}, \text{T}\pm}\nonumber\\
&= -\iu\hbar
\frac{\bm{e}^{(0)}_{\bm{k}, \text{T1}}\mp i\mathrm{sgn}\, (\Delta_{ijl} k_i k_j k_l) \bm{e}^{(0)}_{\bm{k}, \text{T2}}}{\sqrt{2}}\times 
\frac{\bm{e}^{(0)}_{\bm{k}, \text{T1}}\pm i\mathrm{sgn}\, (\Delta_{ijl} k_i k_j k_l) \bm{e}^{(0)}_{\bm{k}, \text{T2}}}{\sqrt{2}} \nonumber\\
&= \pm \hbar\, \mathrm{sgn}\, (\Delta_{ijl} k_i k_j k_l) 
(\bm{e}^{(0)}_{\bm{k}, \text{T1}}\times \bm{e}^{(0)}_{\bm{k}, \text{T2}}), 
\end{align}
whose magnitude is $\hbar$.
Thus, chirality lifts the degeneracy of transverse acoustic phonons at order $O(k^2)$ and simultaneously generates phonon AM $\pm\hbar$.

\subsubsection{Nondegenerate acoustic modes}

Next, we consider wavevectors where the three acoustic modes are nondegenerate. 
This situation generally arises once crystalline anisotropy is taken into account.
In this case, the first-order correction to the phonon dispersion vanishes because
\begin{equation}
\bm{e}^{(0)\, \dagger}_{\bm{k}n} \underline{D}^{(3)}_{ijl} \bm{e}^{(0)}_{\bm{k}n}
= \left(\underline{D}^{(3)}_{ijl}\right)_{\mu\nu}
\left(\bm{e}^{(0)}_{\bm{k}n}\right)_{\mu}
\left(\bm{e}^{(0)}_{\bm{k}n}\right)_{\nu} = 0, 
\end{equation}
which follows from the antisymmetry of $\underline{D}^{(3)}_{ijl}$.

However, the polarization vectors acquire first-order corrections,
\begin{align}
 \bm{e}_{\bm{k}n} &\simeq \bm{e}^{(0)}_{\bm{k}n} + \bm{e}^{(1)}_{\bm{k}n}, \\
\bm{e}^{(1)}_{\bm{k}n} &= \sum_{n' (\neq n)}
\bm{e}^{(0)}_{\bm{k}n'}
\frac{
\bm{e}^{(0)\, \dagger}_{\bm{k}n'}
\underline{D}^{(3)}_{ijl} k_i k_j k_l
\bm{e}^{(0)}_{\bm{k}n}}{\left(\omega^{(0)}_{\bm{k}n}\right)^2 - \left(\omega^{(0)}_{\bm{k}n'}\right)^2}. 
\end{align}
As a result, different acoustic modes become hybridized, and the phonons acquire finite AM,
\begin{equation}
 \bm{S}_{\bm{k}n} =  
-\iu\hbar \left[
\bm{e}^{(0)\, *}_{\bm{k}n}\times \bm{e}^{(1)}_{\bm{k}n} + \bm{e}^{(1)\, *}_{\bm{k}n}\times \bm{e}^{(0)}_{\bm{k}n}
\right] + \hbar \, O(k^2)
=  \hbar \, O(k^1).
\end{equation}
In contrast to the degenerate case discussed above, 
the magnitude of the AM now depends continuously on the wavevector.

\section{Detailed derivation of interfacial transfer of phonon AM}

In this section, we provide the derivation of the analytical expressions for the AM density and AM flux presented in the main text [Eqs.~(10a)--(12b)]. 
We focus on two points. First, we clarify how the chirality-induced splitting of transverse phonons affects the interfacial transfer of AM. 
Second, we explain how the general expressions are simplified to obtain the formulas used in the main text, where chirality enters only through 
the nonequilibrium phonon distribution. 
We also briefly discuss a related problem: AM transfer across a junction between chiral crystals with opposite chiralities.

\subsection{Effect of transverse-mode splitting on interfacial AM transfer}

As a model for transverse phonons in a chiral crystal, 
we assume that the dispersions of the right-handed (RH) and left-handed (LH) circularly polarized modes are split to first order in a small parameter $\chi$ as
\begin{align}
 \Omega_{\bm{k},\text{RH}} &\simeq v_{\text{T}}k + {\chi k^2}/{2}, 
&\Omega_{\bm{k},\text{LH}} &\simeq v_{\text{T}}k - {\chi k^2}/{2}, 
\label{153258_20Jun26}
\end{align}
where $k=|\bm{k}|$.

We first derive analytical expressions for the AM carried by reflected and transmitted phonons 
while fully incorporating the effect of this small splitting. 
Throughout this section, we treat the group velocities of the RH and LH modes separately and write them as 
$v_{\bm{k},\mathrm{RH}}$ and $v_{\bm{k},\mathrm{LH}}$, respectively.
We also distinguish the notation for reflectance and transmittance in the degenerate and split cases. 
For degenerate RH and LH modes, corresponding to the situation considered in the main text, we write
$\mathcal{R}_{nm}(\omega, \bm{k}_{\parallel})$ and $\mathcal{T}_{nm}(\omega, \bm{k}_{\parallel})$. 
For split modes, we instead write
$\mathcal{R}^{(\chi)}_{nm}(\omega, \bm{k}_{\parallel})$ and $\mathcal{T}^{(\chi)}_{nm}(\omega, \bm{k}_{\parallel})$. 

The coefficients $\mathcal{R}_{nm}$ and $\mathcal{T}_{nm}$ satisfy Helmholtz reciprocity,
$\mathcal{R}_{nm}(\omega, \bm{k}_{\parallel}) = \mathcal{R}_{mn}(\omega, \bm{k}_{\parallel})$ and, in addition, satisfy the symmetry relations
\begin{equation}
 \label{231018_15Jun26}
 \begin{aligned}
 \mathcal{R}_{\text{RH}, \text{RH}} & = \mathcal{R}_{\text{LH}, \text{LH}}, 
&\mathcal{R}_{\text{RH}, \text{LH}} & = \mathcal{R}_{\text{LH}, \text{RH}}, 
& \mathcal{R}_{\text{RH}, \text{L}} &= \mathcal{R}_{\text{LH}, \text{L}}, \\
 \mathcal{T}_{\text{RH}, \text{RH}} & = \mathcal{T}_{\text{LH}, \text{LH}}, 
&\mathcal{T}_{\text{RH}, \text{LH}} & = \mathcal{T}_{\text{LH}, \text{RH}}, 
& \mathcal{T}_{\text{RH}, \text{L}} &= \mathcal{T}_{\text{LH}, \text{L}}. 
 \end{aligned}
\end{equation}
All coefficients are evaluated at the same $(\omega,\bm{k}_{\parallel})$.
These relations follow from the degeneracy of the RH and LH modes and from the fact that their polarization vectors are related by complex conjugation.
By contrast, the coefficients $\mathcal{R}^{(\chi)}_{nm}$ and $\mathcal{T}^{(\chi)}_{nm}$ generally do not satisfy the symmetry relations in Eq.~(\ref{231018_15Jun26}) 
because the RH and LH modes are no longer degenerate.

\subsubsection{Preliminaries}

Within the isotropic-body approximation, only the transverse RH and LH modes carry AM. 
Their AM is parallel or antiparallel to the wavevector and has magnitude $\hbar$ (see Sec.~\ref{010350_23Jun26}).
From Eqs.~(4a) and (4b) of the main text, the deviation of the distribution function can be written in terms of the bulk contribution $B_{\bm{k}n}$
and the amplitudes of the interface-localized distributions, $C_{\bm{k}n}$ and $D_{\bm{q}n}$, as
\begin{subequations}
 \begin{align}
 F^{(1)}_{\bm{k}n} &= B_{\bm{k}n} + \Theta_{\text{H}}(-v^z_{\bm{k}n}) \, C_{\bm{k}n}\exp\left[z\Big/({-\tau v^z_{\bm{k}n}})\right],\\ 
 f^{(1)}_{\bm{q}n} &= \Theta_{\text{H}}(c^z_{\bm{q}n}) \, D_{\bm{q}n}\exp\left[-z\Big/{\widetilde{\tau} c^z_{\bm{q}n}}\right].
 \end{align}
\end{subequations}
The AM densities on the chiral side ($z<0$) and the achiral side ($z>0$) are therefore given by
\begin{align}
& \bm{S} (z<0) = \int \frac{\rmd^3\bm{k}}{(2\pi)^3} \sum_{n = \text{L}, \text{RH}, \text{LH}} \bm{S}_{\bm{k}n} F^{(1)}_{\bm{k}n}\nonumber\\
&= \int \frac{\rmd^3\bm{k}}{(2\pi)^3} \hbar \hat{\bm{k}} \left[
\left(B_{\bm{k}, \text{RH}} - B_{\bm{k}, \text{LH}}\right)
+ \Theta_{\text{H}}(-k_z)\left(C_{\bm{k}, \text{RH}} \exp\left({\displaystyle \frac{-z}{\tau v^z_{\bm{k}, \text{RH}}}} \right)
- C_{\bm{k}, \text{LH}} \exp\left({\displaystyle \frac{-z}{\tau v^z_{\bm{k}, \text{LH}}}}\right) \right) \right], \label{190101_15Jun26}\\
&{\bm{S}} (z>0) = 
\int\limits_{q_z > 0} \frac{\rmd^3\bm{q}}{(2\pi)^3} \hbar \hat{\bm{q}}\, \exp\left({\displaystyle \frac{-z}{\widetilde{\tau} c_{\text{T}} \hat{q}_z}}\right)
\left(D_{\bm{q}, \text{RH}}  - D_{\bm{q}, \text{LH}} \right).  \label{190750_15Jun26}
\end{align}
The first term in Eq.~\eqref{190101_15Jun26} contains
\begin{equation}
 B_{\bm{k}n} = - \tau \bm{v}_{\bm{k}n}\cdot \bm{\nabla}T\,  \frac{\partial F^{(0)}_{\bm{k}n}}{\partial T}
\end{equation}
and reproduces the bulk AM density $\bm{S}_0$ given in the main text [Eq.~(9)] after integration.

We next simplify the second term of Eq.~\eqref{190101_15Jun26}. 
The RH and LH modes have slightly different localization lengths because
$\tau v_{\bm{k},\text{RH}} - \tau v_{\bm{k},\text{LH}} = 2\tau \chi k$. We neglect this difference.

The transferred AM extends over a distance of the order of the mean free path, $z\sim \tau v_{\mathrm{T}}$. 
In the following derivation, we focus on the vicinity of the interface, 
$z\ll \tau v_{\mathrm{T}}$, 
where the transferred AM remains appreciable. 
The exponential factor can then be expanded as
\begin{equation}
 \exp\left({\displaystyle \frac{-z}{\tau v^z_{\bm{k}, \text{RH}}}} \right)
 = \exp\left({\displaystyle \frac{-z}{\tau \left(v_{\text{T}} + \chi k \right) \hat{k}_z}} \right)
 \simeq \exp\left({\displaystyle\frac{-z}{\tau v_{\text{T}}\hat{k}_z}}\right) \left(1 + \frac{z}{\tau v_{\text{T}} \hat{k}_z} \frac{\chi k}{v_{\text{T}}}\right). 
\end{equation}
The correction term contains the product of two small parameters, $\chi k/v_{\mathrm{T}}$ and $z/(\tau v_{\mathrm{T}})$.
We therefore retain only the leading term.

Under this approximation, the group velocities
$v_{\bm{k},\mathrm{RH}}$ and $v_{\bm{k},\mathrm{LH}}$ appearing in the exponential factors can be replaced by $v_{\mathrm{T}}$.
Equation~\eqref{190101_15Jun26} then becomes
\begin{align}
 \bm{S}(z<0) - \bm{S}_0 \simeq \int\limits_{k_z <0} \frac{\rmd^3\bm{k}}{(2\pi)^3} \hbar \hat{\bm{k}}  \,
\exp\left({\displaystyle \frac{-z}{\tau v_{\text{T}}\hat{k}_z}}\right) \left(C_{\bm{k}, \text{RH}} - C_{\bm{k}, \text{LH}} \right).  \label{155309_9Jun26}
\end{align}

We next express
$C_{\bm{k},\mathrm{RH}}-C_{\bm{k},\mathrm{LH}}$
and
$D_{\bm{q},\mathrm{RH}}-D_{\bm{q},\mathrm{LH}}$
in terms of the bulk distribution shift
$B_{\bm{k}n}$.
To this end, we impose the following boundary conditions on the phonon distribution functions:
\begin{subequations}
 \begin{align}
 B_{-,n}(\omega, \bm{k}_{\parallel}) + C_{-,n}(\omega, \bm{k}_{\parallel}) 
&= \sum_m \mathcal{R}^{(\chi)}_{nm}(\omega, \bm{k}_{\parallel}) B_{+, m}(\omega, \bm{k}_{\parallel}), \\
D_{+, n}(\omega, \bm{k}_{\parallel}) &= \sum_m \mathcal{T}^{(\chi)}_{nm}(\omega, \bm{k}_{\parallel}) B_{+, m}(\omega, \bm{k}_{\parallel}).
 \end{align}
\end{subequations}
We note that we apply these boundary conditions even outside the strictly linear-dispersion regime of acoustic phonons. 
This is an assumption, and its validity has not yet been established. 
At least within the linear-dispersion regime, we have derived the corresponding boundary conditions with
$\mathcal{R}^{(\chi)}$ and $\mathcal{T}^{(\chi)}$ replaced by $\mathcal{R}$ and $\mathcal{T}$ in our companion paper~\cite{SuzukiSumitaKato2024b}.

Applying the above boundary conditions to
$C_{\bm{k}, \text{RH}} - C_{\bm{k}, \text{LH}} 
= C_{-, \text{RH}}(\Omega_{\bm{k}, \text{RH}}, \bm{k}_{\parallel}) - C_{-, \text{LH}}(\Omega_{\bm{k}, \text{LH}}, \bm{k}_{\parallel})$
and $D_{\bm{q}, \text{RH}} - D_{\bm{q}, \text{LH}} 
= D_{+, \text{RH}}(\omega_{\bm{q}, \text{RH}}, \bm{q}_{\parallel}) - D_{+, \text{LH}}(\omega_{\bm{q}, \text{LH}}, \bm{q}_{\parallel})$
in Eqs.~\eqref{190750_15Jun26} and \eqref{155309_9Jun26}, 
we obtain 
\begin{subequations} 
 \begin{align}
&  \bm{S}(z<0) - \bm{S}_0 \simeq 
- \int\limits_{k_z <0} \frac{\rmd^3\bm{k}}{(2\pi)^3} \hbar \hat{\bm{k}}  \,\exp\left({\displaystyle \frac{-z}{\tau v_{\text{T}}\hat{k}_z}}\right) 
\left(B_{\bm{k},\text{RH}} - B_{\bm{k},\text{LH}}\right) \nonumber\\
& +  \int\limits_{k_z <0} \frac{\rmd^3\bm{k}}{(2\pi)^3} \hbar \hat{\bm{k}}  \,\exp\left({\displaystyle \frac{-z}{\tau v_{\text{T}}\hat{k}_z}}\right) 
\nonumber\\ &\qquad\times 
\sum_{m = \text{L}, \text{RH}, \text{LH}}\left[
\mathcal{R}^{(\chi)}_{\text{RH}, m} (\Omega_{\bm{k}, \text{RH}}, \bm{k}_{\parallel}) B_{+, m} (\Omega_{\bm{k}, \text{RH}}, \bm{k}_{\parallel})
- \mathcal{R}^{(\chi)}_{\text{LH}, m} (\Omega_{\bm{k}, \text{LH}}, \bm{k}_{\parallel}) B_{+, m} (\Omega_{\bm{k}, \text{LH}}, \bm{k}_{\parallel})
\right], \label{193224_15Jun26}\\
&{\bm{S}} (z>0) = 
\int\limits_{q_z > 0} \frac{\rmd^3\bm{q}}{(2\pi)^3} \hbar \hat{\bm{q}}\, \exp\left({\displaystyle \frac{-z}{\widetilde{\tau} c_{\text{T}} \hat{q}_z}}\right) \nonumber\\
&\qquad \times \sum_{m = \text{L}, \text{RH}, \text{LH}} \left[
\mathcal{T}^{(\chi)}_{\text{RH}, m} (\omega_{\bm{q}, \text{RH}}, \bm{q}_{\parallel}) B_{+, m} (\omega_{\bm{q}, \text{RH}}, \bm{q}_{\parallel})
- \mathcal{T}^{(\chi)}_{\text{LH}, m} (\omega_{\bm{q}, \text{LH}}, \bm{q}_{\parallel}) B_{+, m} (\omega_{\bm{q}, \text{LH}}, \bm{q}_{\parallel})
\right].\label{193227_15Jun26}
 \end{align}
\end{subequations}
In Eq.~\eqref{193227_15Jun26}, the relation
$\omega_{\bm{q}, \text{RH}} = \omega_{\bm{q}, \text{LH}} = \omega_{\bm{q}, \text{T}}$
holds when the crystal in the region $z>0$ is achiral.
We nevertheless keep the notation general because we will later consider the case in which the crystal in the region $z>0$ is also chiral.

\subsubsection{Rearrangement of the $O(\chi)$ contributions}

The integrands of Eqs.~\eqref{193224_15Jun26} and \eqref{193227_15Jun26} are already of order $O(\chi)$, reflecting the asymmetry between the RH and LH modes. 
In this subsection, we rewrite these expressions into a form that is more convenient for physical interpretation while retaining all terms up to $O(\chi)$.

The final results are
\begin{align}
& \sum_{m = \text{L}, \text{RH}, \text{LH}}\left[
\mathcal{R}^{(\chi)}_{\text{RH}, m} (\Omega_{\bm{k}, \text{RH}}, \bm{k}_{\parallel}) B_{+, m} (\Omega_{\bm{k}, \text{RH}}, \bm{k}_{\parallel})
- \mathcal{R}^{(\chi)}_{\text{LH}, m} (\Omega_{\bm{k}, \text{LH}}, \bm{k}_{\parallel}) B_{+, m} (\Omega_{\bm{k}, \text{LH}}, \bm{k}_{\parallel})
\right] \nonumber\\
&\simeq  \left[\mathcal{R}^{(\chi)}_{\text{RH}, \text{L}}(\text{RH}) - \mathcal{R}^{(\chi)}_{\text{LH}, \text{L}}(\text{LH})\right] B_{\text{L}}(\text{T})
 + \mathcal{R}_{\text{RH}, \text{L}}(\text{T}) \left[B_{\text{L}}(\text{RH}) - B_{\text{L}}(\text{LH})\right] \nonumber\\
& + \left[\mathcal{R}_{\text{RH}, \text{RH}}(\text{T}) - \mathcal{R}_{\text{RH}, \text{LH}}(\text{T})\right]
\left[B_{\overline{\bm{k}}, \text{RH}} - B_{\overline{\bm{k}}, \text{LH}}\right]
+  2 \mathcal{R}_{\text{RH}, \text{LH}}(\text{T}) \left[ B_{\text{T}}(\text{RH}) - B_{\text{T}}(\text{LH}) \right] \nonumber\\
& + \chi k^2 \left.
\frac{\partial \left[
\mathcal{R}_{\text{RH}, \text{RH}}(\omega, \bm{k}_{\parallel}) + \mathcal{R}_{\text{RH}, \text{LH}}(\omega, \bm{k}_{\parallel})\right]}{\partial \omega}
\right|_{\omega = \Omega_{\bm{k},\text{T}}} B_{\overline{\bm{k}},\text{T}} \nonumber\\
& + \left[ \mathcal{R}^{(\chi)}_{\text{RH}, \text{RH}}(\text{T}) - \mathcal{R}^{(\chi)}_{\text{LH}, \text{LH}}(\text{T})  \right] B_{\overline{\bm{k}},\text{T}}
+ \left[ \mathcal{R}^{(\chi)}_{\text{RH}, \text{LH}}(\text{T}) - \mathcal{R}^{(\chi)}_{\text{LH}, \text{RH}}(\text{T})  \right] B_{\overline{\bm{k}},\text{T}}, 
\label{195218_15Jun26}
\end{align}

\begin{align}
 & \sum_{m = \text{L}, \text{RH}, \text{LH}}\left[
\mathcal{T}^{(\chi)}_{\text{RH}, m} (\omega_{\bm{q}, \text{RH}}, \bm{q}_{\parallel}) B_{+, m} (\omega_{\bm{q}, \text{RH}}, \bm{q}_{\parallel})
- \mathcal{T}^{(\chi)}_{\text{LH}, m} (\omega_{\bm{q}, \text{LH}}, \bm{q}_{\parallel}) B_{+, m} (\omega_{\bm{q}, \text{LH}}, \bm{q}_{\parallel}) \right] \nonumber\\
&\simeq  \left[\mathcal{T}^{(\chi)}_{\text{RH}, \text{L}}(\text{RH}) - \mathcal{T}^{(\chi)}_{\text{LH}, \text{L}}(\text{LH})\right] \check{B}_{\text{L}}(\text{T})
 + \mathcal{T}_{\text{RH}, \text{L}}(\text{T}) \left[\check{B}_{\text{L}}(\text{RH}) - \check{B}_{\text{L}}(\text{LH})\right] \nonumber\\
& + \left[\mathcal{T}_{\text{RH}, \text{RH}}(\text{T}) - \mathcal{T}_{\text{RH}, \text{LH}}(\text{T})\right]
\left[\check{B}_{\text{RH}}(\text{RH}) - \check{B}_{\text{LH}}(\text{LH})\right]
+  2 \mathcal{T}_{\text{RH}, \text{LH}}(\text{T}) \left[ \check{B}_{\text{T}}(\text{RH}) - \check{B}_{\text{T}}(\text{LH}) \right] \nonumber\\
& + \chi k^2 \left.
\frac{\partial \left[
\mathcal{T}_{\text{RH}, \text{RH}}(\omega, \bm{q}_{\parallel}) + \mathcal{T}_{\text{RH}, \text{LH}}(\omega, \bm{q}_{\parallel})\right]}{\partial \omega}
\right|_{\omega = \Omega_{\bm{q},\text{T}}} \check{B}_{\text{T}}(\text{T}) \nonumber\\
& + \left[ \mathcal{T}^{(\chi)}_{\text{RH}, \text{RH}}(\text{T}) - \mathcal{T}^{(\chi)}_{\text{LH}, \text{LH}}(\text{T})  \right] \check{B}_{\text{T}}(\text{T})
+ \left[ \mathcal{T}^{(\chi)}_{\text{RH}, \text{LH}}(\text{T}) - \mathcal{T}^{(\chi)}_{\text{LH}, \text{RH}}(\text{T})  \right] \check{B}_{\text{T}}(\text{T}), 
\label{195231_15Jun26}
\end{align}
where we have introduced the shorthand notation
\begin{equation}
\begin{aligned}
\mathcal{R}^{(\chi)}_{nm}(l) &\equiv \mathcal{R}^{(\chi)}_{nm} (\Omega_{\bm{k}, l}, \bm{k}_{\parallel}), 
&\mathcal{T}^{(\chi)}_{nm}(l) &\equiv \mathcal{T}^{(\chi)}_{nm} (\omega_{\bm{q}, l}, \bm{q}_{\parallel}), 
& B_m(l) & \equiv B_{+, m} (\Omega_{\bm{k}, l}, \bm{k}_{\parallel}), \\
\mathcal{R}_{nm}(l) &\equiv \mathcal{R}_{nm} (\Omega_{\bm{k}, l}, \bm{k}_{\parallel}), 
&\mathcal{T}_{nm}(l) &\equiv \mathcal{T}_{nm} (\omega_{\bm{k}, l}, \bm{q}_{\parallel}), 
& \check{B}_m(l) & \equiv B_{+, m} (\omega_{\bm{q}, l}, \bm{q}_{\parallel}).
\end{aligned}
\label{113717_18Jun26}
 \end{equation}
For a wavevector
$\bm{k}=(k_x,k_y,k_z<0)$ appearing in Eq.~\eqref{193224_15Jun26}, we also define
\begin{equation}
\overline{\bm{k}} \equiv (k_x,k_y,-k_z>0).
\end{equation}
In addition, for an arbitrary quantity $A_{\bm{k}n}$ labeled by the phonon state $(\bm{k},n)$, we introduce the transverse-mode average
\begin{align}
 A_{\bm{k},\text{T}}\equiv \frac{ A_{\bm{k},\text{RH}} + A_{\bm{k},\text{LH}} }{2}. \label{100511_10Jun26}
\end{align}

We first derive Eq.~\eqref{195218_15Jun26}. Starting from
\begin{align}
&\sum_{m = \text{L}, \text{RH}, \text{LH}}\left[
\mathcal{R}^{(\chi)}_{\text{RH}, m} (\Omega_{\bm{k}, \text{RH}}, \bm{k}_{\parallel}) B_{+, m} (\Omega_{\bm{k}, \text{RH}}, \bm{k}_{\parallel})
- \mathcal{R}^{(\chi)}_{\text{LH}, m} (\Omega_{\bm{k}, \text{LH}}, \bm{k}_{\parallel}) B_{+, m} (\Omega_{\bm{k}, \text{LH}}, \bm{k}_{\parallel})
\right] \nonumber\\
&= \mathcal{R}^{(\chi)}_{\text{RH}, \text{L}} (\text{RH}) B_{\text{L}}(\text{RH})
- \mathcal{R}^{(\chi)}_{\text{LH}, \text{L}} (\text{LH}) B_{\text{L}}(\text{LH}) \nonumber\\
&
+ \mathcal{R}^{(\chi)}_{\text{RH}, \text{RH}} (\text{RH}) B_{\text{RH}}(\text{RH})
+ \mathcal{R}^{(\chi)}_{\text{RH}, \text{LH}} (\text{RH}) B_{\text{LH}}(\text{RH})
- \mathcal{R}^{(\chi)}_{\text{LH}, \text{RH}} (\text{LH}) B_{\text{RH}}(\text{LH})
- \mathcal{R}^{(\chi)}_{\text{LH}, \text{LH}} (\text{LH}) B_{\text{LH}}(\text{LH}), 
\end{align}
we rewrite the last four terms by using the identity
\begin{align}
& A\alpha + B\beta - C\gamma -D\delta = 
\left(\frac{A + D}{2} - \frac{B+C}{2}\right)(\alpha - \delta)
+ (B+C)\left(\frac{\alpha + \beta}{2} - \frac{\gamma + \delta}{2}\right) \nonumber\\
&\qquad\qquad + (A + B- C -D) \frac{\alpha + \beta + \gamma + \delta}{4} + \left(\frac{A + C}{2} - \frac{B + D}{2}\right)
\left(\frac{\alpha - \beta}{2} - \frac{\gamma - \delta}{2}\right), 
\end{align}
together with
\begin{subequations}
 \begin{align}
 (A, B, C, D) &= \left(
 \mathcal{R}^{(\chi)}_{\text{RH}, \text{RH}} (\text{RH}), 
 \mathcal{R}^{(\chi)}_{\text{RH}, \text{LH}} (\text{RH}), 
 \mathcal{R}^{(\chi)}_{\text{LH}, \text{RH}} (\text{LH}), 
 \mathcal{R}^{(\chi)}_{\text{LH}, \text{LH}} (\text{LH})
 \right), \\
 (\alpha, \beta, \gamma, \delta)&= \left(
 B_{\text{RH}}(\text{RH}), 
 B_{\text{LH}}(\text{RH}), 
 B_{\text{RH}}(\text{LH}), 
 B_{\text{LH}}(\text{LH}) \right). 
 \end{align}
\end{subequations}
This yields
\begin{align}
& \sum_{m = \text{L}, \text{RH}, \text{LH}}\left[
\mathcal{R}^{(\chi)}_{\text{RH}, m} (\Omega_{\bm{k}, \text{RH}}, \bm{k}_{\parallel}) B_{+, m} (\Omega_{\bm{k}, \text{RH}}, \bm{k}_{\parallel})
- \mathcal{R}^{(\chi)}_{\text{LH}, m} (\Omega_{\bm{k}, \text{LH}}, \bm{k}_{\parallel}) B_{+, m} (\Omega_{\bm{k}, \text{LH}}, \bm{k}_{\parallel})
\right] \nonumber\\
&=  \underbrace{\left[\mathcal{R}^{(\chi)}_{\text{RH}, \text{L}}(\text{RH}) - \mathcal{R}^{(\chi)}_{\text{LH}, \text{L}}(\text{LH})\right]}_{O(\chi^1)}
\frac{B_{\text{L}}(\text{RH}) + B_{\text{L}}(\text{LH})}{2} \nonumber\\
&\quad +
\frac{\mathcal{R}^{(\chi)}_{\text{RH}, \text{L}}(\text{RH}) + \mathcal{R}^{(\chi)}_{\text{LH}, \text{L}}(\text{LH})}{2} 
\underbrace{\left[B_{\text{L}}(\text{RH}) - B_{\text{L}}(\text{LH})\right]}_{O(\chi^1)} \nonumber\\
&\quad + \left[
\frac{\mathcal{R}^{(\chi)}_{\text{RH}, \text{RH}}(\text{RH}) + \mathcal{R}^{(\chi)}_{\text{LH}, \text{LH}}(\text{LH})}{2}
- \frac{\mathcal{R}^{(\chi)}_{\text{LH}, \text{RH}}(\text{LH}) + \mathcal{R}^{(\chi)}_{\text{RH}, \text{LH}}(\text{RH})}{2}
\right] \nonumber\\ &\quad \qquad \times \underbrace{\left[B_{\text{RH}}(\text{RH}) - B_{\text{LH}}(\text{LH})\right]}_{
O(\chi^1)} \nonumber\\
&\quad + \left[\mathcal{R}^{(\chi)}_{\text{LH}, \text{RH}}(\text{LH}) + \mathcal{R}^{(\chi)}_{\text{RH}, \text{LH}}(\text{RH})\right] \nonumber\\
&\quad \qquad \times
\underbrace{\left[\frac{B_{\text{RH}}(\text{RH}) + B_{\text{LH}}(\text{RH})}{2} - \frac{B_{\text{RH}}(\text{LH}) + B_{\text{LH}}(\text{LH})}{2}\right]}_{O(\chi^1)} \nonumber\\
&\quad + \left[
\underbrace{\mathcal{R}^{(\chi)}_{\text{RH}, \text{RH}}(\text{RH}) - \mathcal{R}^{(\chi)}_{\text{LH}, \text{LH}}(\text{LH})}_{O(\chi^1)}
\underbrace{- \mathcal{R}^{(\chi)}_{\text{LH}, \text{RH}}(\text{LH}) + \mathcal{R}^{(\chi)}_{\text{RH}, \text{LH}}(\text{RH})}_{O(\chi^1)}\right]
\nonumber\\ &\quad \qquad \times
\frac{B_{\text{RH}}(\text{RH}) + B_{\text{RH}}(\text{LH}) + B_{\text{LH}}(\text{RH}) + B_{\text{LH}}(\text{LH})}{4} \nonumber\\
&\quad + \underbrace{\left[
\frac{\mathcal{R}^{(\chi)}_{\text{RH}, \text{RH}}(\text{RH}) + \mathcal{R}^{(\chi)}_{\text{LH}, \text{RH}}(\text{LH})}{2}
- \frac{\mathcal{R}^{(\chi)}_{\text{RH}, \text{LH}}(\text{RH}) + \mathcal{R}^{(\chi)}_{\text{LH}, \text{LH}}(\text{LH})}{2}\right]}_{O(\chi^1)}  \nonumber\\
&\quad \qquad \times\underbrace{\left[\frac{B_{\text{RH}} (\text{RH}) - B_{\text{LH}} (\text{RH})}{2}
- \frac{B_{\text{RH}} (\text{LH}) - B_{\text{LH}} (\text{LH})}{2}\right]}_{O(\chi^1)}. \label{004243_10Jun26}
\end{align}
Some factors appearing in this expression can be further simplified as
\begin{subequations}
 \begin{align}
 \frac{B_{\text{L}}(\text{RH}) + B_{\text{L}}(\text{LH})}{2}
 &=  B_{\text{L}}(\text{T}) + O(\chi^2) = B_{-, \text{L}}(\Omega_{\bm{k}, \text{T}}, \bm{k}_{\parallel}) + O(\chi^2), \\
 \frac{\mathcal{R}^{(\chi)}_{\text{RH}, \text{L}}(\text{RH}) + \mathcal{R}^{(\chi)}_{\text{LH}, \text{L}}(\text{LH})}{2}
 &= \mathcal{R}_{\text{RH},\text{L}}(\text{T}) + O(\chi^2), \\
 \frac{\mathcal{R}^{(\chi)}_{\text{RH}, \text{RH}}(\text{RH}) + \mathcal{R}^{(\chi)}_{\text{LH}, \text{LH}}(\text{LH})}{2} &
 = \mathcal{R}_{\text{RH}, \text{RH}}(\text{T}) + O(\chi^2), \\
 \frac{\mathcal{R}^{(\chi)}_{\text{LH}, \text{RH}}(\text{LH}) + \mathcal{R}^{(\chi)}_{\text{RH}, \text{LH}}(\text{RH})}{2} &= 
 \mathcal{R}_{\text{RH}, \text{LH}}(\text{T}) + O(\chi^2), \\
\frac{B_{\text{RH}}(\text{RH}) + B_{\text{LH}}(\text{RH})}{2} &= B_{\text{T}}(\text{RH}), \\
 \frac{B_{\text{RH}}(\text{LH}) + B_{\text{LH}}(\text{LH})}{2} &= B_{\text{T}}(\text{LH}), \\
 \frac{B_{\text{RH}}(\text{RH}) + B_{\text{RH}}(\text{LH})}{2} &= B_{\text{RH}}(\text{T}) + O(\chi^2), \\
 \frac{B_{\text{LH}}(\text{RH}) + B_{\text{LH}}(\text{LH})}{2}&=  B_{\text{LH}}(\text{T}) + O(\chi^2), \\
 \frac{B_{\text{RH}}(\text{RH}) + B_{\text{RH}}(\text{LH}) + B_{\text{LH}}(\text{RH}) + B_{\text{LH}}(\text{LH})}{4}
 &= B_{\text{T}}(\text{T}) + O(\chi^2)
 \end{align}
\end{subequations}
Note that when both the original and approximated expressions are invariant under the interchange of the RH and LH modes, 
equivalently under $\chi\to-\chi$, the terms neglected in the approximation are of order $O(\chi^2)$ or higher.

We next consider the second-to-last term in Eq.~\eqref{004243_10Jun26}. It can be rewritten as
\begin{align}
& \mathcal{R}^{(\chi)}_{\text{RH}, \text{RH}}(\text{RH}) - \mathcal{R}^{(\chi)}_{\text{LH}, \text{LH}}(\text{LH})
- \mathcal{R}^{(\chi)}_{\text{LH}, \text{RH}}(\text{LH}) + \mathcal{R}^{(\chi)}_{\text{RH}, \text{LH}}(\text{RH}) \nonumber\\
&=
\left[ \mathcal{R}^{(\chi)}_{\text{RH}, \text{RH}}(\text{RH}) - \mathcal{R}^{(\chi)}_{\text{RH}, \text{RH}}(\text{T})  \right]
+ \left[ \mathcal{R}^{(\chi)}_{\text{RH}, \text{RH}}(\text{T}) - \mathcal{R}^{(\chi)}_{\text{LH}, \text{LH}}(\text{T})  \right]
+ \left[ \mathcal{R}^{(\chi)}_{\text{LH}, \text{LH}}(\text{T}) - \mathcal{R}^{(\chi)}_{\text{LH}, \text{LH}}(\text{LH})  \right] \nonumber\\
&\quad + 
\left[ \mathcal{R}^{(\chi)}_{\text{RH}, \text{LH}}(\text{RH}) - \mathcal{R}^{(\chi)}_{\text{RH}, \text{LH}}(\text{T})  \right]
+ \left[ \mathcal{R}^{(\chi)}_{\text{RH}, \text{LH}}(\text{T}) - \mathcal{R}^{(\chi)}_{\text{LH}, \text{RH}}(\text{T})  \right]
+ \left[ \mathcal{R}^{(\chi)}_{\text{LH}, \text{RH}}(\text{T}) - \mathcal{R}^{(\chi)}_{\text{LH}, \text{RH}}(\text{LH})  \right] \nonumber\\
&\simeq
\chi k^2 \left.
\frac{\partial \left[
\mathcal{R}_{\text{RH}, \text{RH}}(\omega, \bm{k}_{\parallel}) + \mathcal{R}_{\text{RH}, \text{LH}}(\omega, \bm{k}_{\parallel})\right]}{\partial \omega}
\right|_{\omega = \Omega_{\bm{k},\text{T}}} \nonumber\\
&\qquad + \left[ \mathcal{R}^{(\chi)}_{\text{RH}, \text{RH}}(\text{T}) - \mathcal{R}^{(\chi)}_{\text{LH}, \text{LH}}(\text{T})  \right]
+ \left[ \mathcal{R}^{(\chi)}_{\text{RH}, \text{LH}}(\text{T}) - \mathcal{R}^{(\chi)}_{\text{LH}, \text{RH}}(\text{T})  \right]. 
\end{align}
In the last line, we have used $\Omega_{\bm{k}, \text{RH}} - \Omega_{\bm{k}, \text{T}} =  \Omega_{\bm{k}, \text{T}} - \Omega_{\bm{k}, \text{LH}} = \chi k^2/2$, 
together with the approximation
\begin{subequations}
 \begin{align}
 \mathcal{R}^{(\chi)}_{\text{RH}, \text{RH}}(\text{RH}) - \mathcal{R}^{(\chi)}_{\text{RH}, \text{RH}}(\text{T}) &=
 \mathcal{R}^{(\chi)}_{\text{RH}, \text{RH}}(\Omega_{\bm{k}, \text{RH}}, \bm{k}_{\parallel}) - \mathcal{R}^{(\chi)}_{\text{RH}, \text{RH}}(\Omega_{\bm{k}, \text{T}}, \bm{k}_{\parallel}) \nonumber\\
& \simeq \frac{\chi k^2}{2}
 \left. \frac{\partial  \mathcal{R}_{\text{RH}, \text{RH}}(\omega, \bm{k}_{\parallel})}{\partial \omega}\right|_{\omega = \Omega_{\bm{k},\text{T}}}, \\
 \mathcal{R}^{(\chi)}_{\text{LH}, \text{LH}}(\text{T}) - \mathcal{R}^{(\chi)}_{\text{LH}, \text{LH}}(\text{LH})&\simeq \frac{\chi k^2}{2}
 \left. \frac{\partial  \mathcal{R}_{\text{LH}, \text{LH}}(\omega, \bm{k}_{\parallel})}{\partial \omega}\right|_{\omega = \Omega_{\bm{k},\text{T}}}
 = \frac{\chi k^2}{2}
 \left. \frac{\partial  \mathcal{R}_{\text{RH}, \text{RH}}(\omega, \bm{k}_{\parallel})}{\partial \omega}\right|_{\omega = \Omega_{\bm{k},\text{T}}}, \\
 \mathcal{R}^{(\chi)}_{\text{RH}, \text{LH}}(\text{RH}) - \mathcal{R}^{(\chi)}_{\text{RH}, \text{LH}}(\text{T})&\simeq
 \frac{\chi k^2}{2}
 \left. \frac{\partial  \mathcal{R}_{\text{RH}, \text{LH}}(\omega, \bm{k}_{\parallel})}{\partial \omega}\right|_{\omega = \Omega_{\bm{k},\text{T}}}, \\
 \mathcal{R}^{(\chi)}_{\text{LH}, \text{RH}}(\text{T}) - \mathcal{R}^{(\chi)}_{\text{LH}, \text{RH}}(\text{LH}) &\simeq
 \frac{\chi k^2}{2}\left. \frac{\partial  \mathcal{R}_{\text{LH}, \text{RH}}(\omega, \bm{k}_{\parallel})}{\partial \omega}\right|_{\omega = \Omega_{\bm{k},\text{T}}}
 = \frac{\chi k^2}{2}\left. \frac{\partial  \mathcal{R}_{\text{RH}, \text{LH}}(\omega, \bm{k}_{\parallel})}{\partial \omega}\right|_{\omega = \Omega_{\bm{k},\text{T}}}. 
 \end{align}
\end{subequations}

Furthermore, under the isotropic approximation, the phonon frequencies depend only on $|\bm{k}|$. Therefore,
$\Omega_{\bm{k}, n} = \Omega_{\overline{\bm{k}}, n}\quad (n=\mathrm{L},\mathrm{RH},\mathrm{LH})$, which implies
\begin{equation}
 B_n (n) = B_{+, n} (\Omega_{\bm{k}, n}, \bm{k}_{\parallel}) = B_{+, n} (\Omega_{\overline{\bm{k}}, n}, \bm{k}_{\parallel}) = B_{\overline{\bm{k}},n}. 
\end{equation}

Combining the above relations, we find that the $O(\chi)$ contribution of Eq.~\eqref{004243_10Jun26} reduces to Eq.~\eqref{195218_15Jun26}.
The same procedure can be applied to Eq.~\eqref{193227_15Jun26}. 
The resulting $O(\chi)$ contribution is given by Eq.~\eqref{195231_15Jun26}.

\subsubsection{Spatial profile of the phonon AM}

Substituting Eqs.~\eqref{195218_15Jun26} and \eqref{195231_15Jun26} into 
Eqs.~\eqref{193224_15Jun26} and \eqref{193227_15Jun26}, respectively, we obtain the spatial distribution of the phonon AM density:
\begin{subequations}
\begin{align}
 &\bm{S}(z<0) - \bm{S}_0 \simeq \int\limits_{k_z <0} \frac{\rmd^3\bm{k}}{(2\pi)^3} \hbar \hat{\bm{k}}  
\exp\left({\displaystyle \frac{-z}{\tau v_{\text{T}}\hat{k}_z}}\right) \nonumber\\
&\quad\times \Big\{
- \left(B_{\bm{k},\text{RH}} - B_{\bm{k},\text{LH}}\right)  \nonumber\\
&\quad  + \left[\mathcal{R}^{(\chi)}_{\text{RH}, \text{L}}(\text{RH}) - \mathcal{R}^{(\chi)}_{\text{LH}, \text{L}}(\text{LH})\right] B_{\text{L}}(\text{T})
 + \mathcal{R}_{\text{RH}, \text{L}}(\text{T}) \left[B_{\text{L}}(\text{RH}) - B_{\text{L}}(\text{LH})\right] \nonumber\\
&\quad + \left[\mathcal{R}_{\text{RH}, \text{RH}}(\text{T}) - \mathcal{R}_{\text{RH}, \text{LH}}(\text{T})\right]
\left[B_{\overline{\bm{k}}, \text{RH}} - B_{\overline{\bm{k}}, \text{LH}}\right]
+  2 \mathcal{R}_{\text{RH}, \text{LH}}(\text{T}) \left[ B_{\text{T}}(\text{RH}) - B_{\text{T}}(\text{LH}) \right] \nonumber\\
&\quad + \chi k^2 \left.
\frac{\partial \left[
\mathcal{R}_{\text{RH}, \text{RH}}(\omega, \bm{k}_{\parallel}) + \mathcal{R}_{\text{RH}, \text{LH}}(\omega, \bm{k}_{\parallel})\right]}{\partial \omega}
\right|_{\omega = \Omega_{\bm{k},\text{T}}} B_{\overline{\bm{k}},\text{T}} \nonumber\\
& \qquad + \left[ \mathcal{R}^{(\chi)}_{\text{RH}, \text{RH}}(\text{T}) - \mathcal{R}^{(\chi)}_{\text{LH}, \text{LH}}(\text{T})  \right] B_{\overline{\bm{k}},\text{T}}
+ \left[ \mathcal{R}^{(\chi)}_{\text{RH}, \text{LH}}(\text{T}) - \mathcal{R}^{(\chi)}_{\text{LH}, \text{RH}}(\text{T})  \right] B_{\overline{\bm{k}},\text{T}}
\Big\}, \label{180809_16Jun26}\\[10pt]
&{\bm{S}} (z>0) \simeq
\int\limits_{q_z > 0} \frac{\rmd^3\bm{q}}{(2\pi)^3} \hbar \hat{\bm{q}}\, \exp\left({\displaystyle \frac{-z}{\widetilde{\tau} c_{\text{T}} \hat{q}_z}}\right) \nonumber\\
&\times\Big\{
 \left[\mathcal{T}^{(\chi)}_{\text{RH}, \text{L}}(\text{RH}) - \mathcal{T}^{(\chi)}_{\text{LH}, \text{L}}(\text{LH})\right] \check{B}_{\text{L}}(\text{T})
 + \mathcal{T}_{\text{RH}, \text{L}}(\text{T}) \left[\check{B}_{\text{L}}(\text{RH}) - \check{B}_{\text{L}}(\text{LH})\right] \nonumber\\
& + \left[\mathcal{T}_{\text{RH}, \text{RH}}(\text{T}) - \mathcal{T}_{\text{RH}, \text{LH}}(\text{T})\right]
\left[\check{B}_{\text{RH}}(\text{RH}) - \check{B}_{\text{LH}}(\text{LH})\right]
+  2 \mathcal{T}_{\text{RH}, \text{LH}}(\text{T}) \left[ \check{B}_{\text{T}}(\text{RH}) - \check{B}_{\text{T}}(\text{LH}) \right] \nonumber\\
& + \chi k^2 \left.
\frac{\partial \left[
\mathcal{T}_{\text{RH}, \text{RH}}(\omega, \bm{q}_{\parallel}) + \mathcal{T}_{\text{RH}, \text{LH}}(\omega, \bm{q}_{\parallel})\right]}{\partial \omega}
\right|_{\omega = \Omega_{\bm{q},\text{T}}} \check{B}_{\text{T}}(\text{T}) \nonumber\\
& + \left[ \mathcal{T}^{(\chi)}_{\text{RH}, \text{RH}}(\text{T}) - \mathcal{T}^{(\chi)}_{\text{LH}, \text{LH}}(\text{T})  \right] \check{B}_{\text{T}}(\text{T}) 
+ \left[ \mathcal{T}^{(\chi)}_{\text{RH}, \text{LH}}(\text{T}) - \mathcal{T}^{(\chi)}_{\text{LH}, \text{RH}}(\text{T})  \right] \check{B}_{\text{T}}(\text{T})
\Big\}.
\label{180813_16Jun26}
\end{align} 
\end{subequations}

\subsection{Approximation adopted in the main text}
\label{132237_5Oct26}

\subsubsection{Result of the approximation}

Our primary goal is to estimate the magnitude and characteristic features of the interfacial transport of phonon AM.
To this end, we further simplify Eqs.~\eqref{180809_16Jun26} and \eqref{180813_16Jun26} 
by retaining the contributions included in the main-text treatment and neglecting the remaining first-order corrections. 
As discussed below, this procedure leads to
\begin{align}
 &\bm{S}(z<0) - \bm{S}_0 \simeq \int\limits_{k_z <0} \frac{\rmd^3\bm{k}}{(2\pi)^3} \hbar \hat{\bm{k}}  
\exp\left({\displaystyle \frac{-z}{\tau v_{\text{T}}\hat{k}_z}}\right) \nonumber\\
&\quad\times \left[ - \left(B_{\bm{k},\text{RH}} - B_{\bm{k},\text{LH}}\right) 
+ \left[\mathcal{R}_{\text{RH}, \text{RH}}(\text{T}) - \mathcal{R}_{\text{RH}, \text{LH}}(\text{T})\right]
\left[B_{\overline{\bm{k}}, \text{RH}} - B_{\overline{\bm{k}}, \text{LH}}\right] 
\right]\label{084051_11Jun26}, \\
&{\bm{S}} (z>0) \simeq \int\limits_{k_z > 0} \frac{\rmd^3 \bm{k}}{(2\pi)^3} 
\frac{v_{\text{T}}\cos\theta_{\text{T}}}{c_{\text{T}}\cos\Theta_{\text{T}}(\bm{k})}
\hbar \hat{\bm{q}}(\bm{k})
\exp\left(\frac{- z}{\widetilde{\tau}c_{\text{T}}\hat{q}_z(\bm{k})}\right) \nonumber\\
&\times \left[
\mathcal{T}_{\text{RH}, \text{RH}} (\Omega_{\bm{k}\text{T}}, \bm{k}_{\parallel}) - \mathcal{T}_{\text{RH}, \text{LH}} (\Omega_{\bm{k}\text{T}}, \bm{k}_{\parallel})\right]
\left[B_{\bm{k}, \text{RH}} - B_{\bm{k}, \text{LH}} \right]. 
\label{182750_16Jun26}
\end{align}
For Eq.~\eqref{182750_16Jun26} in the region $z>0$, the definitions of the angles $\theta_{\mathrm{T}}$ and $\Theta_{\mathrm{T}}(\bm{k})$, 
as well as the unit vector $\hat{\bm{q}}(\bm{k})$, will be given later in the discussion of Eqs.~\eqref{140312_18Jun26}--\eqref{142811_18Jun26}.

The above approximation retains only the asymmetry of the phonon distribution functions between the RH and LH modes,
$B_{\bm{k}, \text{RH}} - B_{\bm{k}, \text{LH}}$ and $B_{\overline{\bm{k}}, \text{RH}} - B_{\overline{\bm{k}}, \text{LH}}$, 
which originate from the chirality-induced splitting of the transverse modes. 
All other consequences of the RH-LH splitting, such as the difference in group velocities and the RH-LH asymmetry of the reflectance and transmittance, are neglected.

\subsubsection{Discussion of the approximation}

We first note that the intermode reflection and transmission coefficients,
$\mathcal{R}_{nm}$, $\mathcal{T}_{nm}$, $\mathcal{R}^{(\chi)}_{nm}$, and $\mathcal{T}^{(\chi)}_{nm}$ with $n\neq m$,
are generally much smaller than the intramode coefficients, $\mathcal{R}_{nn}$, $\mathcal{T}_{nn}$, $\mathcal{R}^{(\chi)}_{nn}$, and $\mathcal{T}^{(\chi)}_{nn}$. 
Therefore, the terms proportional to intermode reflection or transmission coefficients are expected to provide only subleading contributions and will be neglected.

One exception is worth noting.
Although the $\mathcal{R}_{\mathrm{RH},\mathrm{LH}}(\mathrm{T})$ term in the final expression~\eqref{084051_11Jun26} is expected to be negligible compared with 
the $\mathcal{R}_{\mathrm{RH},\mathrm{RH}}(\mathrm{T})$ term, we retain both terms. 
These two terms arise together in the derivation, and neglecting only the former would lead to a physically inconsistent approximation.

We next neglect the terms containing $\partial \mathcal{R}_{\text{RH}, \text{RH}}(\omega, \bm{k}_{\parallel})/\partial \omega$ at the frequency
$\omega = \Omega_{\bm{k}, \text{T}} = v_{\text{T}}k$. 
Although $\mathcal{R}_{\mathrm{RH},\mathrm{RH}}$ appears to depend on two variables, $(\omega,\bm{k}_{\parallel})$,
it is in fact a function only of the incidence angle $\theta_{\mathrm{T}}$:
\begin{align}
 \sin \theta_{\text{T}} = \frac{|\bm{k}_{\parallel}|}{k} = \frac{v_{\text{T}}|\bm{k}_{\parallel}|}{\omega}.
\end{align}
Namely, $\mathcal{R}_{\mathrm{RH},\mathrm{RH}} (\omega,\bm{k}_{\parallel}) =  \mathcal{R}_{\text{RH}, \text{RH}}(\theta_{\text{T}})$. 

For fixed $\bm{k}_{\parallel}$, an infinitesimal change in frequency and that in the incidence angle satisfy
\begin{equation}
\cos\theta_{\text{T}} \rmd \theta_{\text{T}}
= - \frac{v_{\text{T} } |\bm{k}_{\parallel}|}{\omega^2}\rmd \omega
= - \frac{\sin\theta_{\text{T}} }{v_{\text{T}} k} \rmd \omega, \qquad\quad \therefore\, 
\frac{\partial \theta_{\text{T}}(\omega, \bm{k}_{\parallel})}{\partial \omega}
= - \frac{\tan\theta_{\text{T}}}{v_{\text{T}} k}, 
\end{equation}
which implies
\begin{equation}
 \chi k^2 \left.
\frac{\partial \mathcal{R}_{\text{RH}, \text{RH}}(\omega, \bm{k}_{\parallel})}{\partial \omega} \right|_{\omega = \Omega_{\bm{k}, \text{T}}}
= \chi k^2 
\frac{\partial \theta_{\text{T}}}{\partial \omega}
\frac{\partial \mathcal{R}_{\text{RH}, \text{RH}}(\theta_{\text{T}})}{\partial \theta_{\text{T}}} 
= - \frac{\chi k}{v_{\text{T}}} \tan\theta_{\text{T}}
\frac{\partial \mathcal{R}_{\text{RH}, \text{RH}}(\theta_{\text{T}})}{\partial \theta_{\text{T}}}. 
\label{001824_11Jun26}
\end{equation}

This contribution can become appreciable only when $\tan\theta_{\mathrm{T}}$ is large, namely for grazing incidence $\theta_{\mathrm{T}}\sim\pi/2$,
and when the reflection coefficient varies appreciably with the incidence angle.
In practice, however, total reflection frequently occurs in this regime.
The reflection coefficient then becomes nearly constant, and the above contribution is expected to be negligible.
For this reason, we neglect the term in Eq.~\eqref{001824_11Jun26}.
The same argument also applies to the terms containing
$\partial\mathcal{T}_{\mathrm{RH},\mathrm{RH}}/\partial\omega$.

Finally, we neglect the terms proportional to
$\left[ \mathcal{R}^{(\chi)}_{\text{RH}, \text{RH}}(\text{T}) - \mathcal{R}^{(\chi)}_{\text{LH}, \text{LH}}(\text{T})  \right]$ and
$\left[ \mathcal{T}^{(\chi)}_{\text{RH}, \text{RH}}(\text{T}) - \mathcal{T}^{(\chi)}_{\text{LH}, \text{LH}}(\text{T})  \right]$, 
for simplicity.
Evaluating these quantities requires a formulation of reflection and transmission coefficients for elastic media whose acoustic dispersion deviates from the linear form; 
if one assumes perfectly linear dispersion, these factors vanish identically
$\mathcal{R}_{\text{RH}, \text{RH}}(\text{T}) - \mathcal{R}_{\text{LH}, \text{LH}}(\text{T}) = 
\mathcal{T}_{\text{RH}, \text{RH}}(\text{T}) - \mathcal{T}_{\text{LH}, \text{LH}}(\text{T}) = 0$. 
However, a proper treatment of chirality-induced mode splitting beyond the linear-dispersion approximation would require a framework 
extending conventional elasticity theory. Such an analysis is beyond the scope of the present work.

With the above approximations, Eq.~\eqref{084051_11Jun26} is obtained for $\bm{S}(z<0)$.
{For $\bm{S}(z>0)$, we make the same approximations and neglect the following $O(\chi)$ contributions in Eq.~\eqref{180813_16Jun26}: 
(i)~transmission between different modes, (ii)~terms involving frequency derivatives of the transmission coefficients, and 
(iii)~contributions involving the full transmission coefficients, evaluated by taking into account the $O(k^2)$ splitting of the modes in the chiral crystal. 
This approximation gives}
\begin{align}
 &{\bm{S}} (z>0) \simeq 
\int\limits_{q_z > 0} \frac{\rmd^3\bm{q}}{(2\pi)^3} \hbar \hat{\bm{q}}\, \exp\left({\displaystyle \frac{-z}{\widetilde{\tau} c_{\text{T}} \hat{q}_z}}\right)
 \left[\mathcal{T}_{\text{RH}, \text{RH}}(\text{T}) - \mathcal{T}_{\text{RH}, \text{LH}}(\text{T})\right]
\left[\check{B}_{\text{RH}}(\text{RH}) - \check{B}_{\text{LH}}(\text{LH})\right] \nonumber\\
&= \int\limits_{q_z > 0} \frac{\rmd^3 \bm{q}}{(2\pi)^3} \hbar \hat{\bm{q}}
\exp\left(\frac{- z}{\widetilde{\tau}c_{\text{T}}\hat{q}_z}\right) \nonumber\\
&\times \left[
\mathcal{T}_{\text{RH}, \text{RH}} (\omega_{\bm{q}\text{T}}, \bm{q}_{\parallel}) - \mathcal{T}_{\text{RH}, \text{LH}} (\omega_{\bm{q}\text{T}}, \bm{q}_{\parallel})\right]
\left[B_{+, \text{RH}}(\omega_{\bm{q}\text{T}}, \bm{q}_{\parallel}) - B_{+, \text{LH}}(\omega_{\bm{q}\text{T}}, \bm{q}_{\parallel})\right], 
\label{124724_18Jun26}
\end{align}
where we have assumed that the crystal occupying the region $z>0$ is achiral and used
$\omega_{\bm{q}, \text{RH}} = \omega_{\bm{q}, \text{LH}} = \omega_{\bm{q}, \text{T}} $. 

We then transform the integration variable from the transmitted transverse-wave wavevector $\bm{q}$ ($q_z>0$)
to the incident transverse-wave wavevector $\bm{k}$ ($k_z>0$) on the chiral side.
The transformation is based on $\omega_{\bm{q}\text{T}} = \Omega_{\bm{k}\text{T}}$ and $\bm{q}_{\parallel} = \bm{k}_{\parallel}$. 

The resulting expression is
\begin{align}
  &{\bm{S}} (z>0) \simeq \int\limits_{k_z > 0} \frac{\rmd^3 \bm{k}}{(2\pi)^3} 
\frac{v_{\text{T}}\cos\theta_{\text{T}}}{c_{\text{T}}\cos\Theta_{\text{T}}(\bm{k})}
\hbar \hat{\bm{q}}(\bm{k})
\exp\left(\frac{- z}{\widetilde{\tau}c_{\text{T}}\hat{q}_z(\bm{k})}\right) \nonumber\\
&\times \left[
\mathcal{T}_{\text{RH}, \text{RH}} (\Omega_{\bm{k}\text{T}}, \bm{k}_{\parallel}) - \mathcal{T}_{\text{RH}, \text{LH}} (\Omega_{\bm{k}\text{T}}, \bm{k}_{\parallel})\right]
\left[B_{+, \text{RH}}(\Omega_{\bm{k}\text{T}}, \bm{k}_{\parallel}) - B_{+, \text{LH}}(\Omega_{\bm{k}\text{T}}, \bm{k}_{\parallel})\right], 
\label{140312_18Jun26}
\end{align}
where the Jacobian factor is given by
\begin{equation}
 \rmd^3\bm{q} = \frac{\rmd \omega_{\bm{q}\text{T}}\rmd^2\bm{q}_{\parallel}}{c_{\text{T}}\cos\Theta_{\text{T}}}
= \frac{\rmd \Omega_{\bm{k}\text{T}}\rmd^2\bm{k}_{\parallel}}{c_{\text{T}}\cos\Theta_{\text{T}}(\bm{k})}
= \rmd^3\bm{k} \frac{v_{\text{T}}\cos\theta_{\text{T}}}{c_{\text{T}}\cos\Theta_{\text{T}}(\bm{k})}. 
\end{equation}
Here, $\theta_{\mathrm{T}}$ denotes the angle between $\bm{k}$ and the $z$ axis, namely the incidence angle.
The angle $\Theta_{\mathrm{T}}(\bm{k})$ is the refraction angle and satisfies Snell's law,
$\sin\Theta_{\text{T}}(\bm{k}) = (c_{\text{T}}/ v_{\text{T}}) \sin\theta_{\text{T}}$. 
The unit vector $\hat{\bm{q}}(\bm{k})$ can also be written as
\begin{equation}
 \hat{\bm{q}}(\bm{k}) = \frac{\bm{q}_{\parallel} + q_z \hat{\bm{z}}}{\omega_{\bm{q}\text{T}}/c_{\text{T}}}
= \frac{\bm{k}_{\parallel} + 
\sqrt{(\Omega_{\bm{k}\text{T}}/c_{\text{T}})^2 - \Abs{\bm{k}_{\parallel}}^2}\, 
\hat{\bm{z}}}{\Omega_{\bm{k}\text{T}}/c_{\text{T}}}
= \frac{c_{\text{T}}}{v_{\text{T}}}
\left(\frac{k_x}{k} \hat{\bm{x}} + \frac{k_y}{k} \hat{\bm{y}}\right)
+ \sqrt{1- \left(\frac{c_{\text{T}}}{v_{\text{T}}}\right)^2 \sin^2\theta_{\text{T}}} \hat{\bm{z}}. 
\label{142811_18Jun26}
\end{equation}

Finally, to make the expression for $\bm{S}(z>0)$ consistent with Eq.~\eqref{084051_11Jun26} for $\bm{S}(z<0)$, 
we further simplify the phonon-distribution imbalance appearing in Eq.~\eqref{140312_18Jun26}:
\begin{align}
& B_{+, \text{RH}}(\Omega_{\bm{k}\text{T}}, \bm{k}_{\parallel}) - B_{+, \text{LH}}(\Omega_{\bm{k}\text{T}}, \bm{k}_{\parallel}) \nonumber\\
&= 
\left[B_{+, \text{RH}}(\Omega_{\bm{k}\text{RH}}, \bm{k}_{\parallel}) - B_{+, \text{LH}}(\Omega_{\bm{k}\text{LH}}, \bm{k}_{\parallel}) \right]
\nonumber\\
& + 
\left[B_{+, \text{RH}}(\Omega_{\bm{k}\text{T}}, \bm{k}_{\parallel}) - B_{+, \text{RH}}(\Omega_{\bm{k}\text{RH}}, \bm{k}_{\parallel}) \right]
+ \left[B_{+, \text{LH}}(\Omega_{\bm{k}\text{LH}}, \bm{k}_{\parallel}) - B_{+, \text{LH}}(\Omega_{\bm{k}\text{T}}, \bm{k}_{\parallel})\right] \nonumber\\
&=  B_{\bm{k}, \text{RH}} - B_{\bm{k}, \text{LH}} 
- \frac{\chi k^2 }{2}
 \left.\frac{\partial B_{+, \text{RH}}(\omega, \bm{k}_{\parallel})}{\partial \omega}\right|_{\omega = \Omega_{\bm{k}\text{T}}}
- \frac{\chi k^2 }{2}
 \left.\frac{\partial B_{+, \text{LH}}(\omega, \bm{k}_{\parallel})}{\partial \omega}\right|_{\omega = \Omega_{\bm{k}\text{T}}}
+ O(\chi^3) \nonumber\\
&=  B_{\bm{k}, \text{RH}} - B_{\bm{k}, \text{LH}} - \chi k^2 
\left.\frac{\partial B_{+, \text{T}}(\omega, \bm{k}_{\parallel})}{\partial \omega}\right|_{\omega = \Omega_{\bm{k}\text{T}}} + O(\chi^3)\label{105635_5Oct26}
\end{align}
by retaining only
\begin{equation}
B_{+, \text{RH}}(\Omega_{\bm{k}\text{T}}, \bm{k}_{\parallel}) - B_{+, \text{LH}}(\Omega_{\bm{k}\text{T}}, \bm{k}_{\parallel})
\simeq B_{\bm{k}, \text{RH}} - B_{\bm{k}, \text{LH}}. \label{105616_5Oct26}
\end{equation}
{Within this additional approximation, we finally obtain the expression in Eq.~\eqref{182750_16Jun26}.}

{We note that the subsequent analysis can also be carried out without making the additional approximation in Eq.~\eqref{105616_5Oct26}, 
i.e., by retaining the expression in Eq.~\eqref{105635_5Oct26}. 
For completeness, we examine this case in Sec.~\hyperref[sec:approximation_assessment]{S2 C 5}, where we derive the resulting AM density in the achiral crystal and discuss the implications for the spatial profiles and the quantitative estimates.}

\subsection{Analytical expressions and upper-bound estimates after the approximation}
\label{105938_5Oct26}

In this subsection, we further simplify Eqs.~\eqref{084051_11Jun26} and \eqref{182750_16Jun26} obtained after the approximation discussed above.
For completeness, we also provide derivations omitted in the main text, 
including the RH--LH imbalance of the phonon distribution function in the bulk [Eq.~(8)] 
and the analytical expressions for the AM flux [Eqs.~(10a) and (10b)].

\subsubsection{RH--LH imbalance of the bulk phonon distribution}

We first evaluate the difference between the RH and LH contributions in the bulk region of the chiral crystal ($z<0$).
The quadratic splitting term $\chi k^2$ leads to the dispersions
$\Omega_{\bm{k},\text{RH}} \simeq v_{\text{T}}k + {\chi k^2}/{2}$ and  
$\Omega_{\bm{k},\text{LH}} \simeq v_{\text{T}}k - {\chi k^2}/{2}$. As a result, the quantity
$B_{\bm{k}n}= - (\tau \bm{v}_{\bm{k}n}\cdot \bm{\nabla}T)\, (\partial F^{(0)}_{\bm{k}n}/\partial T)$
acquires an RH--LH imbalance:
\begin{align}
& B_{\bm{k}, \text{RH}} - B_{\bm{k}, \text{LH}} \nonumber\\
&= -\tau \left(v_{\text{T}} + \chi k\right)\hat{\bm{k}}\cdot \bm{\nabla}T
\frac{\partial f^{\text{eq}}(v_{\text{T}}k + \chi k^2/2, T)}{\partial T} + \tau \left(v_{\text{T}} - \chi k\right)\hat{\bm{k}}\cdot \bm{\nabla}T
\frac{\partial f^{\text{eq}}(v_{\text{T}}k - \chi k^2/2, T)}{\partial T} \nonumber\\
&\simeq \frac{\tau\chi \bm{k}\cdot\bm{\nabla}T}{T}\,Q\left(\frac{\hbar v_{\text{T}} k}{2k_{\text{B}}T}\right). 
\label{112142_19Jun26}
\end{align}
In the last line, we have retained only the terms linear in the small parameter $\chi$.
The function $Q(w)$ is defined by
\begin{equation}
Q(w)\equiv  \frac{2w e^{2w} \left[2w +3 + (2w-3)e^{2w}\right]}{\left(e^{2w}-1\right)^3}  = \frac{\coth w - \frac{3}{2w}}{\sinh^2 w/w^2}.
\end{equation}

\subsubsection{Analytical expression for the AM density in the chiral crystal}

We next simplify Eq.~\eqref{084051_11Jun26} for the AM density in the chiral crystal ($z<0$).
We parameterize the integration variable as
$\bm{k}= (k_x, k_y, k_z < 0) = (k\sin\theta_{\text{T}}\cos\phi, k\sin\theta_{\text{T}}\sin\phi, - k\cos\theta_{\text{T}})$, 
where $0\le\theta_{\mathrm{T}}\le\pi/2$
is the reflection angle of the transverse wave.
Since the reflectance appearing in Eq.~(\ref{084051_11Jun26}) depends only on $\theta_{\mathrm{T}}$, we introduce
\begin{equation}
\mathcal{R}_{\text{RH}, \text{RH}}(\text{T}) - \mathcal{R}_{\text{RH}, \text{LH}}(\text{T})
= \mathcal{R}_{\text{RH}, \text{RH}}(\theta_{\text{T}}) - \mathcal{R}_{\text{RH}, \text{LH}}(\theta_{\text{T}})
\equiv  \Delta \mathcal{R}(\theta_{\text{T}}). 
\end{equation}

Using these definitions, we obtain the $i=x,y,z$ components of the AM density:
\begin{align}
 {S}_i(z<0) - {S}_{0, i} &\simeq \frac{\hbar \tau\chi}{T}
\int_0^{\infty} \frac{4\pi k^2\rmd k}{(2\pi)^3} k Q\left(\frac{\hbar v_{\text{T}} k}{2k_{\text{B}}T}\right) \nonumber\\
&\times \int_0^1 \frac{\rmd \cos\theta_{\text{T}}}{2}
\exp\left(\frac{z/\tau v_{\text{T}}}{\cos\theta_{\text{T}}}\right)\sum_{j = x, y, z}
\left[
- \Braket{\frac{k_i k_j}{k^2}}
+ \Delta \mathcal{R}(\theta_{\text{T}}) \Braket{\frac{k_i \overline{k}_j}{k^2}}
\right]\frac{\partial T}{\partial r_j}.
\end{align}
Here we have introduced the azimuthal average
$\Braket{\cdots} = \int_0^{2\pi} \cdots \rmd \phi/(2\pi)$. 
The coefficients arising from this average are given by
\begin{align}
\Braket{\frac{k_i k_j}{k^2}} &= \delta_{ij}
\begin{cases}
\cos^2\theta_{\text{T}} & (i = z) \\ \frac{1}{2}\sin^2\theta_{\text{T}} & (i = x, y)
\end{cases}, 
&\Braket{\frac{k_i \overline{k}_j}{k^2}} &= \delta_{ij}
\begin{cases}
- \cos^2\theta_{\text{T}} & (i = z) \\ \frac{1}{2}\sin^2\theta_{\text{T}} & (i = x, y)
\end{cases}. 
\label{182319_19Jun26}
\end{align}

The radial integral over $k$ can be evaluated analytically:
\begin{equation}
\frac{4\pi}{(2\pi)^3} \frac{\hbar \tau\chi}{T} \left(\frac{2k_{\text{B}}T}{\hbar v_{\text{T}} } \right)^4
\underbrace{\int_0^{\infty}\rmd w\, w^3 Q(w)}_{\pi^4/30}
= 3\alpha (T) \label{152910_18Jun26}
\end{equation}
Here, $\alpha(T)$
is the coefficient appearing in the bulk AM density, $\bm{S}_0 = \alpha (T) \bm{\nabla}T$, introduced in the main text [Eq.~(9)].

Combining the above results, we obtain
\begin{align}
  &{S}_i(z<0) - {S}_{0, i} \simeq -3 \alpha (T) \frac{\partial T}{\partial r_i}
\int^1_0 \frac{\rmd \cos\theta_{\text{T}}}{2}\exp\left(\frac{z/\tau v_{\text{T}}}{\cos\theta_{\text{T}}}\right)
\begin{cases}
\cos^2\theta_{\text{T}} \left[1 + \Delta \mathcal{R}(\theta_{\text{T}})\right] & (i = z)\\
\frac{1}{2}\sin^2\theta_{\text{T}} \left[1- \Delta \mathcal{R}(\theta_{\text{T}})\right] & (i = x, y)
\end{cases}. 
\end{align}

The AM flux $\bm{j}^{\mathrm{S}}(z<0)$ can then be obtained from the continuity equation
$\partial \bm{j}^{\text{S}} (z)/{\partial z} = - \left[\bm{S}(z) - \bm{S}_{0}\right]/\tau$, 
together with the boundary condition that the flux vanishes far from the interface, $\bm{j}^{\mathrm{S}}(z\to-\infty)=0$.
The result is
\begin{align}
 {j}_i^{\text{S}} (z < 0) = \alpha (T) v_{\text{T}} \frac{\partial T}{\partial r_i}
\begin{cases}
\frac32 \int^{\pi/2}_0 \rmd \theta \sin\theta \cos^3\theta 
\left[1 + \Delta \mathcal{R}(\theta_{\text{T}})\right] \exp\left(- \frac{\Abs{z/\tau v_{\text{T}}}}{\cos\theta_{\text{T}}}\right) & (i = z)\\
\frac34 \int^{\pi/2}_0 \rmd \theta \sin^3\theta \cos\theta \left[
1 - \Delta \mathcal{R}(\theta_{\text{T}})\right] \exp\left(- \frac{\Abs{z/\tau v_{\text{T}}}}{\cos\theta_{\text{T}}}\right) & (i = x, y)
\end{cases}
\end{align}
This expression corresponds to the $z<0$ part of Eqs.~(10a) and (10b) in the main text.

\subsubsection{Analytical expression for the AM density in the achiral crystal}

We now perform an analogous calculation for the achiral crystal occupying the region $z>0$.
We first substitute Eq.~\eqref{112142_19Jun26} into the factor $\left[B_{\bm{k}, \text{RH}} - B_{\bm{k}, \text{LH}} \right]$
appearing in Eq.~\eqref{182750_16Jun26}.
We then parameterize the integration variable as
$\bm{k}= (k_x, k_y, k_z > 0) = (k\sin\theta_{\text{T}}\cos\phi, k\sin\theta_{\text{T}}\sin\phi, k\cos\theta_{\text{T}})$. 
Since the transmittance also depends only on the incidence angle, we introduce
\begin{equation}
\Delta \mathcal{T}(\theta_{\text{T}})\equiv 
\mathcal{T}_{\text{RH}, \text{RH}} (\Omega_{\bm{k}\text{T}}, \bm{k}_{\parallel}) - \mathcal{T}_{\text{RH}, \text{LH}} (\Omega_{\bm{k}\text{T}}, \bm{k}_{\parallel}).
\label{123911_19Jun26}
\end{equation}
Substituting the expressions for $\Theta_{\mathrm{T}}(\bm{k})$ and $\hat{\bm{q}}(\bm{k})$ introduced in Eqs.~\eqref{140312_18Jun26}--\eqref{142811_18Jun26}, we obtain
\begin{align}
 {\bm{S}}_i(z>0)  &\simeq \frac{\hbar \tau\chi}{T}
\int_0^{\infty} \frac{4\pi k^2\rmd k}{(2\pi)^3} k Q\left(\frac{\hbar v_{\text{T}} k}{2k_{\text{B}}T}\right) \nonumber\\
&\times \int_0^1 \frac{\rmd \cos\theta_{\text{T}}}{2}
\exp\left(\frac{-z/\widetilde{\tau} c_{\text{T}}}{\sqrt{1- \left(c_{\text{T}}/v_{\text{T}}\right)^2\sin^2\theta_{\text{T}}}}\right)
\frac{(v_{\text{T}}/c_{\text{T}}) \cos\theta_{\text{T}}}{\sqrt{1- \left(c_{\text{T}}/v_{\text{T}}\right)^2\sin^2\theta_{\text{T}}}}\Delta \mathcal{T}(\theta_{\text{T}}) \nonumber\\
&\times \sum_{j = x, y, z}\left[
\frac{c_{\text{T}}}{v_{\text{T}}}\Braket{\frac{k_x k_j}{k^2}} \hat{\bm{x}}
+ \frac{c_{\text{T}}}{v_{\text{T}}}\Braket{\frac{k_y k_j}{k^2}} \hat{\bm{y}}
+ \sqrt{1- \left(\frac{c_{\text{T}}}{v_{\text{T}}}\right)^2\sin^2\theta_{\text{T}}} \Braket{\frac{k_z}{k}} \hat{\bm{z}}
\right]\frac{\partial T}{\partial r_j}.
\end{align}
The angular region satisfying $\sin\theta_{\text{T}} > v_{\text{T}}/ c_{\text{T}}$ may be excluded from the integration.
Indeed, this corresponds to total reflection of the transverse wave, for which $\Delta \mathcal{T}(\theta_{\text{T}}) = 0$, and therefore does not contribute.
The coefficients obtained after azimuthal averaging are
\begin{align}
\Braket{\frac{k_x k_j}{k^2}} &= \frac{\delta_{xj}}{2}\sin^2\theta_{\text{T}}, 
&\Braket{\frac{k_y k_j}{k^2}} &= \frac{\delta_{yj}}{2}\sin^2\theta_{\text{T}}, 
&\Braket{\frac{k_z}{k}} &= \delta_{zj} \cos\theta_{\text{T}}. 
\end{align}
The radial integral over $k$ again becomes proportional to the coefficient $\alpha(T)$, as in Eq.~\eqref{152910_18Jun26}.

We therefore arrive at
\begin{align}
  {S}_i(z>0)\simeq 3 \alpha (T) \frac{\partial T}{\partial r_i}
\int^1_0 \frac{\rmd \cos\theta_{\text{T}}}{2}
& \exp\left(\frac{- z/\widetilde{\tau} c_{\text{T}}}{\sqrt{1- \left(c_{\text{T}}/v_{\text{T}}\right)^2\sin^2\theta_{\text{T}}}}\right)
\frac{(v_{\text{T}}/c_{\text{T}}) \cos\theta_{\text{T}}}{\sqrt{1- \left(c_{\text{T}}/v_{\text{T}}\right)^2\sin^2\theta_{\text{T}}}}\Delta \mathcal{T}(\theta_{\text{T}}) \nonumber\\
&\times 
\begin{cases}
\cos\theta_{\text{T}} \sqrt{1- \left(\frac{c_{\text{T}}}{v_{\text{T}}}\right)^2\sin^2\theta_{\text{T}}} & (i = z)\\
\frac{c_{\text{T}}}{v_{\text{T}}}\frac{\sin^2\theta_{\text{T}} }{2} & (i = x, y)
\end{cases}. \label{112957_19Jun26}
\end{align}
The AM flux $\bm{j}^{\mathrm{S}}(z>0)$ is obtained by solving
$\partial \bm{j}^{\text{S}} (z)/{\partial z} = - {\bm{S}}(z) /\widetilde{\tau}$
subject to the boundary condition $\bm{j}^{\mathrm{S}}(z\to+\infty)=0$.
The result is
\begin{align}
 {j}_i^{\text{S}} (z > 0) = \frac32\alpha (T) v_{\text{T}} \frac{\partial T}{\partial r_i}
\int^{\pi/2}_0 \rmd \theta &\sin\theta \cos\theta
\Delta \mathcal{T}(\theta) \exp\left(- \frac{\Abs{z/\widetilde{\tau} c_{\text{T}}}}{
\sqrt{1- \left(c_{\text{T}}/v_{\text{T}}\right)^2\sin^2\theta}}\right) \nonumber\\
&\times
\begin{cases}
 \cos\theta \sqrt{1- \left(\frac{c_{\text{T}}}{v_{\text{T}}}\right)^2\sin^2\theta}
 & (i = z)\\
\frac{c_{\text{T}}}{v_{\text{T}}}\frac{\sin^2\theta }{2}
  & (i = x, y)
\end{cases}. 
\label{113003_19Jun26}
\end{align}
This expression corresponds to the $z>0$ part of Eqs.~(10a) and (10b) in the main text.

\subsubsection{Upper-bound estimate for the transmitted AM}

For brevity, we write the right-hand sides of Eqs.~\eqref{112957_19Jun26} and \eqref{113003_19Jun26} as 
$S^{\mathrm{approx}}_i(z>0)$ and $j^{\mathrm{S,approx}}_i(z>0)$, respectively.
Normalizing these quantities by the bulk AM density $\bm{S}_0 = \alpha (T) \bm{\nabla} T$, 
provides a natural measure of the AM transmission efficiency.
We now derive upper bounds for these quantities.

In Eqs.~\eqref{112957_19Jun26} and \eqref{113003_19Jun26}, the exponential factor is bounded from above by its value at the interface ($z=+0$).
Moreover, ${\Delta \mathcal{T}} =  {\mathcal{T}_{\text{RH}, \text{RH}}- \mathcal{T}_{\text{RH}, \text{LH}}} \leq  1$. 
Taking the integration domain into account, we obtain
\begin{align}
\frac{{S}^{\text{approx}}_i(z>0)}{S_{0, i}} &\leq 
\frac{3}{2} \int\limits_{\substack{0 < \theta < \pi/2 \\ \sin\theta < v_{\text{T}}/c_{\text{T}}}}
\rmd \theta \frac{v_{\text{T}}/c_{\text{T}}\sin\theta \cos\theta}{
\sqrt{1 - \left(c_{\text{T}}/ v_{\text{T}}\right)^2\sin^2\theta}}
\begin{cases}
\cos\theta_{\text{T}} \sqrt{1- \left(\frac{c_{\text{T}}}{v_{\text{T}}}\right)^2\sin^2\theta_{\text{T}}} & (i = z)\\
\frac{c_{\text{T}}}{v_{\text{T}}}\frac{\sin^2\theta_{\text{T}} }{2} & (i = x, y)
\end{cases}, \\
\frac{ {j}^{\text{S, approx}}_i (z > 0)}{S_{0, i} v_{\text{T}}}
 &= \frac32 \int\limits_{\substack{0 < \theta < \pi/2 \\ \sin\theta < v_{\text{T}}/c_{\text{T}}}}
 \rmd \theta \sin\theta \cos\theta
\begin{cases}
 \cos\theta \sqrt{1- \left(\frac{c_{\text{T}}}{v_{\text{T}}}\right)^2\sin^2\theta}
 & (i = z)\\
\frac{c_{\text{T}}}{v_{\text{T}}}\frac{\sin^2\theta }{2}
  & (i = x, y)
\end{cases}. 
\end{align}
Introducing the variables
\begin{align}
 w &= \sin^2\theta, &a& = \min\left\{1, \frac{v_{\text{T}}}{c_{\text{T}}}\right\} > 0, 
\end{align}
we find
\begin{subequations}
 \begin{align}
 \frac{{S}^{\text{approx}}_z(z>0)}{S_{0, z}} &\leq \frac{3}{4} \frac{v_{\text{T}}}{c_{\text{T}}} \int^{a^2}_0 \rmd w\sqrt{1-w}, \\
 \frac{{S}^{\text{approx}}_i(z>0)}{S_{0, i}} &\leq \frac{3}{8} \int^{a^2}_0 \rmd w
 \frac{w}{\sqrt{1-\left(c_{\text{T}}/v_{\text{T}}\right)^2w}}, \qquad (i = x, y), \\
 \frac{ {j}^{\text{S, approx}}_z (z > 0)}{S_{0, z} v_{\text{T}}}
 &= \frac34 \int^{a^2}_0 \rmd w\sqrt{(1-w)\left[ 1- \left(c_{\text{T}}/v_{\text{T}}\right)^2 w\right]}, \\
 \frac{ {j}^{\text{S, approx}}_i (z > 0)}{S_{0, i} v_{\text{T}}}
 &= \frac38 \frac{c_{\text{T}}}{v_{\text{T}}} \int^{a^2}_0 \rmd w \, w \qquad (i = x, y). 
 \end{align}
\end{subequations}
Since $a\leq 1$ and $a \leq v_{\text{T}}/c_{\text{T}}$, it follows that
\begin{subequations}
 \begin{align}
 \int^{a^2}_0 \rmd w\sqrt{1-w} & \leq \int^1_0 \rmd w\sqrt{1-w} = \frac{2}{3}, \\
 \int^{a^2}_0 \rmd w \frac{w}{\sqrt{1-\left(c_{\text{T}}/v_{\text{T}}\right)^2w}}
 &\leq \int^{a^2}_0 \rmd w \frac{w}{\sqrt{1-w/a^2}} = \frac{4}{3}a^4 \leq \frac{4}{3}, \\
 \int^{a^2}_0 \rmd w\sqrt{(1-w)\left[ 1- \left(c_{\text{T}}/v_{\text{T}}\right)^2 w\right]}
 &\leq \int^1_0 \rmd w\sqrt{1-w} = \frac{2}{3}, \\
 \frac{c_{\text{T}}}{v_{\text{T}}} \int^{a^2}_0 \rmd w \, w & \leq 
 \int^{a^2}_0 \rmd w \frac{w}{a}  = \frac{a^3}{2} \leq \frac{1}{2}. 
 \end{align}
\end{subequations}
We therefore obtain
 \begin{align}
\frac{\displaystyle S^{\text{approx}}_i(z>0)}{\displaystyle S_{0, i}}& \leq
 \begin{cases}
 v_{\text{T}}/(2 c_{\text{T}}) & i = z\\
 1/2 & i= x, y
 \end{cases}, \\
 \frac{\displaystyle j^{\text{S, approx}}_i(z>0)}{ \displaystyle S_{0, i} v_{\text{T}}} &  \leq 
 \begin{cases}
 1/2 & i =z\\
 3/16 & i =x, y
 \end{cases}, 
\end{align}
which corresponds to Eqs.~(12a) and (12b) of the main text.

{\subsubsection{Assessment of the approximation in Eq.~\eqref{105616_5Oct26}}}
\label{sec:approximation_assessment}

{In Sec.~\ref{132237_5Oct26}, a simplification of the transmitted AM density was introduced in going from Eq.~\eqref{105635_5Oct26} to Eq.~\eqref{105616_5Oct26}. 
It is important to emphasize that this simplification should not be interpreted as a complete expansion to first order in $\chi$. 
In fact, several other $O(\chi)$ contributions have already been neglected when reducing the general expression in Eq.~\eqref{180813_16Jun26} to Eq.~\eqref{105635_5Oct26}, 
including the contributions from transmission between different modes, the frequency derivatives of the transmission coefficients, 
and terms involving the full transmission coefficients evaluated with the $O(k^2)$ splitting of the modes in the chiral crystal retained. 
Equation~\eqref{105616_5Oct26} therefore represents a simplified population-imbalance approximation rather than a systematic evaluation of all terms 
to first order in $\chi$.}

{To assess the effect of the additional approximation in Eq.~\eqref{105616_5Oct26}, we now retain the full expression in Eq.~\eqref{105635_5Oct26}, 
including the derivative term omitted in Eq.~\eqref{105616_5Oct26}. 
In the linear-dispersion regime, the zeroth-order contribution to $B_{+,\mathrm{T}}(\omega,\bm{k}_{\parallel})$ is
\begin{align}
B_{+,\mathrm{T}}(\omega,\bm{k}_{\parallel})
&= - \frac{\tau v_{\mathrm{T}}}{2}
\left[
\sqrt{1-\left(\frac{v_{\mathrm{T}} k_{\parallel}}{\omega}\right)^2}\,\hat{\bm z}
+\frac{v_{\mathrm{T}}\bm{k}_{\parallel}}{\omega}
\right]\cdot\frac{\bm{\nabla}T}{T}
\left.\frac{w}{\sinh^2 w}\right|_{w=\hbar\omega/(2k_{\mathrm B}T)}
+O(\chi^2), 
\end{align}
where we used $B_{\bm{k}n} = - (\tau \bm{v}_{\bm{k}n}\cdot \bm{\nabla}T)\, (\partial F^{(0)}_{\bm{k}n}/\partial T)$ and $\omega = \Omega_{\bm{k}\mathrm{T}}=v_{\mathrm{T}}k$. 
Differentiating this expression gives
\begin{align}
-\chi k^2
\left.
\frac{\partial B_{+,\mathrm{T}}(\omega,\bm{k}_{\parallel})}{\partial\omega}
\right|_{\omega=\Omega_{\bm{k}\mathrm{T}}}
&=\left.
\tau\chi k
\left(\frac{w}{\sinh w}\right)^2
\frac{\bm{\nabla}T}{T}\cdot
\left[
\frac{\hat{\bm z}}{2w(k_z/k)}
-\hat{\bm k}\coth w
\right]\right|_{w=\hbar\omega/(2k_{\mathrm B}T)}.
\end{align}
Combining this contribution with Eq.~\eqref{112142_19Jun26}, i.e., 
\begin{align}
B_{\bm{k},\mathrm{RH}}-B_{\bm{k},\mathrm{LH}}
&=\left.
\tau\chi k
\left(\frac{w}{\sinh w}\right)^2
\frac{\bm{\nabla}T}{T}\cdot\hat{\bm k}
\left(\coth w-\frac{3}{2w}\right)\right|_{w=\hbar\omega/(2k_{\mathrm B}T)}, \label{133543_5Oct26}
\end{align}
we obtain
\begin{equation}
B_{\bm{k},\mathrm{RH}}-B_{\bm{k},\mathrm{LH}}
-\chi k^2
\left.
\frac{\partial B_{+,\mathrm{T}}(\omega,\bm{k}_{\parallel})}{\partial\omega}
\right|_{\omega=\Omega_{\bm{k}\mathrm{T}}}
 = \left. \tau\chi k
\left(\frac{w}{\sinh w}\right)^2
\frac{\bm{\nabla}T}{T}\cdot
\left[
\frac{\hat{\bm z}}{2w(k_z/k)}
-\frac{3}{2w}\hat{\bm k}
\right]\right|_{w=\hbar\omega/(2k_{\mathrm B}T)}.
\label{153647_21Sep26}
\end{equation}
}

{The AM density in the achiral crystal is therefore modified from the result obtained using Eq.~\eqref{105616_5Oct26}
and Eq.~\eqref{112142_19Jun26} (i.e., Eq.~\eqref{133543_5Oct26}). 
The corresponding expression after including the derivative term is 
\begin{align}
S^i_{\mathrm{after}}(z>0)
&=
3\alpha(T)\frac{\partial T}{\partial r_i}
\int_0^1\frac{\mathrm d(\cos\theta_{\mathrm T})}{2}\,
\exp\left[
\frac{-z/\widetilde{\tau}c_{\mathrm T}}
{\sqrt{1-(c_{\mathrm T}/v_{\mathrm T})^2\sin^2\theta_{\mathrm T}}}
\right]
\nonumber\\
&\quad\times
\frac{(v_{\mathrm T}/c_{\mathrm T})\cos\theta_{\mathrm T}}
{\sqrt{1-(c_{\mathrm T}/v_{\mathrm T})^2\sin^2\theta_{\mathrm T}}}
\Delta\mathcal T(\theta_{\mathrm T})
\nonumber\\
&\quad\times
\begin{cases}
\left(
\dfrac{1}{2\cos\theta_{\mathrm T}}
-\dfrac{3}{2}\cos\theta_{\mathrm T}
\right)
\sqrt{1-(c_{\mathrm T}/v_{\mathrm T})^2\sin^2\theta_{\mathrm T}},
& i=z,\\[6pt]
-\dfrac{3}{2}\dfrac{c_{\mathrm T}}{v_{\mathrm T}}
\dfrac{\sin^2\theta_{\mathrm T}}{2},
& i=x,y,
\end{cases}
\label{184349_21Sep26}
\end{align}
where the azimuthal and radial integrals have been performed analytically.}

{For comparison, the result obtained from Eq.~\eqref{105616_5Oct26} is 
\begin{align}
S^i_{\mathrm{before}}(z>0)
&=
3\alpha(T)\frac{\partial T}{\partial r_i}
\int_0^1\frac{\mathrm d(\cos\theta_{\mathrm T})}{2}\,
\exp\left[
\frac{-z/\widetilde{\tau}c_{\mathrm T}}
{\sqrt{1-(c_{\mathrm T}/v_{\mathrm T})^2\sin^2\theta_{\mathrm T}}}
\right]
\nonumber\\
&\quad\times
\frac{(v_{\mathrm T}/c_{\mathrm T})\cos\theta_{\mathrm T}}
{\sqrt{1-(c_{\mathrm T}/v_{\mathrm T})^2\sin^2\theta_{\mathrm T}}}
\Delta\mathcal T(\theta_{\mathrm T})
\begin{cases}
\cos\theta_{\mathrm T}
\sqrt{1-(c_{\mathrm T}/v_{\mathrm T})^2\sin^2\theta_{\mathrm T}},
& i=z,\\[4pt]
\dfrac{c_{\mathrm T}}{v_{\mathrm T}}
\dfrac{\sin^2\theta_{\mathrm T}}{2},
& i=x,y, 
\end{cases}
\label{184951_21Sep26}
\end{align}
which is equivalent to Eq.~\eqref{112957_19Jun26}.}

{For the components parallel to the interface, Eqs.~\eqref{184349_21Sep26} and \eqref{184951_21Sep26} give the simple relation
\begin{equation}
S^i_{\mathrm{after}}(z>0)
= 
-\frac{3}{2}S^i_{\mathrm{before}}(z>0),
\qquad i=x,y.
\end{equation}
Thus, inclusion of the derivative term changes the parallel components by an $O(1)$ factor and reverses their sign.}

{As an instructive limiting case, consider $v_{\mathrm T}=c_{\mathrm T}$ and
\begin{equation}
\mathcal T_{\mathrm{RH},\mathrm{RH}}=1,
\qquad
\mathcal T_{\mathrm{RH},\mathrm{LH}}=0,
\label{185914_21Sep26}
\end{equation}
so that $\Delta\mathcal T(\theta_{\mathrm T})=1$. At the interface, $z=+0$, we then obtain
\begin{align} 
S^z_{\mathrm{before}}(+0)
&=
\frac{\alpha(T)}{2}\frac{\partial T}{\partial z} = \frac{S^z_{0}}{2},
&
S^{x,y}_{\mathrm{before}}(+0)
&=
\frac{\alpha(T)}{2}\frac{\partial T}{\partial r_{x,y}} = \frac{S_0^{x, y}}{2}, \\
S^z_{\mathrm{after}}(+0)
&=0,
&
S^{x,y}_{\mathrm{after}}(+0)
&=
-\frac{3}{2}S^{x,y}_{\mathrm{before}}(+0).
\end{align}
Here, $\bm{S}_0 = \alpha(T) \bm{\nabla}T$ is the bulk AM density in the chiral crystal, introduced in the main text [Eq.~(9)].
Hence, even in this idealized case, the inclusion of the derivative term qualitatively changes the result.}

\begin{figure}[tbp]
\centering
\includegraphics[pagebox=artbox,width=0.99\columnwidth]{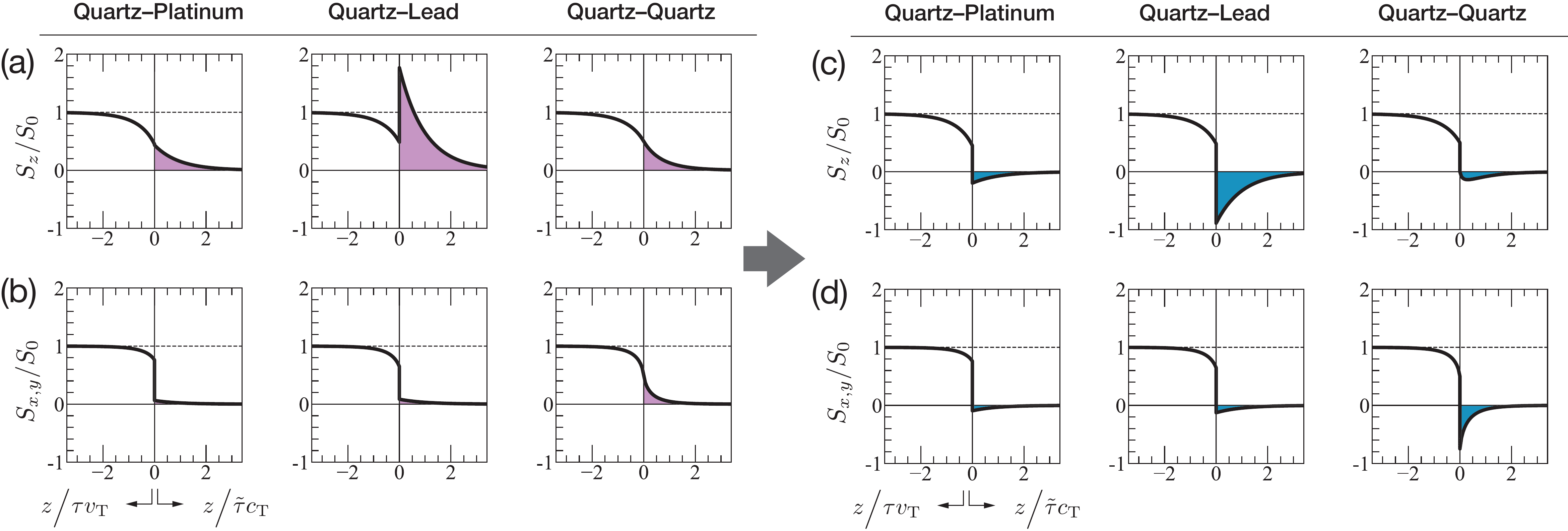}
\caption{
{Spatial profiles of the AM density $\bm S(z)$ before and after including the derivative term omitted in Eq.~\eqref{105616_5Oct26}.
Panels (a) and (b) show the result obtained from Eq.~\eqref{105616_5Oct26}, corresponding to the spatial profiles presented in Fig.~2 of the Letter, 
while (c) and (d) show the result obtained from Eq.~\eqref{184349_21Sep26}.
Here, ``Quartz--Quartz'' denotes a junction between chiral quartz for $z<0$ and an achiral crystal for $z>0$ 
having the same long-wavelength elastic constants as quartz but no transverse-mode splitting.}
}
\label{fig:transmittedSAMdensity_revision}
\end{figure}

{
Performing the numerical angular integration of 
Eqs.~\eqref{184349_21Sep26} and \eqref{184951_21Sep26} 
and using the same parameters as in the calculation to obtain Fig.~2 of the main text, 
we obtain the spatial profiles shown in Fig.~\ref{fig:transmittedSAMdensity_revision}. 
Compared with the result obtained from Eq.~\eqref{105616_5Oct26}, the inclusion of the derivative term produces a substantial change in the AM density. 
In particular, the components parallel to the interface are modified by an $O(1)$ factor and reverse sign relative to the simplified result. 
The revised result exhibits a pronounced tendency toward a sign reversal of the AM density upon transmission into the achiral crystal.}

{These results demonstrate that the derivative term omitted in Eq.~\eqref{105616_5Oct26} is not generically a small quantitative correction. 
At the same time, restoring this term alone does not provide a complete first-order result, because the other $O(\chi)$ contributions already omitted 
in obtaining Eq.~\eqref{105635_5Oct26} remain neglected. 
Consequently, the numerical prefactor, sign, and detailed spatial profile obtained from Eq.~\eqref{105616_5Oct26} should not be regarded as quantitatively robust.}

{The role of the approximation in Eq.~\eqref{105616_5Oct26} is therefore to isolate the contribution associated with the chirality-induced RH--LH population imbalance 
and to obtain a tractable analytic description of the resulting AM transfer. 
The spatial profiles and upper bounds presented in the main text and this Supplemental Material 
should accordingly be regarded as representative results within this simplified population-imbalance approximation, rather than as quantitatively exact predictions.}

{Importantly, this limitation does not affect the general formulation of interfacial phonon AM transfer developed in the Letter. 
The boundary conditions relating the nonequilibrium phonon distributions through mode-resolved reflection and transmission coefficients are established independently 
of the approximation in Eq.~\eqref{105616_5Oct26}. 
The general first-order formulation further shows that the chirality-dependent transferred AM is of order $O(\chi)$, as evident from Eq.~\eqref{180813_16Jun26}. 
Thus, while the quantitative magnitude and spatial profile obtained from Eq.~\eqref{105616_5Oct26}
are sensitive to the neglected $O(\chi)$ terms, the underlying interfacial-transfer mechanism remains intact.}

\subsection{AM transport across junctions of chiral crystals with the same or opposite chiralities}

A chiral crystal and its mirror image cannot be superposed by any proper rotation and are said to possess opposite chiralities.
In the present model, the difference in chirality is reflected in the sign of the parameter $\chi$, which characterizes the deviation from the linear acoustic dispersion.
Equivalently, the ordering of the split RH and LH branches is reversed when the chirality is reversed.

A natural question is therefore whether AM transport across an interface depends qualitatively on whether the two crystals have the same or opposite chiralities.
As we show below, AM can be transmitted across the interface with the same order of magnitude in both cases.
Our starting point is Eq.~\eqref{180813_16Jun26}, which gives the AM density in the region $z>0$ while keeping the RH and LH modes distinct.

\subsubsection{Same chirality}

We first consider two identical chiral crystals with the same chirality joined at $z=0$.
The temperature gradient is assumed to be finite in the region $z<0$ and absent in the region $z>0$.

Although the temperature gradient changes discontinuously at the interface, the elastic medium itself is continuous across $z=0$.
Consequently, there is no reflection of elastic waves or phonons at the interface.
The transmittance is unity only for the same mode and vanishes otherwise.
Since this statement holds at every frequency,
\begin{align}
\mathcal{T}_{nm} &= \mathcal{T}^{(\chi)}_{nm} = \delta_{nm}, 
&\frac{\partial \mathcal{T}_{nm}(\omega, \bm{k}_{\parallel})}{\partial \omega} &= 0.  \label{010002_22Jun26}
\end{align}
Substituting these relations into Eq.~\eqref{180813_16Jun26} immediately gives
\begin{equation}
 {\bm{S}} (z>0) \simeq
\int\limits_{q_z > 0} \frac{\rmd^3\bm{q}}{(2\pi)^3} \hbar \hat{\bm{q}}\, \exp\left({\displaystyle \frac{-z}{\widetilde{\tau} c_{\text{T}} \hat{q}_z}}\right) 
\left[B_{+, \text{RH}}(\omega_{\bm{q}, \text{RH}}, \bm{q}_{\parallel}) - B_{+, \text{LH}}(\omega_{\bm{q}, \text{LH}}, \bm{q}_{\parallel}) \right]. 
\end{equation}

Because the two crystals are identical, $\Omega_{\bm{k}n}=\omega_{\bm{k}n}$ and $v_{\mathrm T}=c_{\mathrm T}$.
Changing the integration variable from $\bm q$ to $\bm k$ and using
$B_{+, l}(\Omega_{\bm{k}, l}, \bm{k}_{\parallel}) = B_{\bm{k}, l} \ (k_z>0)$, we obtain
\begin{align}
{\bm{S}} (z>0) &\simeq
\int\limits_{k_z > 0} \frac{\rmd^3\bm{k}}{(2\pi)^3} \hbar \hat{\bm{k}}\, \exp\left({\displaystyle \frac{-z}{\widetilde{\tau} v_{\text{T}} \hat{k}_z}}\right) 
\left[B_{\bm{k}, \text{RH}} - B_{\bm{k}, \text{LH}}\right]  \nonumber\\
&= \underbrace{\frac{4\pi}{(2\pi)^3} \frac{\hbar \tau\chi }{T}\left(\frac{2 k_{\text{B}}T}{\hbar v_{\text{T}}}\right)^4 \int^{\infty}_0 \rmd w\, w^3 Q(w)}_{3\alpha (T)}
\int^1_0 \frac{\rmd \cos\theta }{2} 
\exp\left[ \frac{-z/\widetilde{\tau} v_{\text{T}}}{\cos\theta}\right] \sum_{j = x, y, z}
\Braket{\frac{\bm{k} k_j}{k^2}} \frac{\partial T}{\partial r_j} \nonumber\\
&= 
\frac{\bm{S}_0}{2} \int^1_0 \rmd w\, \exp \left(\frac{-z/\widetilde{\tau} v_{\text{T}}}{w}\right)
\begin{cases}
3w^2 & (i = z), \\ 
\frac{3}{2} (1-w^2) & (i = x, y)
\end{cases}. 
\label{213924_20Jun26}
\end{align}
In deriving this expression, Eqs.~\eqref{112142_19Jun26} and \eqref{182319_19Jun26} were substituted, together with the bulk relation
$\bm{S}_0=\alpha(T)\bm{\nabla} T$.

At the plane where the temperature gradient is switched on and off ($z=+0$), the AM density is exactly one-half of the bulk value,
${\bm{S}} (z = +0) = \bm{S}_0 /2$, independent of the polarization direction $i=x,y,z$.
This result can also be understood from a simple symmetry argument.
Consider three situations:
(i) the temperature gradient is applied only in $z<0$;
(ii) it is applied only in $z>0$;
and (iii) it is applied uniformly throughout the entire system.
A $180^{\circ}$ rotation that maps $z\rightarrow -z$ interchanges cases (i) and (ii).
Therefore, there exists a common coefficient $\lambda$ such that ${\bm{S}} (z = 0) = \lambda \bm{S}_0$ for both (i) and (ii).
Case (iii) is simply the superposition of (i) and (ii). Within linear response, one therefore has ${\bm{S}} (z = 0) = 2\lambda \bm{S}_0$. 
On the other hand, in case (iii), the uniform temperature gradient generates the bulk value everywhere,
${\bm{S}} (z) =  \bm{S}_0$. Hence $\lambda=1/2$.

\subsubsection{Opposite chiralities}

We next consider two chiral crystals with opposite chiralities joined at $z=0$.
The crystal occupying $z>0$ is the mirror image of that occupying $z<0$.
The phonon dispersions in the region $z>0$ are obtained from those in Eq.~\eqref{153258_20Jun26} by exchanging the RH and LH branches:
\begin{align}
\omega_{\bm{q}, \text{L}} &= \left.\Omega_{\bm{k},\text{T}}\right|_{\bm{k} = \bm{q}} = v_{\text{L}} q, \\
\omega_{\bm{q}, \text{RH}} &= \left.\Omega_{\bm{k},\text{LH}}\right|_{\bm{k} = \bm{q}} \simeq v_{\text{T}}q - {\chi q^2}/{2} , \\
\omega_{\bm{q}, \text{LH}} &= \left.\Omega_{\bm{k},\text{RH}}\right|_{\bm{k} = \bm{q}} \simeq v_{\text{T}}q + {\chi q^2}/{2}.
\label{211842_20Jun26}
\end{align}

Let us first consider the transmittance $\mathcal T_{nm}$ in the limit $\chi\to0$, where the splitting is neglected.
Since the two crystals are otherwise identical, the transmittance satisfies relations [see also Eq.~\eqref{010002_22Jun26}]:
\begin{align}
\mathcal{T}_{nm} &= \delta_{nm}, 
&\frac{\partial \mathcal{T}_{nm}(\omega, \bm{k}_{\parallel})}{\partial \omega} &= 0. \label{211746_20Jun26}
\end{align}

\begin{figure}[tbp]
 \centering
\includegraphics[pagebox=artbox,width=0.99\textwidth]{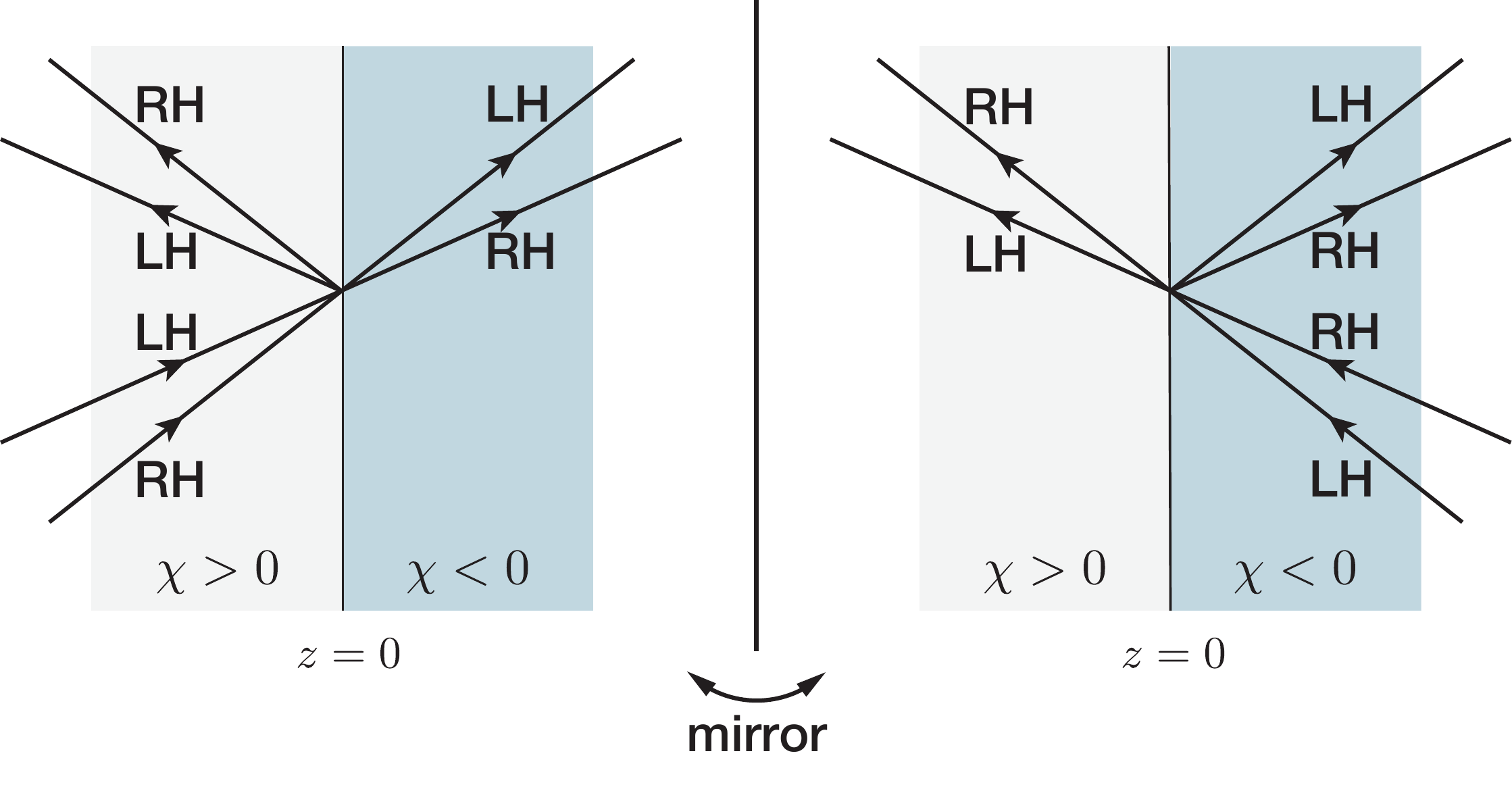}
\caption{Schematic illustration of interfacial scattering at a junction between chiral crystals with opposite chiralities. 
The two panels are related by mirror reflection. 
Owing to the splitting of the RH and LH dispersions, the incidence and refraction angles of the two transverse modes differ slightly.}
\label{181307_20Jun26}
\end{figure}

The transmittance $\mathcal T^{(\chi)}_{nm}$ that includes the splitting no longer satisfies Eq.~\eqref{211746_20Jun26}.
However, they obey a different symmetry.
Figure~\ref{181307_20Jun26} illustrates interface scattering between two crystals with opposite chiralities, characterized by $\chi>0$ and $\chi<0$.
The entire junction is invariant under mirror reflection.
Indeed, the mirror reflection exchanges the two crystals occupying $z<0$ and $z>0$ while simultaneously reversing the sign of $\chi$.

Under the same mirror operation, an RH (LH) mode incident from the $z<0$ side is mapped onto an LH (RH) mode incident from the $z>0$ side.
The scattered states are transformed in the same manner.
This follows because RH and LH phonons are distinguished by the sign of the scalar product between the wavevector and the phonon AM, 
and mirror reflection reverses the sign of this pseudoscalar quantity.

Because mirror-related scattering processes must have identical power reflectance and transmittance, the transmittances for waves incident from the $z<0$ side,
$\mathcal T^{(\chi)}$,
and those for waves incident from the $z>0$ side, $\mathcal T^{\prime(\chi)}$,
satisfy
\begin{equation}
 \mathcal{T}^{(\chi)}_{n m} =  \mathcal{T}^{\displaystyle\prime\, \scriptstyle (\chi)}_{\overline{n}\, \overline{m}}, \qquad \text{with}\quad \overline{n}\equiv 
\begin{cases}
\text{L}  & (n = \text{L}), \\
\text{LH}  & (n = \text{RH}), \\
\text{RH}  & (n = \text{LH}).
\end{cases}
\end{equation}
Combining this relation with Helmholtz reciprocity, $\mathcal{T}^{(\chi)}_{\displaystyle m n} =\mathcal{T}^{\displaystyle\prime\, \scriptstyle (\chi)}_{\displaystyle nm}$, 
one obtains $\mathcal{T}^{(\chi)}_{\displaystyle \overline{n}\, \overline{m}} =\mathcal{T}^{(\chi)}_{\displaystyle m n}$. 
More explicitly, 
\begin{align}
 \mathcal{T}^{(\chi)}_{\text{RH}, \text{RH}} &=  \mathcal{T}^{(\chi)}_{\text{LH}, \text{LH}}, 
&\mathcal{T}^{(\chi)}_{\text{RH}, \text{L}} &=  \mathcal{T}^{(\chi)}_{\text{L}, \text{RH}}, 
&\mathcal{T}^{(\chi)}_{\text{L}, \text{RH}} &=  \mathcal{T}^{(\chi)}_{\text{LH}, \text{L}}. \label{211735_20Jun26}
\end{align}

Using Eqs.~\eqref{211746_20Jun26} and \eqref{211735_20Jun26}, several terms in Eq.~\eqref{180813_16Jun26} cancel, yielding
\begin{align}
 &{\bm{S}} (z>0) \simeq
\int\limits_{q_z > 0} \frac{\rmd^3\bm{q}}{(2\pi)^3} \hbar \hat{\bm{q}}\, \exp\left({\displaystyle \frac{-z}{\widetilde{\tau} c_{\text{T}} \hat{q}_z}}\right) \nonumber\\
&\times\Big\{
 \left[\mathcal{T}^{(\chi)}_{\text{L}, \text{LH},}(\omega_{\bm{q},\text{RH}}, \bm{q}_{\parallel}) 
- \mathcal{T}^{(\chi)}_{\text{L}, \text{RH}}(\omega_{\bm{q},\text{LH}}, \bm{q}_{\parallel}) \right] {B}_{+, \text{L}}(\omega_{\bm{q}, \text{T}}, \bm{q}_{\parallel})
 \nonumber\\
& + B_{+, \text{RH}}(\omega_{\bm{q},\text{RH}}, \bm{q}_{\parallel}) - B_{+, \text{LH}}(\omega_{\bm{q},\text{LH}}, \bm{q}_{\parallel}) \nonumber\\
& + \left[ \mathcal{T}^{(\chi)}_{\text{RH}, \text{LH}} (\omega_{\bm{q},\text{T}}, \bm{q}_{\parallel})
- \mathcal{T}^{(\chi)}_{\text{LH}, \text{RH}} (\omega_{\bm{q},\text{T}}, \bm{q}_{\parallel}) \right] 
B_{+, \text{T}}(\omega_{\bm{q},\text{T}}, \bm{q}_{\parallel}) 
\Big\}.
\end{align}
Using Eq.~\eqref{211842_20Jun26}, together with $c_{\mathrm T}=v_{\mathrm T}$,
and changing the integration variable from $\bm q$ to $\bm k$, the expression can be rewritten as
\begin{align}
 {\bm{S}} (z>0) &\simeq
\int\limits_{k_z > 0} \frac{\rmd^3\bm{k}}{(2\pi)^3} \hbar \hat{\bm{k}}\, \exp\left({\displaystyle \frac{-z}{\widetilde{\tau} v_{\text{T}} \hat{k}_z}}\right) \nonumber\\
&\times\Big\{
B_{\bm{k}, \text{RH}} - B_{\bm{k}, \text{LH}} 
- 2\chi k^2 \left.\frac{\partial B_{+, \text{T}}(\omega, \bm{k}_{\parallel})}{\partial \omega}\right|_{\omega = \Omega_{\bm{k}, \text{T}}} 
\nonumber\\
& + \left[\mathcal{T}^{(\chi)}_{\text{L}, \text{LH}}(\Omega_{\bm{k},\text{LH}}, \bm{k}_{\parallel}) 
- \mathcal{T}^{(\chi)}_{\text{L}, \text{RH}}(\Omega_{\bm{k},\text{RH}}, \bm{k}_{\parallel}) \right] {B}_{+, \text{L}}(\Omega_{\bm{k}, \text{T}}, \bm{k}_{\parallel})
\nonumber\\
& + \left[ \mathcal{T}^{(\chi)}_{\text{RH}, \text{LH}} (\Omega_{\bm{k},\text{T}}, \bm{k}_{\parallel})
- \mathcal{T}^{(\chi)}_{\text{LH}, \text{RH}} (\Omega_{\bm{k},\text{T}}, \bm{k}_{\parallel}) \right] 
B_{+, \text{T}}(\Omega_{\bm{k},\text{T}}, \bm{k}_{\parallel}) 
\Big\}. 
\label{214119_20Jun26}
\end{align}
The following relation has also been used:
\begin{align}
& B_{+, \text{RH}}(\Omega_{\bm{k},\text{LH}}, \bm{k}_{\parallel}) - B_{+, \text{LH}}(\Omega_{\bm{k},\text{RH}}, \bm{k}_{\parallel}) \nonumber\\
&=
\left[
B_{+, \text{RH}}(\Omega_{\bm{k},\text{LH}}, \bm{k}_{\parallel}) - B_{+, \text{RH}}(\Omega_{\bm{k},\text{RH}}, \bm{k}_{\parallel}) 
\right]
+ \left[
B_{+, \text{RH}}(\Omega_{\bm{k},\text{RH}}, \bm{k}_{\parallel}) - B_{+, \text{LH}}(\Omega_{\bm{k},\text{LH}}, \bm{k}_{\parallel}) 
\right] \nonumber\\
&\qquad + \left[
B_{+, \text{LH}}(\Omega_{\bm{k},\text{LH}}, \bm{k}_{\parallel}) - B_{+, \text{LH}}(\Omega_{\bm{k},\text{RH}}, \bm{k}_{\parallel}) 
\right] \nonumber\\
&= 
(\Omega_{\bm{k},\text{LH}} - \Omega_{\bm{k},\text{RH}})
 \left.\frac{\partial B_{+, \text{RH}}(\omega, \bm{k}_{\parallel})}{\partial \omega}\right|_{\omega = \Omega_{\bm{k}, \text{T}}}
+ B_{\bm{k}, \text{RH}} - B_{\bm{k}, \text{LH}} \nonumber\\
&\qquad + (\Omega_{\bm{k},\text{LH}} - \Omega_{\bm{k},\text{RH}}) 
\left.\frac{\partial B_{+, \text{LH}}(\omega, \bm{k}_{\parallel})}{\partial \omega}\right|_{\omega = \Omega_{\bm{k}, \text{T}}} 
+ O(\chi^3) \nonumber\\
&= B_{\bm{k}, \text{RH}} - B_{\bm{k}, \text{LH}}
- 2\chi k^2 
\left.\frac{\partial B_{+, \text{T}}(\omega, \bm{k}_{\parallel})}{\partial \omega}\right|_{\omega = \Omega_{\bm{k}, \text{T}}} + O(\chi^3). 
\end{align}

An important conclusion follows.
Even when the chirality changes sign across the interface, a finite phonon AM density diffuses into 
the neighboring crystal according to Eq.~\eqref{214119_20Jun26}.
Inspection of the integrand of Eq.~\eqref{214119_20Jun26} shows that this transport originates from two distinct mechanisms:
(i)~the RH--LH imbalance of the phonon distribution generated in the bulk region subjected to the temperature gradient, and
(ii)~the RH--LH asymmetry of the mode-resolved transmission coefficients.
In particular, the contribution proportional to
$B_{\bm{k}, \text{RH}} - B_{\bm{k}, \text{LH}} $
is exactly the same as that appearing in the same-chirality junction [Eq.~\eqref{213924_20Jun26}]. 
Therefore, reversing the chirality across the interface does not significantly reduce the AM transfer. 
The transmitted AM remains of the same order as that in the same-chirality junction.

\end{document}